\documentclass[aps,pra,twocolumn,superscriptaddress,longbibliography]{revtex4-1}%
\usepackage{graphicx}
\usepackage{float}
\usepackage{dcolumn}
\usepackage{bm,color}
\usepackage{amsmath,amssymb,dsfont,amstext,amsfonts}
\usepackage{extarrows}
\usepackage[colorlinks=true,linkcolor=blue,urlcolor=blue,citecolor=blue]{hyperref}
\usepackage{xcolor}
\usepackage{wasysym}
\usepackage{mathtools,tabstackengine}
\usepackage{bbold}
\usepackage{amsthm}
\usepackage{mathtools}
\usepackage{tikz-cd}

\usepackage[normalem]{ulem} %makes underlining possible: \uline \uwave \sout
\usepackage{color}%makes \textcolor{grey}{text} available

\newcommand{\parag}[1]{\vspace{0.3em}{\it #1.}}

\def\bra#1{\mathinner{\langle{#1}|}}
\def\ket#1{\mathinner{|{#1}\rangle}}
\def\inner#1#2{\mathinner{\langle{#1}|{#2}\rangle}}

\def\Bra#1{\left\langle#1\right|}
\def\Ket#1{\left|#1\right\rangle}

\def\re{\mathrm{Re}\,}

\def\tr{\mathrm{Tr}}
\def\dd{\mathrm{d}}
\def\id{\mathbb{1}} % need bbold,bm package

\def\mod{\:\mathrm{mod}\:}
\def\bs#1{\boldsymbol{#1}}

\newcommand{\ie}{i.e.\;}
\newcommand{\eg}{e.g.\;}
\newcommand{\imi}{\,\text{i}}

  \newcommand\SmallMatrix[1]{{%
\tiny\arraycolsep=0.3\arraycolsep\ensuremath{\begin{bmatrix}#1\end{bmatrix}}}}

\makeatletter
\newcommand{\subalign}[1]{%
  \vcenter{%
    \Let@ \restore@math@cr \default@tag
    \baselineskip\fontdimen10 \scriptfont\tw@
    \advance\baselineskip\fontdimen12 \scriptfont\tw@
    \lineskip\thr@@\fontdimen8 \scriptfont\thr@@
    \lineskiplimit\lineskip
    \ialign{\hfil$\m@th\scriptstyle##$&$\m@th\scriptstyle{}##$\hfil\crcr
      #1\crcr
    }%
  }%
}
\makeatother

\begin{document}
\title{Quantum geometry beyond projective single bands}
\author{Adrien Bouhon}
\email{adrien.bouhon@gmail.com} 
\affiliation{TCM Group, Cavendish Laboratory, University of Cambridge, J.J.~Thomson Avenue, Cambridge CB3 0HE, United Kingdom}
\author{Abigail Timmel}
\affiliation{TCM Group, Cavendish Laboratory, University of Cambridge, J.J.~Thomson Avenue, Cambridge CB3 0HE, United Kingdom}
\email{ant28@cam.ac.uk}
\author{Robert-Jan Slager}
\email{rjs269@cam.ac.uk}
\affiliation{TCM Group, Cavendish Laboratory, University of Cambridge, J.J.~Thomson Avenue, Cambridge CB3 0HE, United Kingdom}

%%TC:ignore
\date{\today}
\begin{abstract}
The past few years have seen a revived interest in quantum geometrical characterizations of band structures due to the rapid development of topological insulators and semi-metals. Although the metric tensor has been connected to many geometrical concepts for single bands, the exploration of these concepts to a multi-band paradigm still promises a new field of interest. Formally, multi-band systems, featuring in particular degeneracies, have been related to projective spaces, explaining also 
the success of relating quantum geometrical aspects of flat band systems, albeit usually in the single band picture. Here, we propose 
a different route involving Pl\"ucker embeddings to represent arbitrary classifying spaces, being the essential objects that encode $all$ the relevant topology.
This paradigm allows for the quantification of geometrical quantities directly in readily manageable vector spaces that a priori do not involve projectors or the need of flat band conditions. As a result, our findings are shown to pave the way for identifying new geometrical objects and defining metrics in arbitrary multi-band systems, especially beyond the single flatband limit, promising a versatile tool that can be applied in contexts that range from response theories to finding quantum volumes and bounds on superfluid densities as well as possible quantum computations.
\end{abstract}
\maketitle
%%TC:endignore

\parag{Introduction}
The past decade has witnessed an ever increasing interest in topological matter. Topological insulators and semi-metals provide~\cite{Rmp1,Rmp2,Weyl} in this regard a direct route to mimic intrinsic topological features and, owing to the inclusion of (crystalline) symmetries, a broad range of results have been achieved~\cite{Clas1, Clas2, Clas3, Clas4, Clas5, Codefects2, unsupcom, afmmagnetic,UnifiedBBc, Kitaev, Schnyder08,song2018quantitative, magnetic, mtqc, mSI,Kemp_nested_2022,volovik2018investigation,wannieroptical,SchnyderClass}. 
While on a case by case bases the synergy between topology and geometry has been rather well established, the past years have seen a reinvigorated interest in the geometrical side.
Notably, following the earlier construction of quantum geometric tensors~\cite{provost1980riemannian, resta_2011_metric}, with a distance measuring real part and imaginary part that encapsulates generalized Berry phases, this object has been in particular useful in the context of topological band theory. Examples include probing single band invariants, such as Chern numbers, in which circularly-polarized light couples to the right elements of the metric tensor to render a response proportional to the determining invariant~\cite{ozawa_2018,Tran_2017,palumbo_goldman2020,deJuan_2017,meraozawa, Palumbo_tensor_2021} and recent relations with quantum metrology~\cite{YuUnal_2022,Li_2022_metrology}. The latter is particularly appealing in the context of quantum simulators and quenched systems and can even be related to e.g. Cramér-Rao information bounds. In addition, given the surge of interest in flatband physics, notably in the context of twisted multi-layered Van der Waals systems, also relations that bound the superfluid density by general correspondences to the Fubini-Study metric at zero temperature have been receiving increasing interest as well as recent pursuits that relate this density to Wannier properties and real space invariants~\cite{peotta2015superfluidity,torma2021superfluidity,bergholtz2013,rhim2020quantum,herzog2022superfluid, parker2021field,julku2016,peri2021}.

In view of all these physical implications a general multi-band approach promises a widely applicable and powerful tool. This relevance is not in the least place also motivated by recent progress on the topological side where recent studies have shown the existence of multi-gap dependent topological phases that culminate in new invariants, such as Euler class~\cite{bouhon2019nonabelian, BJY_nielsen, bouhonGeometric2020}. Multi-gap topologies have been seen in several meta-material studies~\cite{Jiang1Dexp, Guo1Dexp, Jiang_meron, Jiang2021, zhao2022observation, Park2021}, and the proposed unusual quench dynamical behavior of Euler class phases~\cite{Eulerdrive} has been observed in trapped ion insulators~\cite{zhao2022observation}. Similarly, recent proposals in phonon bands and electronic structures under strain have shown to provide feasible routes towards real materials~\cite{peng2022multi, konye2021, Bouhon_4d2023, Peng2021, Lange2021, Lange2022} and novel anomalous phases~\cite{AnEuler2022}.

While there has been recent progress on multi-gap geometrical characterizations, such as the observation that general dipole transitions can be seen as tangent vectors of a Riemannian manifold~\cite{Ahn_2021rio}, we here point out a general {\it explicit} route towards formulating geometric aspects of multi-band systems. Similarly, our approach has no inherent redundancy that comes from taking a specific basis of eigenstates that can be relabeled and thus impose a huge gauge degree of freedom as inherent to recent other setups~\cite{ma_2010}. Namely, our construction thrives on the key insight that one can directly use the classifying space, the minimal manifold that encodes all topological information without redundant degrees of freedom \footnote{More precisely, the classifying space is the homogeneous space $\mathcal{C}=\mathsf{G}/\mathsf{H}$ that captures the gauge structure of the spectral decomposition of the Bloch Hamiltonian, with $\mathsf{G}$ the Lie group of diagonalizing matrices and $\mathsf{H}$ the group of gauge transformations that preserve the spectral decomposition between disconnected bands. Then, the homotopy classes of Bloch Hamiltonian of $d$-dimensional systems, $[\mathbb{T}^d,\mathcal{C}]$, can be decomposed into the homotopy groups $\bigoplus_n \pi_n[\mathcal{C}]$.}, and specific characterizations, going by the name of Pl\"ucker embeddings, to formulate a general approach to geometric signatures.
A Pl\"ucker embedding, in essence, is a mapping from a Grassmannian or Flag manifold, which will be shown underneath to be the effective objects to define topology for multi-band settings, into a higher dimensional vector space. We have already shown in recent work~\cite{bouhonGeometric2020, Bouhon2022braiding2} that Pl\"ucker mappings can be employed to define arbitrary multi-gap topological models. Here, however, we take full profit from the geometrical power of this approach that allows for a characterization in terms of straightforward vector spaces.
Indeed, as these vector spaces come with very natural geometrical identifications, these embeddings provide for a direct and universal route to define geometrical tensors and properties, providing opportunity to generalize the above mentioned physical impact of the single-band metric to the multi-band context.

This paper proceeds as follows. First we review the single-band metric in the case of band systems having Chern numbers. This provides a good starting point to introduce the Pl\"ucker formalism and address the true generalization of single band geometrical notions to a multi-band context, notably in the Chern number context. We then outline how these characterizations offer new routes to model arbitrary topologies, the content of which is then concretely analyzed from a Riemannaian geometrical perspective. We finally highlight the tractable nature of our framework by addressing recently discovered multi-gap topologies in specific models that necessitate descriptions that go beyond single-band formulations and even showcase various general applications that pave the way for future pursuits that will eminently profit from our general perspective.

\section{Review of single band metric and Chern number} To set the stage we first recall the single band metric formalism in the case of Chern bands. Each band can be viewed as a submanifold of the projective Hilbert space $\mathbb{C}\mathbb{P}^{N-1}$, which is the complex space $\mathbb{C}^N$ where all vectors related by complex scalar multiplication are identified.  We will use $\Ket{\tilde{u}}$ to denote a vector in $\mathbb{C}\mathbb{P}^{N-1}$ and $\Ket{u}$ to denote a normalized but gauge-dependent representative in $\mathbb{C}^N$.  This manifold of quantum states is parametrized by crystal momentum $k$, thus assigning a state $\Ket{\tilde{u}(k)} \in \mathbb{C}\mathbb{P}^{N-1}$ to each point of the Brillouin zone.  On this manifold one may define a natural metric $g_{ij}$ and symplectic form $\omega_{ij}$,
\begin{eqnarray}
g_{ij} &=& \inner{\partial_i \tilde{u}}{ \partial_j \tilde{u}} + c.c. \nonumber\\
\omega_{ij} &=& \inner{\partial_i \tilde{u}}{ \partial_j \tilde{u}} - c.c
\end{eqnarray}
The tangent space of $\mathbb{C}\mathbb{P}^{N-1}$ at a point $\Ket{\tilde{u}}$ differs from that of $\mathbb{C}^N$ at $\Ket{u}$ by the absence of a vector along $\Ket{u}$ due to the identification of all complex multiples of $\Ket{u}$, so we can write these quantities for a representative $\Ket{u} \in \mathbb{C}^N$ by subtracting this component.  Defining $Q = \id - \Ket{u}\Bra{u}$ as the projector into the tangent space of $\mathbb{C}\mathbb{P}^{N-1}$,  we then obtain   
\begin{eqnarray}\label{eq:singlebandmetric}
g_{ij} &=& \Bra{ \partial_i u}Q\Ket{\partial_j u} + c.c. \nonumber\\                                
\omega_{ij} &=& \Bra{ \partial_i u}Q\Ket{\partial_j u} - c.c.
\end{eqnarray}
These form the symmetric and anti-symmetric components of the Fubini-Study metric, respectively.  Both the Berry curvature $\omega$ and the quantum metric $g$ describe geometric features of the band with far-reaching consequences on electronic behavior that we have already alluded to above. Indeed, examples include bounds on superfluidity, optical responses when light couples to the right elements to render a response proportional to the Chern number and quantum metrology setups.  

When considered from a more mathematical perspective, the curvature $\omega$ has a further description in terms of a connection on $\mathbb{C}^N$ viewed as a principle $U(1)$-bundle.  A principle $G$-bundle consists of a total space, in this case $\mathbb{C}^N$, a projection map $\pi$ to a base space $\mathbb{C}\mathbb{P}^{N-1}$, and a group $G$ acting on the total space, here $U(1)$ by scalar multiplication.  Further requirements are that the pre-image of each neighborhood $U$ in the base space is diffeomorphic to $U \times G$, and the group action translates purely within the $G$ component of this trivialization.  A connection $A$ is a Lie algebra-valued one-form that maps vertical tangent vectors, those which translate purely within the fiber $G$, to their corresponding element of the Lie algebra.  The kernel of $A$ then defines the horizontal section of the tangent space, i.e. vectors that translate purely along base space.  In the case of our $U(1)$ bundle, we have the Berry connection
\begin{equation}
A = \inner{ u }{ \partial_i u}
\end{equation}                                                                                      
This indeed gives the component of the tangent vector $\Ket{\partial_i u} \in T\mathbb{C}^N$ along the $U(1)$ fiber, defining an element of the one-dimensional Lie algebra.  The Berry curvature is the exterior derivative $\omega = \dd A$ of this connection.  It is a purely horizontal two-form, describing features of the gauge-invariant base space. Integrating this curvature form then renders the first Chern number associated with the U(1) bundle.

\section{Multi-band setup and the fundamental role of Grassmannians and Flag manifolds as classifying spaces} A general extension of the above framework to multiple bands necessitates a promotion of these scalar geometric quantities to matrices.  If we let $U$ be a matrix whose columns are orthonormal $\Ket{u}$ spanning an occupied collection of bands, we can define
\begin{eqnarray}
{\bf g}_{ij} &=& \partial_i U^\dagger Q\partial_j U + H.c.\nonumber\\
\boldsymbol{\omega}_{ij} &=& \partial_i U^\dagger Q\partial_j U - H.c.,
\end{eqnarray}
where $Q$ generalizes to the projector out of the occupied manifold, $Q = \id - UU^\dagger$. Similar to how the $\mathbb{C}^N$ representation of the single-band case involved a subtraction of gauge-dependent degree of freedom associated with derivatives along $|u\rangle$, here we remove all components lying within the occupied manifold.  This corresponds to a $U(k)$ gauge redundancy, where $k$ is the number of occupied bands.  We can now define $\tilde{U}$ to be the identification of all $U(k)$ rotations of $U$, and we obtain
\begin{eqnarray}
{\bf g}_{ij} = \partial_i \tilde{U}^\dagger \partial_j \tilde{U} + H.c.\nonumber \\                                         
\boldsymbol{\omega}_{ij} = \partial_i \tilde{U}^\dagger \partial_j \tilde{U} - H.c.
\end{eqnarray}
The space of $\tilde{U}$ is best expressed by considering a complete orthonormal basis of $N$ states arranged as columns of $U \in U(N)$ with $k$ states designated as the occupied sector and $N-k$ as the unoccupied sector.  $\tilde{U}$ then corresponds to such a $U \in U(N)$ under the identification of all $U(k)$ rotations of the occupied sector and all $U(N-k)$ rotations of the unoccupied sector, i.e. $U(N)/(U(k)\times U(N-k))$.  This is the definition of the Grassmanian $Gr^{\mathbb{C}}_{k,N}$.                                                                                                           
As before, we can view this as a principle $G$-bundle.  Though it may seem most natural to consider a $U(k) \times U(N-k)$ bundle, we recall that the $U(N-k)$ component of the gauge group was only introduced as a convenience for considering $U(N)$ as the total space.  The relevant part of the gauge group for the occupied manifold is $U(k)$.  Since the group is a product, we can easily map to a principle $U(k)$-bundle with total space $U(N)/U(N-k)$.  The tangent space to $U(N)$ is the space of anti-Hermitian $N\times N$ matrices, and by quotienting out $U(N-k)$ we remove those with components $\inner{ u_n }{ \partial u_m }$ for $u_n$, $u_m$ unoccupied.  The vertical section consists of those matrices with only $\inner{ u_n }{ \partial_i u_m }$ components where $u_n$, $u_m$ are both occupied.  Thus the Lie algebra-valued connection is the anti-Hermitian $k\times k$ matrix with components $ A^{nm} = \inner{ u_n }{ \partial_i u_m } $, that is
\begin{equation}
A = U^\dagger \partial_i U.
\end{equation}
As before, the curvature $\boldsymbol{\omega}$ is the exterior derivative $\dd A$.  However, the matrix-valued curvature is not invariant under $U(k)$ rotations.  To get something invariant, we must construct a scalar.  One useful scalar is the trace, which for the Berry curvature gives the sum of the single-band curvatures over bands $\Ket{u_\ell}$ in the occupied manifold.
\begin{equation}
\omega_{ij} \equiv \tr(\boldsymbol{\omega}_{ij}) = \sum_\ell \Bra{\partial_i u_\ell}Q\Ket{\partial_j u_\ell} - c.c.    
\end{equation}
This quantity integrated over the Brillouin zone gives the Chern number of the occupied manifold.  The trace of the symmetric part of the Fubini-Study metric defines a metric on the Grassmanian,
\begin{equation}
g_{ij} \equiv \tr({\bf g}_{ij}) = \sum_\ell \Bra{\partial_i u_\ell}Q\Ket{\partial_j u_\ell} + c.c. ,
\end{equation}
which then defines the usual notion of distance between two collections of states.

We wish to emphasize the principal topological role of the emergent Grassmannian structure as acting classifying space. The topology of a multi-band model is in all generality set by the relevant classifying space, that, heuristically put, arises by flattening the bands (as dispersion does not affect topological properties), partitioning the system (e.g. in a valence and conduction sector, defining the gap about which topology considered) and removing the redundant degrees of freedom (the permutations of bands in the valence or conduction sector). 
From a K-theory perspective, meaning a classification perspective that is stable under the addition of trivial bands~\cite{Clas3,SchnyderClass}, one usually proceeds by analyzing the Clifford algebra extension problem $Cl_d\rightarrow Cl_{d+1}$~\cite{Schnyder08,Kitaev}. The set of representations, denoted by ${\cal C}_q$ or ${\cal R}_q$ for complex and real case respectively, precisely entails the classifying space, being physically speaking all mass terms that can be added without breaking the assumed symmetries of the system. The topological characterization is then obtained by considering the distinct components, or zeroth homotopy group $\pi_0 (X)$, where $X={\cal C}_q$ or $X={\cal R}_q$~\cite{Clas3,SchnyderClass,Kitaev}.

From a homotopy perspective, where one is interested in the topological invariants of a $n$-band system, the classifying space plays a similar fundamental role as it determines the homotopy charges of the relevant Bloch Hamiltonian directly~\cite{bouhonGeometric2020}.
An illustrative example entails a simple 2D two-band system with Hamiltonian $H=\mathbf{d}(\mathbf{k})\cdot\boldsymbol{\sigma}$ in terms of the Pauli matrices $\boldsymbol{\sigma}$.
General maps from the Brillouin zone into the Bloch space of gapped Hamiltonians are in this case determined by the classifying space $Gr^{\mathbb{C}}_{1,2}=U(2)/(U(1)\times U(1)$), being the protective space $\mathbb{CP}^1$ or the Riemann sphere. The second homotopy group $\pi_2(\mathbb{CP}^1)=\mathbb{Z}$ then coincides with the opposite Chern number ${\cal C}$ of each band that can take integer values. This is usually rephrased as a 'wrapping of the sphere' by the $\mathbf{d}(\mathbf{k})$-vector,  as quantified using the skyrmion formula ${\cal C}=\frac{1}{4\pi}\int \hat{\mathbf{d}}(\mathbf{k})\cdot\partial_{k_x}\hat{\mathbf{d}}(\mathbf{k})\times \partial_{k_y}\hat{\mathbf{d}}(\mathbf{k})$ in terms of $\hat{\mathbf{d}}(\mathbf{k})=\mathbf{d}(\mathbf{k})/|\mathbf{d}(\mathbf{k})|$. We stress however that the general point of view on the classifying will underpin richer and more general topologies.

Not in the least place one may note that a general handle on arbitrary classifying space is essential to understand and characterize recently discovered multi-gap topologies as will be detailed in the subsequent sections. These phases arise due to more refined gap conditions that generalize Grassmannians to Flag manifolds~\cite{bouhonGeometric2020, Bouhon2022braiding2}. Considering a three-band system, for example, one may conventionally partition the bands as a two-band and single band subspace, that is $2+1$, or more esoterically as $1+1+1$. This has direct topological consequences when a reality condition due to the presence of $C_2{\cal T}$ (two-fold rotations and time reversal symmetry) or ${\cal PT}$ (inversion and time reversal symmetry) is assumed. Indeed, the partition into two sectors then relates to the real Grassmannian $Gr^{\mathbb{R}}_{2,3}=O(3)/(O(2)\times O(1))$ (or oriented forms thereof that give the same homotopy results~\cite{bouhonGeometric2020}), being the real projective plain $\mathbb{RP}^2$, while the `full Flag limit' corresponds to $Fl^{\mathbb{R}}_{1,1,1}=O(3)/(O(1)\times O(1) \times O(1)$). Interestingly, the first homotopy group $\pi_1(Fl_{1,1,1})$, conveying the possible charges of band nodes, entails the non-Abelian quaternion group $\mathbb{Q}$~\cite{bouhon2019nonabelian, Wu1273} akin to how vortices act in certain nematics~\cite{Kamienrmp, Prx2016,volovik2018investigation, Beekman20171}. As a result, braiding band node charges between two bands with other nodes residing between two adjacent bands renders phase factors and may lead to similarly valued band node charges within a single two-band subspace. The resulting obstruction to annihilate these nodes is subsequently characterized by a real analogue of the Chern number, the Euler class, that can defined on patches in the Brillouin that include the nodes of this two-band subspace zone to quantify the stability~\cite{bouhon2019nonabelian}. From a topological point of view, we thus again see the principal role of the classifying space and the relation between different partitions. Indeed, when the third band is subsequently gapped, as is possible for a double braid~\cite{bouhon2019nonabelian,Peng2021, bouhonGeometric2020}, this stability of the Euler class in the isolated nodal two-band subspace is precisely captured by the second homotopy of the gapped $2+1$ system that indeed has $\pi_2(\mathbb{RP}^2)=\mathbb{Z}$. In other words, when considering how a system evolves into a multi-gap topological phase characterized by Euler class, we see that starting from a Flag limit in which non-Abelian nodal charges can be braided, we can eventually end up in a gapped system with classifying space $\mathsf{Gr}^{\mathbb{R}}_{2,3}=\mathbb {R}P^2$, whose second homotopy group coincides with the Euler class that quantifies the obstruction to annihilate the nodal charges in the isolated two-band subspace due to the previous braiding process~\cite{bouhon2019nonabelian, Eulerdrive, bouhonGeometric2020}.

\section{Fubini-Study metric in Plucker embedding}
The above outlined role of the classifying space manifestly shows that in order to define a general quantum geometric setup, a universal handle on project manifolds as Grassmannians and Flag manifolds is required. Indeed, a pitfall in naively studying multi-band systems using the matrix-valued connection $A$ is that there is a substantial gauge redundancy which makes it easy to consider quantities which may not be observable.  One must be careful when considering band off-diagonal quantities, as these may not have physical relevance.  More specifically, in the Chern context, the consideration of a collection of states where one does not wish to make any distinction among bands in the occupied sector should ideally result in the mapping of the $U(k)$ gauge theory to a $U(1)$ gauge theory. Similarly, for the multi-gap topologies as analyzed in the subsequent, one requires a general universal handle that notwithstanding does not include physically meaningless unitary or orthogonal rotations of bands in the partitioned subspaces. An essential insight in this regard is that one can depart from the specific relevant classifying space, being the minimal object that encodes all topology, and employ Pl\"ucker embeddings as direct universal scheme to concretely parameterize all quantities of interest.
To introduce the Pl\"ucker embedding we will first motivate it in the familiar Chern context described above, before profiting from its generalizable nature to any Grassmannian or Flag manifold to come to a fully unverisal framework.

As said above, for the characterization of a Chern number of a $k$-band valence sector, the Pl\"ucker embedding allows for direct map from the $U(k)$ gauge theory to a $U(1)$ gauge theory. 
The Pl\"ucker embedding is an embedding of $Gr^{\mathbb{C}}_{k,N}$ into $\mathbb{C}^d$ where $d = \binom{N}{k}$.  Put differently, within this context we map the matrix $U$ representing $k$ states to a single vector in a $d$-dimensional Hilbert space.  This single state vector can then be treated using single-band techniques.
The Pl\"ucker embedding $\iota$ is defined using the wedge product of vectors, which maps $k$ vectors in $N$ dimensions to a vector in $d=\binom{N}{k}$ dimensions and is fully antisymmetric.  Given a basis $e_1,\ldots,e_N$ of the original Hilbert space, we define a new basis $\check{e}_1,\ldots,\check{e}_d$ from the $d$ distinct collections $I_\ell$ of $k$ indices $\in 1,\ldots,N$ as 
\begin{equation}
\check{e}_\ell = \bigwedge_{i \in I_\ell} e_i 
\end{equation}
Explicitly, a matrix $U$, with elements $u_i^j$ labeling the $j$th component of the $i$th state, maps to the vector with components
\begin{equation}
V^\ell = \epsilon_{i_1\ldots i_k} u_1^{i_1}\ldots u_k^{i_k} \quad i_1,\ldots, i_k \in I_\ell, 
\end{equation}
where $\epsilon$ is the Levi-Civita symbol and repeated indices are assumed to be summed over.  This defines an explicit way of calculating $P^\ell$, however for analytical purposes it will be more useful to work with the object
\begin{equation}
V  = u_1 \wedge \ldots \wedge u_k,    
\end{equation}
representing the groundstate manifold in the Plucker embedding, and also the analog to the energy eigenbasis
\begin{equation}
E_n = \bigwedge_{i \in I_n} u_i.  
\end{equation}

Important algebraic properties of $V$ are the inner product
\begin{equation}
V_1^\dagger V_2 = (u_1 \wedge \ldots \wedge u_k)^\dagger (v_1 \wedge \ldots \wedge v_k) = \det(u_i^\dagger v_j) 
\end{equation} 
and the Leibniz rule
\begin{equation}
\partial V = \partial u_1 \wedge u_2 \wedge \ldots \wedge u_k + \ldots + u_1 \wedge \ldots \wedge u_{k-1} \wedge \partial u_k 
\end{equation}
Furthermore, using the inner product property, one can show that if $u_1,\ldots, u_k$ are orthonormal, then $V$ is normalized.  $V$ additionally inherits a $U(1)$ gauge redundancy from the product of $U(1)$ gauge degrees of freedom on the $u_i$.

Using these properties, one can show that, see Appendix \ref{app:pluckercherneq},
\begin{eqnarray}\label{eq:Chern_genplucker}
\Bra{\partial_i V}Q\Ket{\partial_j V} - c.c. &=& \omega_{ij}\nonumber \\
\Bra{\partial_i V}Q\Ket{\partial_j V} + c.c. &=&  g_{ij},
\end{eqnarray}
where $Q$ is now $\id - \Ket{V}\Bra{V}$.  Thus the natural Fubini-Study metric on the Pl\"ucker embedding $\iota(U)$ is exactly the trace of the matrix-valued ${\bf g}$ and $\boldsymbol{\omega}$.  Similarly, the connection on this $U(1)$ bundle is
\begin{equation}
\inner{V}{ \partial_i V} = \tr{A} 
\end{equation}
We thus observe that, as anticipated, under the Pl\"ucker embedding the multi-band generalization of the metric and Chern number map to the single-band version defined on a new Hilbert space, showing its universal power. Naively, the Pl\"ucker map constructs a vector parameterization that is invariant under the gauge degrees of freedom (the permutations in the partitioned band subspaces), meaning that it manifestly paramaterizes the right classifying space in a generic manner, giving direct access to all topological invariants. This becomes even more apparent in $n$-band models, where the interplay between partitioning and emergent topological structures is directly tractable in simple models, providing a further intuitive understanding of the Pl\"ucker embedding as detailed in the subsequent two Sections.

\section{Multi-band systems modeling through Pl{\"u}cker embedding I: Chern phases}\label{sec_complex_fewband}
The Pl{\"u}cker embedding of Grasmmannians cannot only used in the characterization of the quantum geometry of Bloch Hamiltonians, but can also  be used to define arbitrary topological models involving the associated classifying space directly. We here review the use of the Pl{\"u}cker embedding in the systematic modeling of multi-band topological phases. By removing the redundant gauge degrees of freedom, the embedding in fact determines all relevant variables, called principal angles, that most naturally parametrize the points of the Grassmannian. 

In this Section and the next, we treat these characterizations in detail. In particular, we will analyze the (``complex") two-band and three-band Chern phases here, while the (``real") Euler and Stiefel-Whitney phases are presented in the next section. 

\subsection{Two-band Chern models}
As outlined above, the classifying space of gapped two-band complex Hamiltonians is the Riemann sphere $\mathsf{Gr}_{1,2}^{\mathbb{C}}=\mathbb{C}P^1\cong \mathbb{S}^2$. The corresponding Bloch Hamiltonians take the simple form $H^{\mathbb{C},1+1}(\bs{k})=\bs{d}(\bs{k})\cdot \bs{\sigma} + d_0(\bs{k}) \mathbb{1}_2$, where the unit vector $\hat{\bs{d}} \in \mathbb{S}^2$ parametrizes the points of the classifying space and its winding determines the Chern number via ${\cal C}=\frac{1}{4\pi}\int \hat{\mathbf{d}}(\mathbf{k})\cdot\partial_{k_x}\hat{\mathbf{d}}(\mathbf{k})\times \partial_{k_y}\hat{\mathbf{d}}(\mathbf{k})$ in terms of $\hat{\mathbf{d}}(\mathbf{k})=\mathbf{d}(\mathbf{k})/|\mathbf{d}(\mathbf{k})|$. Alternatively, we can start by defining the matrix of Bloch eigenvectors $U^B=(u_1\,u_2)$, \ie entering in the $1+1$-spectral decomposition of the Bloch Hamiltonian
\begin{equation}
    H^{\mathbb{C},1+1} = U^B \cdot \text{diag}[E_1,E_2] \cdot U^{B\dagger}\,,
\end{equation}
where we assume $E_1<E_2$, in the form of a generic $\mathsf{U}(2)$ matrix parametrized by 4 angles, \ie $U^B = e^{\imi \varphi} e^{-\imi \alpha/2 \sigma_z} e^{-\imi \beta/2 \sigma_y} e^{-\imi \gamma/2 \sigma_z}$. Clearly the phase $e^{\imi \varphi}$ and the diagonal factor $e^{-\imi \gamma/2 \sigma_z}$ only act as gauge phases of $u_1$ and $u_2$. We are thus left with a minimal form \begin{equation}
    U^B(\alpha,\beta) = e^{-\imi \alpha/2 \sigma_z} e^{-\imi \beta/2 \sigma_y}\,,   
\end{equation}
that is parametrized by the two angles that determine a point of the sphere through $\bs{d}(\alpha,\beta) = -\epsilon (\cos \alpha \sin \beta,\sin \alpha \sin \beta,\cos\beta)$, for which we have set the energy eigenvalues to $E_1=-E_2=-\epsilon$. We write occupied eigenvector $u_1$ as a reference for the 3-band case treated below,
\begin{equation}
\label{eq_2B_C_vec}
    u_1(\alpha,\beta) = \left[ 
        e^{\imi \alpha/2} \cos\dfrac{\beta}{2} \,,\, e^{-\imi \alpha/2} \sin \dfrac{\beta}{2}
    \right]\,.
\end{equation}

\subsection{Three-band Chern phases}
While the two-band Chern case entails a paradigmatic model, the universal character of the Pl\"ucker embedding allows for a direct generalization of this intuitive understanding to novel arbitrary settings. In this regard, we now turn to the systematic modeling of three-band Chern phases where we assume the presence of two occupied bands and one unoccupied one~\cite{Kemp_nested_2022}. Writing the spectral form, 
\begin{equation}
    H^{\mathbb{C},2+1} = U^B \cdot \text{diag}[E_1,E_2,E_3] \cdot U^{B\dagger}\,,
\end{equation}
where $U^B=(u_1\,u_2\,u_3)\in\mathsf{U}(3)$ and we have assumed $E_1\leq E_2 < E_3$, the classifying space is now the complex projective plane $\mathsf{Gr}_{2,3}^{\mathbb{C}} = \mathsf{U}(3)/[\mathsf{U}(2)\times \mathsf{U}(1)] = \mathsf{SU}(3)/\mathsf{S}[\mathsf{U}(2)\times \mathsf{U}(1)] = \mathbb{C}P^2$. This is a four-dimensional manifold that can be parametrized by four principal angles. (We refer here to its real dimension. The complex Grassmannian is also a complex manifold of complex dimension 2.) 

Similarly to the two-band case, we start with the matrix of Bloch eigenvectors in the form of a generic $\mathsf{SU}(3)$ matrix, \ie \cite{guise2018factorization}
\begin{equation}
\label{eq_3B_chern_U3}
    U^B =  U_{23}(\alpha_1,\beta_1,\gamma_1)\cdot U_{12}(\alpha_2,\beta_2,\alpha_2)
    \cdot U_{23}(\alpha_3,\beta_3,\gamma_3)\,,
\end{equation}
in terms of the $\mathsf{SU}(2)$ block matrices
\begin{equation}
    [U_{i\,i+1}(\alpha,\beta,\gamma)]_{kl} = [e^{-\imi \alpha/2 \sigma_z} e^{-\imi \beta/2 \sigma_y} e^{-\imi \gamma/2 \sigma_z}]_{kl}\,,
\end{equation}
whenever $k,l\in\{i,i+1\}$ and by the identity, \ie $[U_{i\,i+1}]_{kl}=\delta_{kl}$, when $k,l \not\in \{i,i+1\}$. Out of the eight variables of this generic form, four are redundant to capture the complex projective plane target. Writing $(e_1,e_2,e_3)$ the basis vectors of the ambient complex vector space $\mathbb{C}^3$, we take the wedge product of the two occupied Bloch eigenvectors (the first two columns of $U^B$) in the spirit of accounting for redundant permutations of bands in the occupied band subspace and verify that
\begin{equation}
\begin{aligned}
       V = u_1 \wedge u_2  = \check{\bs{e}}^{\top} \cdot \left[ 
        \begin{array}{c}
            e^{\imi \alpha_2} \cos \beta_2/2 \\
            e^{\imi (\alpha_1+\gamma_1)/2} 
            \cos \beta_1/2
            \sin \beta_2/2 \\
            e^{\imi (-\alpha_1+\gamma_1)/2}
            \sin \beta_1/2
            \sin \beta_2/2
        \end{array}
       \right]\,.
\end{aligned}
\end{equation}
In the above $\check{\bs{e}} = (e_1\wedge e_2,e_1\wedge e_3,e_2\wedge e_3)$ is the basis of the exterior power space $\bigwedge^2(\mathbb{C}^3)\cong \mathbb{C}^3$. We hence see that the angles $(\alpha_3,\beta_3,\gamma_3)$ in Eq.\,(\ref{eq_3B_chern_U3}) are redundant. After inspection, it turns out that neither the angle $\gamma_1$ plays a role in the topology. This leaves the four angles $(\alpha_1,\beta_1,\alpha_2,\beta_2)$ that fully parametrize the points of the complex projective plane. The minimal form of the Bloch Hamiltonian for all complex $2+1$-partitioned phases is then obtained as
\begin{equation}
\label{eq_H_3B_complex}
\begin{aligned}
    H^{\mathbb{C},2+1}&(\alpha_1,\beta_1,\alpha_2,\beta_2) = U^B \cdot \text{diag}[E_1,E_2,E_3] \cdot U^{B\dagger}\,,\\
    H^{\mathbb{C},2+1}_{11} &= \dfrac{1}{2}(-1-\cos\beta_1-\cos\beta_2+\cos\beta_1\cos\beta_2)\,\\
    H^{\mathbb{C},2+1}_{22} &= \dfrac{1}{2}(-1+\cos\beta_1-\cos\beta_2-\cos\beta_1\cos\beta_2)\,\\
    H^{\mathbb{C},2+1}_{33} &= \cos\beta_2\,\\
    H^{\mathbb{C},2+1}_{12} &= -e^{\imi \alpha_1} \sin\beta_1 (\sin\beta_2/2)^2\,\\
    H^{\mathbb{C},2+1}_{13} &= e^{\imi (\alpha_1+2\alpha_2-\gamma_1)/2} \sin\beta_1/2 \sin\beta_2\,\\
    H^{\mathbb{C},2+1}_{12} &= -e^{-\imi (\alpha_1-2\alpha_2+\gamma_1)/2} \cos\beta_1/2 \sin\beta_2\,
\end{aligned}
\end{equation}
where we have taken the energy eigenvalues $E_1=E_2=-E_3=-1$. 

The Chern phases are determined by which 2D region of the complex Grassmannian space is covered by $u_1\wedge u_2$. We find that all the $2+1$-Chern phases are entirely controlled by the two angles $(\alpha_2,\beta_2)$. In particular, by setting $(\alpha_1,\beta_1,\gamma_1)=(0,0,0)$, we obtain the form $\left[e^{\imi \alpha_2}\cos\beta_2/2,\sin\beta_2/2,0\right]$, which is equivalent to Eq.\,(\ref{eq_2B_C_vec}). 

We finally recover the Berry curvature form from the Pl{\"u}cker vector $V(\alpha_1,\beta_1,\alpha_2,\beta_2)=u_1\wedge u_2$ through
\begin{equation}
\begin{aligned}
    \Omega^B &= -\imi \left[(\partial_{\alpha_2} V^{\dagger})\cdot 
    (\partial_{\beta_2} V) - (\partial_{\beta_2} V^{\dagger})\cdot 
    (\partial_{\alpha_2} V) \right]d\alpha_2\wedge\beta_2\,,\\
    &= \dfrac{\sin\beta_2}{2} d\alpha_2\wedge d\beta_2\,,
\end{aligned}
\end{equation}
from which we get
\begin{equation}
       \mathcal{C}_{\mathbb{S}^2} = \dfrac{1}{2\pi} \int_{\mathbb{S}^2} \Omega^B = 1 \,,
\end{equation}
assuming that the angles $(\alpha_2,\beta_2)$ cover a base sphere $\mathbb{S}^2$ with $\alpha_2\in[0,2\pi)$ the azimuthal and $\beta_2\in[0,\pi]$ the polar angles (more precisely, such that the unit vector $\bs{n}(\alpha_2,\beta_2) = (\cos\alpha_2\sin\beta_2,\sin\alpha_2\sin\beta_2,\cos\beta_2)$ wraps the sphere). The above result confirms that when $\mathbb{S}^2$ is covered one time, the Pl{\"u}cker vector $V=u_1\wedge u_2$ covers a nontrivial two-dimensional sphere-image within the Grassmannian, \ie the image $V(\mathbb{S}^2)\subset \iota_P(\mathbb{C}P^2)$ is not null-homotopic (it cannot be continuously shrunk to a point within $\mathbb{C}P^2$). 
 
Pulling back to the torus Brillouin zone and allowing multiple wrappings of the sphere-image $V(\mathbb{S}^2)$, we define the mapping
\begin{subequations}
\label{eq_pullback}
\begin{equation}
    H^{W}:\mathbb{T}^2\rightarrow \mathbb{S}^2:\bs{k}\mapsto \bs{n}(\alpha_2(\bs{k}),\beta_2(\bs{k}))\,,    
\end{equation}
with a degree $W\in \mathbb{Z}$. It is convenient to split this higher winding map into two steps as
\begin{equation}
    \begin{aligned}
        H^W :&\mathbb{T}^2\xrightarrow{f_{TS}}\mathbb{S}^2_0 \xrightarrow{f^W_{SG}} \mathbb{S}^2:\\
        &\bs{k}\mapsto (\phi_0,\theta_0)\mapsto (\alpha_2,\beta_2)\,,
    \end{aligned}
\end{equation}
\end{subequations}
such that, reminding that the degree of a map is the number of times it wraps the target space as one scan through its domain, the first map from the torus to the sphere $f_{TS}$ has degree 1 and the second map from the sphere to the Grassmannian $f^W_{SG}$ has degree $W\in\mathbb{Z}$ (in particular $f_{SG} = id$), and the resulting map $H^W=f^W_{SG} \circ f_{TS}$ has degree $W$. Then, the Chern number is nothing but (via pullbacks)
\begin{equation}
\begin{aligned}
    \mathcal{C}_{\mathbb{T}^2} &= H^{W*}\mathcal{C}_{\mathbb{S}^2} = f^*_{TS}\left[f^{W*}_{SG}\mathcal{C}_{\mathbb{S}^2}\right] \\
    &= \text{deg}(f_{TS})\text{deg}(f^W_{SG}) = W   \in \mathbb{Z} \,. 
\end{aligned}
\end{equation}
We conclude that all the $2+1$-Chern phases are realized by the Bloch Hamiltonian $H^{\mathbb{C},2+1}(\bs{k})$ of Eq.\,(\ref{eq_H_3B_complex}) with a Chern number $W$ readily fixed by the winding number of the map $f^W_{SG}$ from the sphere to its image within the Grassmannian. (While the above pullback structure may seem like a tedious exercise in the context of the three-band Chern phases, it will lie at the basis of our exposition of generalized metrics for an arbitrary number of bands in Section \ref{sec_general_metric}.)

There remains the question of whether another choice of a pair of angles among the set of variables $\{\alpha_1,\beta_1,\alpha_2,\beta_2\}$ would also lead to non-trivial Chern topology. The answer in fact is negative and can be traced back to the so-called CW structure.  Although a full treatment of CW complexes is beyond the scope of the present work, we note that the CW structure of a topological space $X$, say of dimension $m$, corresponds to a decomposition of the space into cells $e^n$ of increasing dimensions, such that each cell of a certain dimension $n$ is obtained from the inclusion of an $n$-dimensional disc $\mathbb{D}^n$ in $X$, and such that the boundary of the disc, $\partial\mathbb{D}^n$, is mapped continuously on the cells of dimensions $l\leq n$. Notably for the evaluation of the Chern topology one may deduce that that the CW complex structure of the complex projective plane contains only one sub-Grassmannian sphere, $\mathbb{S}^2\cong \mathbb{C}P^1 \subset \mathbb{C}P^2$, parametrized by the angles $(\alpha_2,\beta_2)$ (The CW structure of $\mathbb{C}P^2 $ comprises of a point, the sphere $\mathbb{C}P^1\cong \mathbb{S}^2$ and a four-dimensional disc $\mathbb{D}^4$ whose $\mathbb{S}^3$ boundary is glued on the sphere via the Hopf fibration.) In other words, whenever the Pl{\"u}cker vector $V$ does not fully wrap $\mathbb{C}P^1$ within $\mathbb{C}P^2$, the topology is necessarily trivial.

We crucially note that the topological invariant is insensitive to the local details of the mapping $H^W$ from the torus Brillouin zone to the sphere image within the Grassmannian. On the contrary, we will see that such details do affect the quantum geometry of the Bloch eigenstates. We will revisit this point from a more general perspective in the subsequent.

We discuss in Section \ref{sec_TB_model} a general procedure to obtain explicit tight-binding Hamiltonians from the Pl{\"u}cker ansatz. 

\section{Multi-band systems modeling through Pl{\"u}cker embedding II: Euler and Stiefel-Whitney phases}\label{sec_real_fewband}
We continue our demonstration of systematic modeling of topological phases via the Pl{\"u}cker embedding, now focusing on the other fundamental topological class in 2D, namely the (``real") Euler and Stiefel-Whitney topology~\cite{bouhon2019nonabelian, Jiang2021, Ahn2019, Eulerdrive,BJY_nielsen}. We first review the three-band and four-band Euler phases, and then present the $\mathbb{Z}$ Euler to $\mathbb{Z}_2$ Stiefel-Whitney topological reduction within five-band phases. Since Euler topology is only realized in orientable two-band subspaces (corresponding to rank-2 real orientable vector bundles), we briefly review the question of orientability that is controlled by the 1D sub-dimensional topology and is indicated by the $\mathbb{Z}_2$ first Stiefel-Whitney class. A systematic treatment of the modeling of 1D non-Abelian multi-gap topology will be reported elsewhere~\cite{Bouhon1D_2022}. 

While the results in the sections on three-band and four-band Euler phases coincide with results of previous work \cite{bouhonGeometric2020,Bouhon2022braiding2} on the homotopy classification of these models, we propose here an alternative Pl{\"u}cker-based derivation of the Bloch Hamiltonians that captures more closely the inclusion relation between Grassmannians of increasing dimensions, \ie $ \mathsf{Gr}_{2,3}^{\mathbb{R}}\hookrightarrow \mathsf{Gr}_{2,4}^{\mathbb{R}} \hookrightarrow \mathsf{Gr}_{2,5}^{\mathbb{R}} $. This turns out to be particularly useful for the construction of the five-band models addressed shortly. We also reveal the common geometric origin of topology in complex Chern and real Euler classes that will be useful in the characterization of the quantum geometry of these phases exposed in Sections \ref{sec_general_metric} and \ref{sec_applications}. 

\subsection{Reality condition}
Contrary to the Chern class, which, within the crystalline context (\ie symmetric under discrete translations), does not require any additional symmetry, the Euler and Stiefel-Whitney classes are protected by an anti-unitary symmetry (\ie including complex conjugation) that squares to $+1$ and leaves the quasi-momentum invariant. A ubiquitous example in electronic crystalline systems is the combination of time-reversal symmetry (TRS) ${\cal T}$ and $\pi$-rotation $C_{2z}$ around the axis perpendicular to the system's basal plane, \ie here chosen as $\hat{z}$. TRS acts on spin-$1/2$ Bloch orbital states $\vert \bs{\varphi},\bs{k}\rangle = ( \vert\varphi_{A,\uparrow},\bs{k} \rangle\, \vert\varphi_{A,\downarrow},\bs{k} \rangle \, 
\vert\varphi_{B,\uparrow},\bs{k} \rangle\, \vert\varphi_{B,\downarrow},\bs{k} \rangle\,\dots
)$, where $\{A,B,\dots\}$ label the electronic orbital and sub-lattice degrees of freedom which are both of bosonic type, as $^{{\cal T}}\vert \bs{\varphi},\bs{k}\rangle = \vert \bs{\varphi},-\bs{k}\rangle (U_{T}\otimes -\imi \sigma_y) \mathcal{K}$, with $U_{\cal T}$ the unitary part acting on the bosonic degrees of freedom and $\mathcal{K}$ is complex conjugation. Similarly, $C_{2z}$ acts as $^{C_{2z}}\vert \bs{\varphi},\bs{k}\rangle=\vert \bs{\varphi},C_{2z}\bs{k}\rangle (U_2\otimes -\imi \sigma_z)$, such that the combination with TRS gives $^{C_{2z}{\cal T}}\vert \bs{\varphi},\bs{k}\rangle=\vert \bs{\varphi},-C_{2z}\bs{k}\rangle (U_2U_{{\cal T}}\otimes \imi \sigma_x) \mathcal{K}$. We recall that for any basis function formed by the (tensor) product of bosonic factors and an {\it odd} number of fermionic factors (here the spin $1/2$), we may infer the properties ${\cal T}^2=-1$ implying $U_{\cal T}U_{\cal T}^*=\mathbb{1}$, $C_{2z}^2=-1$ implying $U_2U_2=\mathbb{1}$ and $C_{2z}{\cal T}={\cal T}C_{2z}$ implying $U_{\cal T}U_2^*=U_2U_{\cal T}$, that lead to $[C_{2z}{\cal T}]^2 = +1$ since $^{[C_{2z}{\cal T}]^2}\vert \bs{\varphi},\bs{k}\rangle=\vert \bs{\varphi},-C_{2z}\bs{k}\rangle (U_2U_{{\cal T}}U_2^*U_{{\cal T}}^*\otimes\mathbb{1}_2) =\vert \bs{\varphi},-C_{2z}\bs{k}\rangle$, where $-C_{2z}\bs{k} = m_z\bs{k}=\bs{k}$ whenever $\bs{k}\cdot \hat{z} =0$. Interestingly, spinless TRS combined with spinless $C_{2z}$ also gives $[{\cal T}C_{2z}]^2=+1$, leaving basal momenta similarly invariant as in the case of spinless ${\cal PT}$ symmetry (with ${\cal P}$ inversion), such that crystalline bosonic systems are also subjected to Euler and Stiefel-Whitney topological classes. In order to concretize ideas and models, we only refer to the $C_2T$ symmetry in the following and take $\bs{k}$ within the $C_2T$-invariant plane. 

Writing a Bloch Hamiltonian in the general basis $\vert \bs{\varphi},\bs{k}\rangle$
\begin{equation}
    \mathcal{H} = \sum\limits_{\bs{k}} \vert \bs{\varphi},\bs{k}\rangle H(\bs{k}) \langle \bs{\varphi},\bs{k} \vert \,,
\end{equation}
and defining $U_{2{\cal T}} = U_2 U_{\cal T}$, the condition of $C_2{\cal T}$ symmetry $^{C_{2z}{\cal T}}\mathcal{H} = \mathcal{H}$ leads to the constraint
\begin{equation}
    U_{2{\cal T}} \cdot H(\bs{k})^* \cdot U_{2{\cal T}}^{\dagger} = H(\bs{k}) \,.
\end{equation}

The condition $[C_{2z}{\cal T}]^2=+1$ implies $U_{2{\cal T}}^{\top} = U_{2{\cal T}}$, \ie it is unitary and symmetric, in which case the Takagi factorization $U_{2{\cal T}} = V_{\text{TF}} V_{\text{TF}}^{\top}$ is readily obtained from the singular value decomposition $U_{2{\cal T}} = U_{svd}\mathbb{1} V_{svd}$ as $U_{\text{TF}} = U_{svd}\sqrt{U_{svd}^{\dagger} V_{svd}^*}$ \cite{chen2021manipulation,CHEBOTAREV2014380,Peng2021}. As a consequence, performing a change of basis through $\vert \bs{\varphi},\bs{k}\rangle = \vert \bs{\phi},\bs{k}\rangle U_{\text{TF}}^{\top}$, we obtain the new Hamiltonian matrix 
\begin{equation}
    \widetilde{H}(\bs{k}) = U_{\text{TF}}^{\dagger} \cdot H(\bs{k}) \cdot U_{\text{TF}}\,.
\end{equation}
that now satisfies the $C_{2}{\cal T}$-symmetry constraint
\begin{equation}
    \widetilde{H}(\bs{k})^*=  \widetilde{H}(\bs{k})\,,
\end{equation}
\ie it must be real. In the following, we assume that the appropriate basis has been adopted and we assume that $H(\bs{k})$ is real without marking it with a tilde.

\subsection{1D topology and orientability}
The Euler class only exists for rank-2 oriented vector bundles in 2D \cite{Hatcher_2}, \ie here the Bloch bundle of a two-band subspace $\mathcal{B}_{n,n+1} = \bigcup_{\bs{k}\in\mathbb{T}^2} \langle u_n(\bs{k}),u_{n+1}(\bs{k})\rangle $ spanned by the Bloch eigenvectors $\{u_n(\bs{k}),u_{n+1}(\bs{k})\}$ \cite{bouhonGeometric2020,Panati_Chern}. We thus restrict our discussion to {\it orientable} phases. The obstruction to orientability is measured by the first Stiefel-Whitney class, which is practically computed by the Berry phase factor $e^{\imi \gamma_B[l]}$ over the two non-contractible directions $\{l_1,l_2\}$ of the torus Brillouin zone $\mathbb{T}^2$ \cite{Ahn2018b}. We thus require zero Berry phases. This is simply enforced by modeling the Euler phases through Bloch Hamiltonians that are periodic over the first Brillouin zone, \ie such that $H(\bs{k}+\bs{b}_i) = H(\bs{k})$ for the two primitive reciprocal lattice vectors $\bs{b}_1$ and $\bs{b}_2$. 

We must distinguish oriented vector bundles from orientable phases because there is no canonical choice of orientation for Bloch Hamiltonians such that the orientation of the associated Bloch bundle can be reversed through an adiabatic deformation, see \cite{bouhonGeometric2020,Bouhon2022braiding2} for a detailed discussion. This technical distinction has the consequence of reducing the $\mathbb{Z}$ counting of 2D phases by a signed Euler class to the $\mathbb{N}$ homotopy classification of Euler phases indicated by unsigned Euler classes.   

For simplicity, we also assume that the system is compatible with an orientable atomic flag limit, \ie such that the band structure can be fully trivialized with every band disconnected one-from-another by an energy gap and such that each band has zero first Stiefel-Whitney class. (We refer to it as the {\it flag} limit, because when all the bands are disconnected in energy the most general classifying space of the Bloch Hamiltonian becomes the flag manifold $\mathsf{Fl}_{1,1,1,\dots}^{\mathbb{R}} = \mathsf{O}(N)/\mathsf{O}(1)^N$.) This condition is necessarily satisfied when all the atomic orbitals are located at the center of the unit cell (see \cite{Jiang_meron} for the presentation of an Euler phase realized in the kagome lattice for which there is no orientable atomic flag limit).

\subsection{Three-band Euler phases}\label{sec_3B_real}
Having set the necessary preliminaries, we return to the analysis of the Pl\"ucker embedding in the concrete context of $n$-band models, specifying now to the case of a real $n=3$ system. We start with a $C_2T$-symmetric (thus ``real") three-band system possessing two occupied bands and one unoccupied band. As alluded to above, the classifying space of the Bloch Hamiltonian
corresponds to the real Grassmannian $\mathsf{Gr}_{2,3}^{\mathbb{R}} = \mathsf{O}(3)/[\mathsf{O}(2)\times \mathsf{O}(1)] = \mathsf{SO}(3)/\mathsf{S}[\mathsf{O}(2)\times \mathsf{O}(1)] = \mathbb{R}P^2$, being the real projective plane. Given the above condition of orientability and of orientable atomic flag limit, it is convenient to build the Bloch Hamiltonians through the Pl{\"u}cker embedding of the {\it oriented} Grassmannian $\widetilde{\mathsf{Gr}^{\mathbb{R}}}_{2,3} = \mathsf{SO}(3)/\mathsf{SO}(2) = \mathbb{S}^2$. 

Our strategy is again to start from the matrix of Bloch eigenvectors $R^B=(u_1\,u_2\,u_3)$ in a generic $\mathsf{SO}(3)$ form, \ie $R^B(\alpha,\beta,\gamma)=e^{\imi (\alpha L_{x}+\beta L_y + \gamma L_z)}$ with the angular momentum matrices $ L_x$=\SmallMatrix{0&0&0\\0&0&-1\\0&1&0} $L_y$=\SmallMatrix{0&0&1\\0&0&0\\-1&0&0} and $L_z$=\SmallMatrix{0&-1&0\\1&0&0\\0&0&0} that form a basis of the Lie algebra $\mathsf{so}(3)$, and write the $2+ 1$-partitioned Bloch Hamiltonian form as
\begin{equation}
    H^{\mathbb{R},2+1} = R^B(\alpha,\beta,\gamma) \cdot \text{diag}[E_1,E_2,E_3] \cdot R^{B}(\alpha,\beta,\gamma)^{\top} \,,
\end{equation}
assuming the ordering of the eigenenergies $E_1\leq E_2<E_3$. 
    
Setting $(\alpha,\beta)=(-\pi/2+\phi,\pi/2+\theta)$ and taking the wedge product of the two first eigenvectors as for the three-band Chern model above, we get a Pl\"ucker vector of the form 
\begin{equation}
\label{eq_L_Euler_3B}
\begin{aligned}
    V(\theta,\phi) &= u_1\wedge u_2 = \check{\bs{e}}^{\top}\cdot \bs{n}(\phi,\theta)\,,\\
    \bs{n}(\phi,\theta) &=(\cos\phi\sin\theta,\sin\phi\sin\theta,\cos\theta)^{\top}\in \mathbb{S}^2\,,
\end{aligned}
\end{equation}
where the column of vectors $\check{\bs{e}}=(e_1\wedge e_2,e_3\wedge e_1,e_2\wedge e_3)$ is the basis of $\bigwedge^2(\mathbb{R}^3)\cong \mathbb{R}^3$ formed from the basis $(e_1,e_2,e_3)$ of $\mathbb{R}^3$. (We incidentally note that $u_3=\bs{n}(\phi,\theta)$, as is expected since by taking the Hodge star $u_3=*(u_1\wedge u_2)$.) We conclude that the phase $\gamma$ is redundant and $V$ wraps the target sphere whenever the angles $(\phi,\theta)$ cover a base sphere. (This approach is slightly more general than the derivation in \cite{bouhonGeometric2020} [based on the special spherical frame representing the tangent bundle of the sphere] and allows us to point out the methodological similarity with the Chern phases shortly.)

The Euler class is defined from the real $\mathsf{SO}(2)$-connection of the two-band subspace considered. Using the principal angles that parameterize the classifying space, the 2-by-2 Berry connection 1-form for the two lower bands is 
\begin{subequations}
\label{eq_Euler_connection_form_class}
\begin{multline}
    \mathcal{A}[\{u_1,u_2\}] = \\
    \left[\begin{array}{cc}
        0 & u_1^{\top} \partial_{\phi} u_2\, d\phi + u_1^{\top} \partial_{\theta} u_2\, d\theta\\
        -u_1^{\top} \partial_{\phi} u_2\, d\phi - u_1^{\top} \partial_{\theta} u_2 \, d\theta & 0
    \end{array}\right] ,
\end{multline}
from which we define the Euler connection $\mathtt{a}=\text{Pf}\mathcal{A}[\{u_1,u_2\}] = u_1^{\top} \partial_{\phi} u_2\, d\phi + u_1^{\top} \partial_{\theta} u_2\, d\theta$ \cite{BJY_nielsen,bouhon2019nonabelian,xie2020bounded}. The Euler two-form is then obtained as
\begin{equation}
\label{eq_euler_form}
\begin{aligned}
    \mathtt{Eu} = d\mathtt{a} &=    (\partial_{\phi} u_1^{\top} \partial_{\theta} u_2 - \partial_{\theta} u_1^{\top} \partial_{\phi} u_2) \,
    d\phi\wedge d\theta\,,\\
    &= -\sin\theta\, d\phi\wedge d\theta \,,
\end{aligned}
\end{equation}
leading to the Euler number (or Euler class) 
\begin{equation}
\label{eq_euler_number}
    \chi_{\mathbb{S}^2}[\{u_1,u_2\}] = \dfrac{1}{2\pi}\int_{\mathbb{S}^2} \mathtt{Eu} = -2\,.
\end{equation}
\end{subequations}
The minus sign here simply comes from our convention in the definition of the Euler form. Furthermore, as noted above, it is the unsigned Euler class that is in one-to-one correspondence with the homotopy classes of (orientable) Euler Bloch Hamiltonians, \ie
\begin{equation}
    [\mathbb{S}^2_0,\mathbb{R}P^2] = 2\mathbb{N} \ni \left\vert \chi[\{u_1,u_2\}]\right\vert \,.
\end{equation}
It remains to explain the factor $2$. A first route involves noting that the form of the Bloch Hamiltonian, if we flatten the spectrum as $E_1=E_2=-E_3=-1$, relates to the canonical form $H^{\mathbb{R},2+1} = 2 \bs{n}(\phi,\theta)\cdot \bs{n}(\phi,\theta)^{\top} - \mathbb{1}_3$. As a result, we observe that the Hamiltonian ``winds" twice when the unit vector $\bs{n}(\phi,\theta)$ wraps the sphere once. A deeper explanation is that the Euler class matches with the Euler characteristics of the sphere, thus realizing the Gauss-Bonnet theorem
\begin{equation}
    \left\vert \chi_{\mathbb{S}^2}[\{u_1,u_2\}] \right\vert= \chi[\mathbb{S}^2] = 2\,,
\end{equation}
for which the Euler form plays the role of the Gauss curvature. 

We note the direct agreement between the above evaluation of the Euler number and the winding number of the Pl{\"u}cker vector represented by the unit vector $\bs{n}(\phi,\theta)$ [Eq.\,(\ref{eq_L_Euler_3B})]. We thus conclude once again that the generality of the Pl{\"u}cker embedding acts as a unification principle, revealing the essentially similar origin of topology in Chern and in Euler phases. 

We conclude with generating all $2+1$-Euler phases from the minimal Bloch Hamiltonian form $H^{\mathbb{R},2+1}(\phi,\theta)$, in the same way we did for the Chern phases, namely through the pullback to the torus Brillouin zone by the same map $H^W=f^W_{SG}\circ f_{TS}$ [Eq.\,(\ref{eq_pullback})]. Via a similar line of reasoning, we then obtain
\begin{equation}
\begin{aligned}
    \left\vert \chi_{\mathbb{T}^2} \right\vert &= \left\vert H^{W*} \chi_{\mathbb{S}^2} \right\vert \\
    &=2 \left\vert \text{deg}(H^{W}) \right\vert \\
    &= 2\vert W\vert \in 2\mathbb{N}\,.
\end{aligned}
\end{equation}
 
\subsubsection{Correspondence with Chern phases}

The similarity in the origin of nontrivial topology in Chern and Euler phases noted above should not hide their qualitatively different physical manifestations. We only mention here the ``leading order" differences between the two topologies. On one hand for the Chern phases we note that $(i)$ a nonzero Chern number can be carried by a single band disconnected in energy from all other bands, $(ii)$ the sum of the Chern number of all the bands below an energy gap, $C_{I}$, and the Chern of all the bands above the gap, $C_{II}$, must add to zero, \ie $C_{II} = -C_{II}$, $(iii)$ the topological phase transition between Chern phases are generically mediated (discarding crystalline symmetry constraints) through a band inversion happening at a single point of the 2D Brillouin zone (this can be seen as the result of an embedding of the 2D phase in the 3D Brillouin zone of a Weyl semimetallic phase and a subsequent sweep of the two-torus section through one Weyl point). On the other hand for Euler phases we note $(i)$ a nonzero Euler class can only be carried by a pair of bands connected by a number $2\vert \chi\vert$ of stable nodal points distributed over the 2D Brillouin zone~\cite{bouhon2019nonabelian, bouhonGeometric2020, Bouhon2022braiding2, BJY_nielsen}, $(ii)$ the parity of the sum of the Euler classes of multiple two-band subspaces located within different energy windows must be zero, \ie $\sum_{J=I,II,\dots} \vert\chi_{J}\vert \mod 2 = 0$ (which guarantees that the total second Stiefel-Whitney class is zero, see the five-band phases below) \cite{bouhonGeometric2020,Bouhon2022braiding2}, and finally as a most striking difference $(iii)$ we recall that the topological phase transition between distinct gapped Euler phases is mediated by the braiding of nodal points belonging to adjacent energy gaps~\cite{Wu1273,bouhon2019nonabelian, bouhonGeometric2020, Peng2021, chen2021manipulation,Bouhon2022braiding2, BJY_nielsen}.

\subsection{Four-band Euler phases}\label{sec_4B_model}
We now move to the four-band Euler phases, assuming the spectral separation into two occupied and two unoccupied bands. The classifying space entials the non-oriented Grassmannian $\mathsf{Gr}_{2,4}^{\mathbb{R}} = \mathsf{SO}(4)/\mathsf{S}[\mathsf{O}(2)\times \mathsf{O}(2)]$. As for the three-band case, it is convenient, to build the Bloch Hamiltonian, to instead use the oriented Grassmannian $\widetilde{\mathsf{Gr}}_{2,4}^{\mathbb{R}} = \mathsf{SO}(4)/[\mathsf{SO}(2)\times \mathsf{SO}(2)]$ which enjoys a diffeomorphism to the product of two spheres, as will become apparent from the Pl{\"u}cker embedding. The derivation here starts from the standard rotation matrices that span $\mathsf{SO}(4)$, which differs from the approach of \cite{bouhonGeometric2020,Bouhon2022braiding2} (see also below).

The dimension of the Grassmannian, now seen as a manifold (see Section \ref{sec_PE}), is $\dim\widetilde{\mathsf{Gr}}_{2,4}^{\mathbb{R}}=\dim \mathsf{SO}(4)-2\dim \mathsf{SO}(2)=4\times3/2-2\times1=4$, and we symbolically parametrize a representing element $R^B\in\mathsf{SO}(4)$ of the coset $[R^B]$ with four angles, \ie $R^B=R^B(\theta^1,\theta^2,\theta^3,\theta^4)$ (from now on, we use superscripts for labeling the coordinates). We then write the real $2+2$-partitioned Bloch Hamiltonian in the canonical form
\begin{subequations}
\label{eq_H_22}
\begin{multline}
\label{eq_41}
    H^{\mathbb{R},2+2} =\\  R^{4B}(\theta^1,\theta^2,\theta^3,\theta^4)  \,\text{diag}\left[\begin{array}{c}E_1\\E_2\\E_3\\E_4\end{array}\right] R^{4B}(\theta^1,\theta^2,\theta^3,\theta^4)^{\top} \, ,
\end{multline}
assuming the energy ordering $E_1\leq E_2<E_3\leq E_4$, and give the frame of column eigenvectors $R^B=(u_1~u_2~u_3~u_4)$ the generic form (see below)
\begin{equation}
\label{eq_frame_parametrization_main}
    R^{4B}(\theta^1,\theta^2,\theta^3,\theta^4) = e^{\theta^1 L_{14}}
    e^{\theta^2L_{23}}
    e^{\theta^3L_{24}}  
    e^{\theta^4L_{13}}\,,
\end{equation}
\end{subequations}
formed by the rotation matrices $ e^{\theta \,L_{ij}}$,
with the angular momentum matrices $[L_{ij}]_{ab} = -\delta_{a i}\delta_{b j}+\delta_{a j}\delta_{b i}$ indexed by the pairs 
\begin{equation}
    \begin{aligned}
        (i,j)\in I_{2,4} &= \{(a,b)\vert 1\leq a< b\leq 4\} \,,\\
        &= \{(1,2),(1,3),(1,4),(2,3),(2,4),(3,4)\}\,,
    \end{aligned}
\end{equation}
that form a basis of the Lie algebra $\mathsf{so}(4)$ (dim\,$\mathsf{so}(4)$=6). Crucially, the form Eq.\,(\ref{eq_frame_parametrization_main}) is deduced from the local parametrization of the Grassmannian manifold in terms of the {\it normal coordinates}, see Section \ref{sec_norm_coo}. 

We now use the Pl{\"u}cker embedding to extract the Euler topology of Eq.\,(\ref{eq_41}), starting with the wedge products 
\begin{equation}
\begin{aligned}
    V_{I} (\theta^1,\theta^2,\theta^3,\theta^4) &=u_1\wedge u_2 \,,\\
    V_{II}(\theta^1,\theta^2,\theta^3,\theta^4) &= u_3\wedge u_4 \,,
\end{aligned}
\end{equation}
that both define a six-dimensional vector in $\bigwedge^2\left(\mathbb{R}^4\right)\cong \mathbb{R}^6$ and written in the basis 
\begin{equation}
    \{\check{e}_{m}\}_{m=1}^{N(N-1)/2} = \{e_i\wedge e_j\}_{(i,j)\in I_{2,4}} \,,    
\end{equation}
(again with $\{e_i\}_{i=1}^4$ the Cartesian basis of $\mathbb{R}^4$), \ie, using the Einstein summation convention,
\begin{equation}
    V_{I}=  V_I^{m} \check{e}_{m} \,,~
        V_{II}=  V_{II}^{m} \check{e}_{m} \,.
\end{equation}

We now perform an orthonormal change of basis, 
\begin{equation}
    \check{\bs{e}}^{\top} V_{\alpha}  = \check{\bs{e}}^{\top} M V'_{\alpha} = \check{\bs{e}}^{'\top} V_{\alpha}'\,,~\text{for}~\alpha = I,II\,,
\end{equation}
where
\begin{equation}
    M = \dfrac{1}{\sqrt{2}}\left[
        \begin{array}{rrrrrr}
             0 & 1 & 0 & 0 & 1 & 0  \\
             1 & 0 & 0 & 1 & 0 & 0  \\
             0 & 0 & 1 & 0 & 0 & 1  \\
             0 & 0 & 1 & 0 & 0 & -1  \\
             -1 & 0 & 0 & 1 & 0 & 0  \\
             0 & 1 & 0 & 0 & -1 & 0  
        \end{array}
    \right]\,,
\end{equation}
giving the rotated wedge vectors 
\begin{equation}
    V_{\alpha}'=\left[ 
        \begin{array}{r}
             V_{\alpha}^2-V_{\alpha}^5\\
              V_{\alpha}^1+V_{\alpha}^6\\
              V_{\alpha}^3+V_{\alpha}^4\\
              V_{\alpha}^2+V_{\alpha}^5\\
              V_{\alpha}^1-V_{\alpha}^6\\
              V_{\alpha}^3-V_{\alpha}^4
        \end{array}
    \right]\,,~\alpha=I,II\,.
\end{equation}

In order to make the geometry of the Grassmannian more apparent, we finally make the change of coordinates
\begin{equation}
\label{eq_Gr24_param_main}
\begin{alignedat}{4}
    \theta^1 & = \frac{1}{2}(2\pi+ \phi_- -\phi_+ ) \,,~ 
    &&\theta^3&&=  \frac{1}{2}(\pi+ \theta_- +\theta_+ )\,,\\
    \theta^2 &= \frac{1}{2}(\pi - \phi_- -\phi_+)\,,~ 
    &&\theta^4 &&=  \frac{1}{2}(\theta_- -\theta_+ )\,.
\end{alignedat}
\end{equation}
We then deduce that
\begin{subequations}
\label{eq_pluckerfour}
\begin{equation}
    \begin{aligned}
    V_I'(\theta_+,\phi_+,\theta_-,\phi_-) & = \dfrac{1}{\sqrt{2}} (n^1_{+},n^2_{+},n^3_{+},n^1_{-},n^2_{-},n^3_{-}) \\
    &= \dfrac{1}{\sqrt{2}}\left(\bs{n}_+ \oplus \bs{n}_-\right)\,,\\
    V_{II}'(\theta_+,\phi_+,\theta_-,\phi_-) & = \dfrac{1}{\sqrt{2}} (n^1_{+},n^2_{+},n^3_{+},-n^1_{-},-n^2_{-},-n^3_{-}) \\
    &= \dfrac{1}{\sqrt{2}}\left(\bs{n}_+ \oplus -\bs{n}_-\right)\,,\\
    \end{aligned}
\end{equation}
with the unit vectors 
\begin{equation}
\label{eq_unit_vec}
\begin{aligned}
    \bs{n}_{\pm}(\theta_{\pm},\phi_{\pm})&=(n^1_{\pm},n^2_{\pm},n^3_{\pm})_{(\theta_{\pm},\phi_{\pm})}\,,\\
    &= (\cos\phi_{\pm}\sin\theta_{\pm},\sin\phi_{\pm}\sin\theta_{\pm},\cos\theta_{\pm}) \in \mathbb{S}^2_{\pm} \,,
\end{aligned}
\end{equation}
such that 
\begin{equation}
\label{eq_4B_diff}
V_{\alpha}'(\theta_+,\phi_+,\theta_-,\phi_-) \in \mathbb{S}^2_+(\frac{1}{\sqrt{2}}) \times \mathbb{S}^2_-(\frac{1}{\sqrt{2}})\,,~\alpha=I,II\,.
\end{equation}
\end{subequations}
We have thus recovered the well-known diffeomorphism $\widetilde{\mathsf{Gr}}^{\mathbb{R}}_{2,4}\cong \mathbb{S}^2 \times \mathbb{S}^2$ \footnote{We note that 
\begin{equation}
     V'_{\pm} = V'_I\pm V_{II}' = \sqrt{2}\bs{n}_{\pm} \,,
\end{equation}
are the $\pm1$-eigenvectors of the Hodge star, \ie $*(V'_{\pm}) = \pm V'_{\pm}$.}. Although our results may seem cumbersome, they will prove to be very useful in the next section where we study in detail the Riemannian structures of the Euler phases. 

Crucially, we emphasize that our ansatz Eq.\,(\ref{eq_frame_parametrization_main}) of the frame of eigenvectors covers the whole of the four-dimensional Grassmannian which is the universal space for all real Bloch Hamiltonians with the $2+2$ spectral decomposition. In other words, every (real) four-band system is fully captured by Eq.\,(\ref{eq_H_22}).

We are now in the position to seek the Euler number of two-dimensional phases represented by $\widetilde{\mathsf{Gr}}_{2,4}^{\mathbb{R}}$. Focusing on orientable phases (assuming that the Bloch Hamiltonian is strictly periodic in the reciprocal space), the Bloch Hamiltonian maps continuously the two-torus Brillouin zone to a compact two-dimensional region of the Grassmannian. The structure of $\widetilde{\mathsf{Gr}}_{2,4}^{\mathbb{R}}$ comprises two embedded copies of $\widetilde{\mathsf{Gr}}_{2,3}^{\mathbb{R}}\cong \mathbb{S}^2$ \cite{Hatcher_2}. We thus conclude that the Bloch Hamiltonian maps the two-torus Brillouin zone either to one of the two spheres inside the Grassmannian, or it wraps the two spheres at once. The number of wrapping of the two two-spheres inside the Grassmannian (the degree of the Bloch Hamiltonian map) determines the topology of the system. 

As done previously, it is convenient to decompose the Bloch Hamiltonian map $H^{(W_+,W_-)}$ into a degree-1 map of the torus Brillouin zone to a sphere, $f_{TS}:\mathbb{T}^2\rightarrow \mathbb{S}^2_0$, followed by a map of the sphere to the Grassmannian, $f^{(W_+,W_-)}_{SG}:\mathbb{S}^2_0\rightarrow \mathbb{S}^2_+ \times \mathbb{S}^2_-$ of variable degrees $(W_+,W_-)$. While the first map $f_{TS}$ depends on the details of the system, the topology is only determined by the winding of the second map. It is therefore sufficient to consider the second map, which we simply define through 
\begin{subequations}
\label{eq_winding_ansatz}
\begin{multline}
    f^{(W_+,W_-)}_{SG}\left(\bs{n}(\theta,\phi)\right) = 
    \left\{\begin{aligned}
    &\bs{n}_+\left(\,[1-\delta_{W_+,0}]\,\theta,W_+\,\phi\right)\\
    &\bs{n}_-\left(\,[1-\delta_{W_-,0}]\,\theta,W_-\,\phi\right)
    \end{aligned}\right.\,,
\end{multline}
obtained from the map on the coordinates
\begin{equation}
\tilde{f}^{(W_+,W_-)}_{SG}(\theta,\phi) \mapsto \left\{
        \begin{aligned}
            \theta_{\pm} &= [1-\delta_{W_{\pm},0}]\,\theta \\
            \phi_{\pm} &= W_{\pm}\,\phi
        \end{aligned}
    \right.\,,
\end{equation}
and we write the resulting frame as
\begin{equation}
    R^{4B}(\theta^1,\theta^2,\theta^3,
    \theta^4)\rightarrow R^{4B}_{(W_+,W_-)}(\theta,\phi)\,.  
\end{equation}
\end{subequations}

We now show that the winding (wrapping) numbers $(W_\pm)$ of the unit vectors $\bs{n}_{\pm} \in \mathbb{S}^2_{\pm}$ readily determine the Euler numbers $(\chi_I,\chi_{II})$ of the two-band subspaces. Indeed, we find through direct computation (by integrating the Euler form $\mathsf{F}(\theta,\phi)$ over the sphere as in Eq.\,(\ref{eq_Euler_connection_form_class}))
\begin{equation}
\label{eq_euler_winding}
\begin{aligned}
        \chi_I &= \chi[\{u_1,u_2\}] = W_- - W_+ \,,\\
        \chi_{II} &= \chi[\{u_3,u_4\}] = -W_- -W_+\,.
\end{aligned}
\end{equation}
We thus recover the homotopy result $\pi_2[\mathsf{Gr}_{2,4}^{\mathbb{R}}] = \mathbb{Z}^2$. However, since the Hamiltonian homotopy classes are actually not sensitive to the orientation of the frame (see Appendix \ref{sec_orientability} and \cite{bouhonGeometric2020}), which is captured by the signs of the Euler numbers, the strict homotopy invariants are given the unsigned numbers
\footnote{Notwithstanding a subtle refined homotopy inequivalence $(\chi_I,\chi_{II})\sim (-\chi_I,-\chi_{II}) \not\sim (\chi_I,-\chi_{II})\sim (-\chi_I,\chi_{II})$ discussed in detail in \cite{Bouhon2022braiding2}.}
\begin{equation}
\label{eq_4B_classification}
\begin{aligned}
    \vert \chi_I \vert = \vert W_+ - W_- \vert \,,&~
        \vert \chi_{II} \vert = \vert W_+ +W_-\vert \,,\\
        W_+,&W_-\in\mathbb{Z}\,.
\end{aligned}
\end{equation} 

We note the following condition that is automatically satisfied 
\begin{equation}
    (\chi_{I} + \chi_{II}) \mod 2 = 0\,,
\end{equation}
guaranteeing that the total (stable) second Stiefel-Whitney class, \ie $\sum_{J=I}^{II} w_{2,J}$ with $w_{2,J} = \chi_{J}\mod 2$, vanishes. 

We conclude this section with the explicit form of the Bloch Hamiltonian in the above parametrization. Substituting the parameters Eq.\,(\ref{eq_Gr24_param_main}) in the $2+2$-Hamiltonian form Eq.\,(\ref{eq_H_22}), we get, after setting $E_1=E_2=-E_3=-E_4=-1$,
\begin{subequations}
\begin{equation}
\begin{aligned}
    H^{\mathbb{R},2+2}[\bs{n}_+,\bs{n}_-] =&\,  
        n^3_{-}\left(-n^3_{+} \Gamma_{33} -
        n^2_{+} \Gamma_{10} 
        + n^1_{+} \Gamma_{31}\right) \\
         +&\,
        n^1_{-} \left(- n^3_{+} \Gamma_{01} 
        -n^2_{+} \Gamma_{22} 
        -n^1_{+} \Gamma_{03}\right) \\
        +&\,
        n^2_{-} \left(+ n^3_{+} \Gamma_{13} -
        n^2_{+} \Gamma_{30} -
        n^1_{+} \Gamma_{11} \right) \,,\\
        =&\, 
        \bs{n}_+^{\top}\cdot \underline{\Gamma} \cdot \bs{n}_-\,,
\end{aligned}
\end{equation}
with the four-by-four matrices $\Gamma_{ij}=\sigma_i \otimes \sigma_j$ for $i,j=0,1,2,3$ and $\sigma_0=\mathbb{1}_2$ (not to be confused with the Christoffel symbols introduced in the next section), with the tensor
\begin{equation}
    \underline{\Gamma} = \left(\begin{array}{rrr}
        -\Gamma_{03} & -\Gamma_{11} & \Gamma_{31} \\
        -\Gamma_{22} & -\Gamma_{30} & -\Gamma_{10} \\
        -\Gamma_{01} & \Gamma_{13} & -\Gamma_{33}
    \end{array}\right)\,.
\end{equation}
\end{subequations}

We note that the above derivation differs from the previous ones exposed in \cite{bouhonGeometric2020,Bouhon2022braiding2} by the initial choice of the parametrization [Eq.\,(\ref{eq_frame_parametrization_main})] of the $\mathsf{SO}(4)$ matrix representing the frame of eigenvectors. The present approach in the parametrization of the Grassmannian is more systematic as it allows us to generalize it analytically to an even higher number of bands, as we show below. We note however that the Pl{\"u}cker embedding can always be found in any system numerically. (For yet a different choice of parametrization of the element $R\in\mathsf{SO}(4)$, see Appendix \ref{app_SO4_B}.)

\subsection{Five-band Euler-to-Stiefel-Whitney phases}\label{sec_5B_model}
%To exemplify this we close this section of representative models by 

We now turn to five-band phases which, beyond demonstrating the high flexibility of the Pl{\"u}cker approach, allows us to address the role of yet another invariant, namely the second Stiefel-Whitney class. In particular, we will consider systems with three occupied and two unoccupied bands (or equivalently, two occupied and three unoccupied bands, by simply reversing the sign of the Bloch Hamiltonian). As we will show, adding one extra band to a two-band subspace allows us to capture the $\mathbb{Z}\rightarrow \mathbb{Z}_2$ reduction of the homotopy classes, corresponding to the transition from the Euler of a rank-2 Bloch bundle to the second Stiefel-Whitney class of a rank-3 Bloch bundle.

We start with the frame of eigenvectors $R^B=(u_1~u_2~u_3~u_4~u_5)\in \mathsf{SO}(5)$ representing a generic coset $[R^B]\in \widetilde{\mathsf{Gr}}_{3,5}^{\mathbb{R}}$ which, since $\dim \widetilde{\mathsf{Gr}}_{3,5}^{\mathbb{R}}=\dim \mathsf{SO}(5)-\dim\mathsf{SO}(3)-\dim\mathsf{SO}(2)=5\times 4/2-3-1=6$, can be fully parameterized by six angles. We chose a form that directly embeds the above modeling of four-band systems, \ie (see also Section \ref{sec_norm_coo} on the local normal coordinates)
\begin{subequations}
\begin{multline}
\label{eq_R5_ansatz}
R^{5B}(\theta^1,\theta^2,\theta^3,\theta^4,\theta^5,\theta^6) = 
e^{\theta^6 L_{14}}e^{\theta^5 L_{15}} \\
\cdot e^{\theta^1 L_{25}}e^{\theta^2 L_{34}}e^{\theta^3 L_{35}}e^{\theta^4 L_{24}}\,,
\end{multline}
where the angular momentum matrices $[L_{ij}]_{ab} = -\delta_{a i}\delta_{b j}+\delta_{a j}\delta_{b i}$ indexed by the pairs 
\begin{equation}
    \begin{aligned}
        (i,j)\in I_{2,5} &= \{(a,b)\vert 1\leq a< b\leq 5\} \,,
    \end{aligned}
\end{equation}
now form a basis of the Lie algebra $\mathsf{so}(5)$ (dim\,$\mathsf{so}(5)$=10), entering the $3+2$-Bloch Hamiltonian form
\begin{multline}
\label{eq_H_32}
    H^{\mathbb{R},3+2} = R^{5B}(\theta^1,\theta^2,\theta^3,\theta^4,\theta^5,\theta^6) \, \text{diag}\left[\begin{array}{c}
        E_1\\E_2\\E_3\\E_4\\E_5
    \end{array}\right]\\
     \cdot R^{5B}(\theta^1,\theta^2,\theta^3,\theta^4,\theta^5,\theta^6)^{\top}\,,
\end{multline}
\end{subequations}
with $E_1\leq E_2 \leq E_3 < E_4 \leq E_5$. If we set $\theta^5=\theta^6=0$ in Eq.\,(\ref{eq_R5_ansatz}) the $(2{:}5\times2{:}5)$-block is identical to the four-band ansatz Eq.\,(\ref{eq_frame_parametrization_main}). The physical interpretation is, starting from the four-band case, that we have added one atomic orbital leading to an additional occupied band with eigenenergy $E_1$ and Bloch eigenvector $u_1$. Allowing the hybridization between the extra orbital and the four others, which is accounted for by nonzero values of $(\theta_5,\theta_6)$, while preserving the energy gap between the bands $3$ and $4$, the dimension of the Grassmannian increases from $4$ to $6$. 

Let us first impose the structure of the 4-band model to the product of matrix factors depending only on $\{\theta^1,\theta^2,\theta^3,\theta^4\}$, \ie we set 
\begin{multline}
R^{5B}(\theta^1,\theta^2,\theta^3,\theta^4,\theta^5,\theta^6) \rightarrow  
R^{5B}_{(W_+,W_-)}(\theta,\phi,\theta^5,\theta^6) \\
  = e^{\theta^6 L_{14}}e^{\theta^5 L_{15}} \left(\begin{array}{cc}
     1 & \mathbb{0}  \\
     \mathbb{0} & R^{4B}_{(W_+,W_-)}(\theta,\phi)
\end{array}
\right)\,,  
\end{multline} 
obtained from the change of coordinates $(\theta^1,\theta^2,\theta^3,\theta^4)\rightarrow(\theta_+,\phi_+,\theta_-,\phi_-)$ of Eq.\,(\ref{eq_Gr24_param_main}) and the parametrization $\{\theta_{\pm}(\theta,\phi),\phi_{\pm}(\theta,\phi)\}$ of Eq.\,(\ref{eq_winding_ansatz}) which fixes the winding of the two-dimensional Bloch Hamiltonian. Interestingly, we find the same Euler numbers for the bands $(2,3)$ and $(4,5)$ independently of the values of $(\theta^5,\theta^6)$. More remarkably, even after generalizing the frame of eigenvectors to
\begin{multline}
\label{eq_R32_extension}
    R^{5B}_{(W_+,W_-)}(\theta,\phi,\theta^5,\theta^6) \rightarrow \\
    e^{\theta^{10} L_{45}}e^{\theta^{9} L_{23}}
    e^{\theta^{8} L_{13}}e^{\theta^{7} L_{12}}
    R^{5B}_{(W_+,W_-)}(\theta,\phi,\theta^5,\theta^6)\,,
\end{multline}
corresponding to the complete hybridization of the five orbitals, the analytical Euler numbers remain unchanged, \ie we find
\begin{equation}
\begin{aligned}
    \chi_I&=\chi[\{u_2,u_3\}] = W_- - W_+\,,\\
    \chi_{II}&=\chi[\{u_4,u_5\}] = -W_- - W_+\,,
\end{aligned}\forall (\theta^5,\theta^6,\theta^7,\theta^8,\theta^9,\theta^{10})\,.
\end{equation}
In other words, the extension Eq.\,(\ref{eq_R32_extension}) preserves the homotopy classes of the four-band Bloch Hamiltonian. (This result does not depend on the ordering of the matrix factors depending on the angles $\{\theta^5,\dots,\theta^{10}\}$.) Furthermore, the above discussion (in the 4-band case) on the absence of an intrinsic orientation of the Hamiltonian (contrary to the frame that can be consistently oriented), still applies here and the homotopy classes are again classified by Eq.\,(\ref{eq_4B_classification}).
We however remark that in practice, we can compute an Euler number for the bands $2$ and $3$ only when band 1 is disconnected from them, \ie if $E_1<E_2\leq E_3$. 

We now address the question of the effect of mixing the eigenstates below the energy gap between bands 3 and 4, \ie mixing the occupied states ($u_1,u_2,u_3$), and the unoccupied states ($u_4,u_5$). This is accounted for through the extension
\begin{multline}
\label{eq_ext_2}
    R^{5B}_{(W_+,W_-)}(\theta,\phi,\theta^5,\theta^6) \rightarrow \\
    R^{5B}_{(W_+,W_-)}(\theta,\phi,\theta^5,\theta^6) 
    e^{\theta_{a}L_{23}}e^{\theta_{b}L_{45}}   e^{\theta_{c}L_{12}}e^{\theta_{d}L_{13}}\,, 
\end{multline}
corresponding to a general $\mathsf{SO}(3)\times \mathsf{SO}(2)$ transformation of the frame \footnote{We remark that the Bloch Hamiltonian Eq.\,(\ref{eq_H_32}) is strictly invariant under such a transformation
only if the energy levels of the two separated band subspaces are degenerate, \ie if $E_1=E_2=E_3$ and $E_4=E_5$. At a topological level, all such transformations are allowed as long as the principal energy gap remains open.}. As can be expected, the factors in $\theta_a$ (that mixes the eigenstates $2$ and $3$) and in $\theta_b$ (that mixes the eigenstates $4$ and $5$) do not affect the above homotopy classification. The factors in $\{\theta_c,\theta_d\}$ (mixing eigenstates 1 with 2 and 3), on the contrary, make the Euler class $\chi_I$ undefined, while $\chi_{II}$ remains unchanged. We review below that the homotopy characterization of the $3+2$-phase, when the mixing between the three occupied bands cannot be neglected (\ie due to band inversions among the three bands), undergoes a $\mathbb{Z}^2\rightarrow \mathbb{Z}$ reduction, such that the $\mathbb{Z}$ Euler class of the occupied band subspace (Bloch bundle) is reduced to the $\mathbb{Z}_2$ second Stiefel-Whitney class of the rank-3 occupied band subspace.   

We conclude that the two types of extensions, Eq.\,(\ref{eq_R32_extension}) and Eq.\,(\ref{eq_ext_2}), while keeping fixed the winding form inherited from the 4-band ansatz [Eq.\,(\ref{eq_winding_ansatz})] may reduce the homotopy classes of the $1+2+2$-gapped system (only through the specific factors in $\{\theta_c,\theta_d\}$) but have no effect on the homotopy class of the $3+2$-gapped Bloch Hamiltonian. We claim that any concrete 5-band $3+2$-real Bloch Hamiltonian can be modeled by the above extended ansatz of the frame of eigenvectors. 

We finally remark that the rationale for the homotopy features of the 4-band model to be preserved within the 5-band model is a manifestation of the successive embeddings of Grassmannians, \ie $\mathsf{Gr}^{\mathbb{R}}_{2,3} \hookrightarrow \mathsf{Gr}^{\mathbb{R}}_{2,4} \hookrightarrow \mathsf{Gr}^{\mathbb{R}}_{3,5}$ \cite{Hatcher_2}. For the modeling of orientable phases, we have the corresponding embeddings of oriented Grassmannians $\widetilde{\mathsf{Gr}}^{\mathbb{R}}_{2,3} \hookrightarrow \widetilde{\mathsf{Gr}}^{\mathbb{R}}_{2,4} \hookrightarrow \widetilde{\mathsf{Gr}}^{\mathbb{R}}_{3,5}$. In particular, both $\widetilde{\mathsf{Gr}}^{\mathbb{R}}_{2,4}$ and $\widetilde{\mathsf{Gr}}^{\mathbb{R}}_{3,5}$ contain two (homotopy equivalent) copies of $\widetilde{\mathsf{Gr}}^{\mathbb{R}}_{2,3}\cong \mathbb{S}^2$ (as CW subcomplexes \cite{Hatcher_2}), which explains that two winding numbers are sufficient to exhaustively model all two-dimensional topologies of five-band systems with a single energy gap. Indeed, there is no more two-dimensional CW subcomplex in $\widetilde{\mathsf{Gr}}^{\mathbb{R}}_{3,5}$ (see the discussion in Section \ref{sec_TB_model}). Very concretely, we have verified that the Euler numbers of the five-band model are completely independent of the angles $\{\theta^5,\theta^6\}$, \ie we cannot associate any further winding number with these dimensions. (This reflects that we have exhausted the two-dimensional CW subcomplexes. We keep a more formal analysis of this for later as this goes beyond the scope of the work.)

\subsubsection{Reduction of $\mathbb{Z}$-Euler to $\mathbb{Z}_2$-second Stiefel Whitney numbers}\label{sec_ESSW_reduction}
When more than two bands are connected together, the (2D) Euler number of the band subspace is not defined. The second Stiefel Whitney invariant should be used instead \cite{Ahn2018b} to characterize the $\mathbb{Z}_2$ topology. This can be readily explained in the following way. Let us consider the progressive hybridization of a two-band subspace hosting an Euler number $\chi$ with a third band (single bands of real Bloch Hamiltonian have a trivial 2D topology$-$discarding the 1D non-orientable topology). Before hybridization, the Euler number indicates the number of pairs of stable nodal points (NP) formed by the crossing of the two bands, \ie $\vert \chi\vert = 2\#\text{NP}$. Then switching on the hybridization with a third band facilitates band inversions among the three bands and the process of braiding the NP within one gap (say between bands 1 and 2) around the NP of the adjacent gap (between bands 2 and 3)~\cite{Bouhon2022braiding2,Ahn2018b,Ahn2019SW,Tiwari,bouhon2019wilson}. Then, the braiding process among three bands can only create or annihilate an even number of NP-pairs per gap \cite{bouhon2019nonabelian}. Therefore, whenever the two-band Euler number is even, the hybridization with a third band permits the complete annihilation of the stable nodes (a process we call ``debraiding'' \cite{bouhonGeometric2020,Bouhon2022braiding2}), leading to a trivial topology. On the other hand, when the two-band Euler number is odd, there is a minimum of one pair of stable nodes that remains irreducible upon hybridization with a third band. It is precisely this even-odd feature in the stability of NP-pairs that is captured by the second Stiefel-Whitney class $w_{2,I}$ of the lower three-band subspace, \ie 
\begin{equation}
    w_{2,I}=w_{2}[\{u_1,u_2,u_3\}] = \chi_{I} \mod 2 \in \mathbb{Z}_2\,,
\end{equation}
where $\chi_{I} = \chi[\{u_2,u_3\}]$. Again, the global consistency of the system (\ie the five bands taken together must realize a trivial vector bundle) requires 
\begin{equation}
    w_{2,I} + \chi_{II} \mod 2 = (w_{2,I} + w_{2,II}) \mod 2 = 0\,,
\end{equation}
where $w_{2,II} = w_{2}[\{u_4,u_5\}] = \chi_{II} \mod 2 $. 

The above implies that there exists a homotopy transformation from any $1+ 2+ 2$-phase $(\chi_{I},\chi_{II})$, \ie such that there are $\vert \chi_{I}\vert $ many NP-pairs formed by bands 2 and 3, and $\vert \chi_{II}\vert $ many NP-pairs formed by bands 4 and 5 before hybridization with the fifth orbital, to the maximally debraided phase, under the adiabatic constraint $E_3<E_4$. The maximally debraided phase is a new $1+2+2$-phase with the number of NP-pairs of $\vert \chi_{I}\vert\mod 2$ in the lower band subspace, while the number $\vert \chi_{II}\vert $ of NP-pairs in the upper band subspace is preserved. This homotopy deformation thus connects the above 5-band ansatz with the winding numbers 
\begin{equation}
\label{eq_5_1}
    W_+=-\dfrac{\chi_I+\chi_{II}}{2}\,,~
    W_- = \dfrac{\chi_I-\chi_{II}}{2}\,,
\end{equation}
to the 5-band ansatz with the winding numbers
\begin{equation}
\label{eq_5_2}
    W_+ = -\dfrac{\chi_I\,\text{mod}\,2+\chi_{II}}{2}\,,~
    W_- = \dfrac{\chi_I\,\text{mod}\,2-\chi_{II}}{2}\,.
\end{equation}
Thus, by allowing band inversions between band $1$ and the two-band subspace with bands $2$ and $3$, there exists be a deformation of the Bloch Hamiltonian from Eq.\,(\ref{eq_5_1}) to Eq.\,(\ref{eq_5_1}), while the energy gap between bands 3 and 4 remains open. This transition from the split band subspaces $\mathcal{B}_1 \cup \mathcal{B}_2 \oplus\mathcal{B}_3$ (we mean here the direct sum $\oplus$ of vector bundles \cite{Hatcher_2}) to $\mathcal{B}_1 \oplus \mathcal{B}_2 \oplus\mathcal{B}_3$, hence explicitly realizes the $\mathbb{Z}\rightarrow \mathbb{Z}_2$ reduction from the Euler to the second Stiefel-Whitney topology. While we can readily build such a homotopy deformation numerically, we leave it as an interesting future problem of finding a corresponding analytical homotopy path.

Proceeding further with the addition of one extra unoccupied band, thus realizing the Grassmannian $\mathsf{G}_{3,6}^{\mathbb{R}}$, the $\mathbb{Z}$ Euler number above the gap, $\chi_{II}=\chi[\{u_4,u_5\}]$, is now also reduced to a $\mathbb{Z}_2$ second Stiefel-Whitney number, \ie
\begin{equation}
    w_{2,II}=w_{2}[\{u_4,u_5,u_6\}]=\chi_{II}\mod 2\,,
\end{equation}
under the global constraint
\begin{equation}
    (w_{2,I} + w_{2,II} ) \mod 2 =0\,.
\end{equation}
Therefore, there is a reduction of the second homotopy group of $\mathbb{Z}^2\rightarrow \mathbb{Z}_2$ from $\widetilde{\mathsf{G}}_{2,4}^{\mathbb{R}}$ to $\widetilde{\mathsf{G}}_{3,6}^{\mathbb{R}}$, since now both band subspaces (occupied and unoccupied) are characterized by a single $\mathbb{Z}_2$ second Stiefel-Whitney number. 

We will come back to the phenomenology of the second Stiefel-Whitney topology in Section \ref{sec:5-band and sw} where we reveal its manifestation within the geometric properties of the system.

\subsection{Derivation of tight-binding models}\label{sec_TB_model}
As is evident from the general nature of the Pl\"ucker framework as outlined above, it provides a concrete route towards defining models in all generality. We wish to nonetheless comment on the general case with a few remarks. We reiterate that our approach has been to decompose the winding Hamiltonian map from the torus Brillouin zone to the Grassmannian into two steps, \ie
$H^W=f^W_{SG}\circ f_{TS}$ [Eq.\,(\ref{eq_pullback})]. In the general context we are seeking the expression of an explicit tight-binding Hamiltonian that realizes any given homotopy class, \ie with a prefixed topology. Considering the above few-band examples, the topological sector is fixed by $W$, or $\bs{W}=(W_+,W_-)$, independently of the local details of the maps $f^W_{SG}$ and $f_{TS}$. 

While we have defined the map $f^W_{SG}$ ($f^{\bs{W}}_{SG}$) for the few-band models, we now need to specify $f_{TS}:\bs{k}\mapsto (\theta,\phi)$. An obvious choice is
\begin{equation}
\label{eq_torus2sphere}
\left\{
\begin{aligned}
    \theta(\bs{k}) &= \max\{\vert k_1\vert,\vert k_2\vert\}\,,\\
    \phi(\bs{k}) &= \arg(k_1+\imi k_2) \,.
    \end{aligned}\right.
\end{equation}
We then expand each element of the Bloch Hamiltonian as the truncated Fourier series
\begin{equation}
    H_{\alpha\beta}(\bs{k}) = 
    \sum\limits_{n_1,n_2=-N_{\text{max}}}^{N_{\text{max}}} t_{\alpha\beta}^{(n_1,n_2)}
    e^{\imi \bs{k}\cdot \bs{\delta}^{(n_1,n_2)} }\,,
\end{equation}
with the hopping vectors $\bs{\delta}^{(n_1,n_2)}=n_1 \bs{a}_1+n_2 \bs{a}_2$ (where $n_1,n_2\in\mathbb{Z}^2$ and $\{\bs{a}_1,\bs{a}_2\}$ are the primitive vectors), and where the hopping parameters are obtained through the discrete Fourier transform
\begin{equation}
    t_{\alpha\beta}^{(n_1,n_2)} =
    \dfrac{1}{N_{k}^2} 
    \sum\limits_{\kappa_1,\kappa_2=-N_{k}}^{N_{k}} e^{-\imi \dfrac{\kappa_1 \bs{b}_1+\kappa_2 \bs{b}_2}{2 N_{k}}\cdot \bs{\delta}^{(n_1,n_2)} } H_{\alpha\beta}(\bs{k})\,,
\end{equation}
with the primitive reciprocal lattice vectors $\{\bs{b}_1 = 2\pi \bs{a}_2 \times \bs{a}_3/\vert \bs{a}_1\cdot \bs{a}_2 \times \bs{a}_3\vert   ,\bs{b}_2= 2\pi \bs{a}_3 \times \bs{a}_1/\vert \bs{a}_1\cdot \bs{a}_2 \times \bs{a}_3\vert\}$ with $\bs{a}_3=\bs{a}_1\times \bs{a}_2/\vert \bs{a}_1\times \bs{a}_2\vert$. By setting a finite sampling of the Bloch Hamiltonian over the Brillouin zone, \ie taking $N_k$ finite, and setting the maximum range of the hopping processes as $N_{\text{max}} \lesssim 3 N_k$ we guarantee the analyticity of each term $H_{\alpha\beta}(\bs{k})$.

An important remark is in place here. By construction, any gapped Bloch Hamiltonian must realize a point of its associated (unoriented, see Appendix \ref{sec_orientability}) Grassmannian. However, the Bloch Hamiltonians have additional structures beyond topology. Indeed, the set of all explicit Bloch Hamiltonians at a given momentum is in one-to-one correspondence with the set of all orthonormal frames and ordered energy eigenvalues, \ie (say in the real case)
\begin{multline}
    \left\{\left.\mathcal{H}^{p+(N-p)}\right\vert_{\bs{k}}\right\} = \mathsf{SO}(N) \times 
    \left\{
    (E_1,\dots,E_N)_{\bs{k}}\in\mathbb{R}^N \big\vert \right.\\
    \left. E_1\leq \cdots\leq E_p < E_{p+1}\leq\cdots\leq E_N  
    \right\}\,.
\end{multline}
For the complex case, we have the unitary group instead. Therefore, the map from a Pl{\"u}cker-based ansatz to an explicit tight-binding model is one-to-many. While we are here only interested in a representative model of a given homotopy class, our approach can be combined with further selection criteria pertaining to concrete physical systems. For instance, a higher winding is less likely in solid-state contexts since it typically requires long-range hopping, while the corresponding parameters are typically exponentially suppressed for increasing distances. The question of modeling optimization is a very promising one, but this goes beyond the scope of the present work which instead focuses on the general Pl{\"u}cker framework to capture topological but also, as shown below, geometrical features in the many-band context.

\subsection{Generalization to arbitrarily-many-band systems}\label{sec_Nband}
Our Pl{\"u}cker approach is strongly based on the homotopy classification of the topological phases of specific gapped Bloch Hamiltonians. We first consider the effect of including more bands, while preserving the condition of a single energy gap, in which case the classifying space is always a Grassmannian. Above, we have discussed the effect on the homotopy classification of two-dimensional phases of successively adding single additional bands first to the occupied, then to the unoccupied two-band subspaces, corresponding to the embedding of $\mathsf{G}_{2,4}^{\mathbb{R}}$ in $\mathsf{G}_{3,5}^{\mathbb{R}}$, and then of $\mathsf{G}_{3,5}^{\mathbb{R}}$ in $\mathsf{G}_{3,6}^{\mathbb{R}}$. We have argued that only two winding numbers are needed to exhaustively model all the (orientable) topological gapped phases. The rationale behind this is the fact that $\widetilde{\mathsf{G}}_{2,4}^{\mathbb{R}}$, $\widetilde{\mathsf{G}}_{3,5}^{\mathbb{R}}$, and $\widetilde{\mathsf{G}}_{3,6}^{\mathbb{R}}$, each contains only two two-dimensional CW subcomplexes that are both homotopy equivalent to $\widetilde{\mathsf{G}}_{3,4}^{\mathbb{R}}=\mathbb{S}^2$. It can be easily checked (through the counting of the echelon forms of the rectangular matrix $(u_1~u_2)$, see \cite{Hatcher_2}) that every oriented Grassmannian with $p\geq 2$ and $N-p\geq 2$ does contains only two two-dimensional CW subcomplexes. Then, through the sub-complex inclusions $\widetilde{\mathsf{G}}_{2,4}^{\mathbb{R}}\hookrightarrow\widetilde{\mathsf{G}}_{3,5}^{\mathbb{R}}\hookrightarrow\widetilde{\mathsf{G}}_{3,6}^{\mathbb{R}}\hookrightarrow\cdots$, these two-dimensional CW subcomplexes are homotopy equivalent to $\widetilde{\mathsf{G}}_{2,3}^{\mathbb{R}}\cong \mathbb{S}^2$. We thus conclude that two winding numbers are sufficient to exhaustively model all the homotopy classes of two-dimensional Bloch Hamiltonians with a single energy gap. (We note that the CW subcomplex structure is a topological notion in essence, while we rely on the smoothness of the Bloch Hamiltonian map to relate it to geometric structures.)

Generalizing further, we have given in Ref. \cite{bouhonGeometric2020} the complete homotopy classification of real Bloch Hamiltonian of multi-gap phases, \ie when several energy gaps are specified in the spectrum. The topology of these generalized gapped phases is fully captured by the real (unoriented) partial flag manifolds  
\begin{multline}
\label{eq_partial_flag}
    \mathsf{Fl}_{p_1,\dots,p_{n_g},N-p_1\dots-p_{n_g}} 
    = \\
    \dfrac{\mathsf{O}(N)}{\mathsf{O}(p_1)\times\cdots\times \mathsf{O}(p_{n_g})\times \mathsf{O}(N-p_1\dots-p_{n_g})}\,,
\end{multline}
here for a system with $n_g$ energy gaps and $n_g+1$ successive band subspaces of rank $p_1$,$\dots$, $p_{n_g}$, and $N-p_1\dots-p_{n_g}$.

Focusing on orientable phases (\ie excluding $\pi$-Berry phase polarizations), in which case the multi-gap Bloch Hamiltonian defines a (maximally) two-dimensional region of the {\it oriented} flag manifold 
\begin{multline}
\label{eq_partial_flag}
    \widetilde{\mathsf{Fl}}_{p_1,\dots,p_{n_g},N-p_1\dots-p_{n_g}} 
    = \\
    \dfrac{\mathsf{SO}(N)}{\mathsf{SO}(p_1)\times\cdots\times \mathsf{SO}(p_{n_g})\times \mathsf{SO}(N-p_1\dots-p_{n_g})}\,,
\end{multline}
we showed in \cite{bouhonGeometric2020,Bouhon2022braiding2} that any $n'$-th two-band subspace (\ie rank$\,\mathcal{B}_{n'}=2$) is characterized by an Euler number $\chi_{n'}\in \mathbb{Z}$ and that any $n''$-th band subspace with more than two bands (\ie rank$\,\mathcal{B}_{n''}>2$) is characterized by a second Stiefel-Whitney number $w_{2,n''}\in \mathbb{Z}_2$, under the global constraint that $\left(\sum_{n'} \chi_{n'} + \sum_{n''} w_{2,n''}\right) \mod 2 = 0 $. We hence conclude, generalizing our discussion of the five-band model, that the generic Pl{\"u}cker ansatz of a multi-gap phase is characterized by a vector of winding numbers $\bs{W}=(W_1,\dots,W_{n_W})$ with one winding number per band-subspace of rank $r\geq 2$. 

We importantly note that the homotopy classification of the gapped phases is over-determined by the winding numbers used in the modeling. First, the models are homotopy equivalent upon the inversion of pairs of Euler numbers, $(\chi_{n_1},\chi_{n_2})\rightarrow (-\chi_{n_1},-\chi_{n_2})$ \cite{bouhonGeometric2020,Bouhon2022braiding2}. Then, there is the $\mathbb{Z}\rightarrow\mathbb{Z}_2$ (Euler-to-Stiefel-Whitney) topological reduction for every band-subspace with more than two bands, as we have shown above for the five-band model.

The above examples highlight the highly versatile nature of the Pl{\"u}cker description to obtain analytical models of few-band systems. We emphasize that at no point we had to solve the eigenvalue problem from the Bloch Hamiltonian. Given that there is no analytical solution to the eigenvalue problem in the five-band case and that the ansatz for the three-band and four-band cases would not allow the analytical integration of the topological invariants, the simplicity of our approach based on the Pl{\"u}cker embedding to fully capture the topology is rather striking. In its essence, our method naturally generalizes to the general $N$-band context and to the multi-gap phases addressed above. Indeed, this applicability goes beyond formulating analytical descriptions as the approach also provides a general route to numerically formulate realistic $N$-band systems beyond the analytical tractable limit. 

In the following sections, we show that our Pl{\"u}cker approach is not only useful for the characterization of topology but, moreover, gives a natural framework to obtain new geometric signatures of topological condensed matter systems.

\section{Riemanniann structure of many-band systems through the Pl{\"u}cker representation of Grassmannians}
The Pl\"ucker approach does not only provide for a direct modeling of phases and a concise non-redundant description of the Fubini-Study metric as outlined above for the many band Chern case but more importantly allows for a general (and moreover directly calculable) formulation of the metric and quantum geometric tensor. Due to the above demonstrated general applicability of the embedding, this route provides an efficient tool to address the generalization of single band metric to arbitrary systems, that is generalizing Eqs.\,\eqref{eq:Chern_genplucker} beyond the $n$-band Chern case, which formed the true generalization of the single band expressions Eqs.\,\eqref{eq:singlebandmetric}.

Here, the emphasis will be on the case of real Bloch Hamiltonians, \ie hosting Euler and Stiefel-Whitney topologies, as these have not been systematically considered under the Riemannian viewpoint before while they exhibit rich geometric features. We moreover note that with our Pl\"ucker formalism, tractable results for the complex case can be readily obtained from the real ones simply by treating the complex Grassmannian as a combination of two real manifolds generated by the separation of the unitary eigenvector-frames as $U=R_r + \imi \,R_i $ with $R_r=\text{Re}\,U $ and $R_i=\text{Im} \,U$.

\subsection{General Pl{\"u}cker framework for multi-band Bloch Hamiltonians}\label{sec_general_metric}

\subsubsection{Pl{\"u}cker embedding}\label{sec_PE}

Since the above Grassmannians originate from the gauge structure of the spetral decomposition of the gapped Bloch Hamiltonian, \ie$H(\bs{k}) = U(\bs{k})\cdot \text{diag}[E_1(\bs{k}),\dots,E_p(\bs{k}),E_{p+1}(\bs{k}),\dots,E_N(\bs{k})] \cdot U(\bs{k})^{\dagger}$, with the ordered energy eigenvalues 
\begin{equation}
    E_1(\bs{k})\leq \cdots\leq E_p(\bs{k})< E_{p+1}\leq\cdots\leq E_N(\bs{k})\,,    
\end{equation}
and the corresponding matrix of Bloch (column) eigenvectors $U(\bs{k})=\left[u_1(\bs{k})~\cdots~ u_{N}(\bs{k})\right]\in \mathsf{U}(N)$, we first define the classifying Grassmannian as the set of left cosets 
\begin{equation}
\begin{aligned}
    [U] &= \{U\cdot [G_I\oplus G_{II}]\vert G_I\oplus G_{II}\in\mathsf{U}(p)\times\mathsf{U}(N-p)\}\,,\\
    &\in \mathsf{Gr}^{\mathbb{C}}_{p,N} = \mathsf{U}(N)/[\mathsf{U}(p)\times \mathsf{U}(N-p)]\,,
\end{aligned}
\end{equation}
\ie the Grassmannian is here defined as a homogeneous space. In the case of real Bloch Hamiltonians we replace the unitary matrix and unitary groups ($U(\bs{k}),\mathsf{U}(N),\mathsf{U}(p),\mathsf{U}(N-p)$) by an orthogonal matrix and orthogonal groups ($R(\bs{k}),\mathsf{O}(N),\mathsf{O}(p),\mathsf{O}(N-p)$). Since we will focus in this work on the physical signatures coming from the geometric features of {\it orientable} phases, we will mainly work with the oriented Grassmannians, \ie $\widetilde{\mathsf{Gr}}^{\mathbb{R}}_{p,N} = \mathsf{SO}(N)/[\mathsf{SO}(p)\times \mathsf{SO}(N-p)]$ and $\widetilde{\mathsf{Gr}}^{\mathbb{C}}_{p,N} = \mathsf{SU}(N)/[\mathsf{SU}(p)\times \mathsf{SU}(N-p)]$.

The Pl{\"u}cker embedding then allows us to represent a point $[R]\in\mathsf{G}$ of the real oriented Grassmannian as a vector in the $p$-th external power space, the elements of which we call $p$-vectors, \ie
\begin{equation}
\begin{aligned}
    \iota_P : \widetilde{\mathsf{Gr}}_{p,N}^{\mathbb{R}} &\hookrightarrow \iota_P\left(\widetilde{\mathsf{Gr}}_{p,N}^{\mathbb{R}}\right) \equiv \mathsf{G} \subset \bigwedge^p\mathbb{R}^N \cong \mathbb{R}^{\left(\substack{N\\p}\right)} : \\
    [R] &\mapsto V = u_1\wedge \cdots\wedge u_p = V^{m} \check{e}_m
\,,
\end{aligned}
\end{equation}
where $(\check{e}_1,\dots,\check{e}_{\left(\substack{N\\p}\right)})$ is a Cartesian basis of the (Euclidean) exterior power space $\bigwedge^p\mathbb{R}^N$, \ie
\begin{equation}
    \check{e}_m = e_{i^m_1}\wedge \cdots \wedge e_{i^m_p}\,,~
    (i^m_1,\dots,i^m_p) \in I_{p,N} \,,
\end{equation}
with $(e_1,\dots,e_N)$ the Cartesian basis of the underlying $\mathbb{R}^N$ Euclidean space, and with the set of $\left(\substack{N\\p}\right)$ possible ordered $p$-tuples of indices 
\begin{equation}
    I_{p,N} = \left\{(i_1,\dots,i_p) \vert 1\leq i_1 < \cdots< i_p \leq N\right\} \,.
\end{equation}

The general conditions for a $p$-vector $V \in \bigwedge^p \mathbb{R}^N$ to belong to $\iota_P \left( \widetilde{\mathsf{Gr}}_{p,N}^{\mathbb{R}}\right)$ are that it must be (i) a {\it simple} $p$-vector, \ie such that there exists a basis $(e'_i)_{i=1,\dots,N}$ of $\mathbb{R}^N$ in which the $p$-vector takes the form $V=e'_1\wedge \cdots \wedge e'_N$, and $(ii)$ a unit vector, \ie $V\in \mathbb{S}^{\left(\substack{N\\p}\right)-1}\subset \mathbb{R}^{\left(\substack{N\\p}\right)}$ \cite{Kozlov_Gr}.
By defining the $p$-vector $V$ as a wedge product of Bloch eigenvectors, \ie that are columns of an orthogonal matrix, both conditions are readily satisfied. It follows from the above considerations that $\mathsf{G}$ is a submanifold of the exterior power space \cite{Kozlov_Gr} with dimension $\dim \widetilde{\mathsf{Gr}}_{p,N}^{\mathbb{R}} = \dim \mathsf{SO}(N) - \dim \mathsf{SO}(p) - \dim \mathsf{SO}(N-p) = p(N-p)$. From now on we will use the short notations $\wedge_p  = \bigwedge^p \mathbb{R}^N$ and  $\mathsf{G}=\iota_P \left( \widetilde{\mathsf{Gr}}_{p,N}^{\mathbb{R}}\right)$ for the submanifold in $\wedge_p$. 

%\begin{itemize}
%    \item a {\it simple} $p$-vector, \ie there exists a basis $(e'_i)_{i=1,\dots,N}$ of $\mathbb{R}^N$ in which the $p$-vector takes the form $V=e'_1\wedge \cdots \wedge e'_N$
 %   \item a unit vector, \ie $V\in \mathbb{S}^{\left(\substack{N\\p}\right)-1}\subset \mathbb{R}^{\left(\substack{N\\p}\right)}$~\cite{Kozlov_Gr}.
%\end{itemize}

%$(i)$ a {\it simple} $p$-vector, \ie such that there exists a basis $(e'_i)_{i=1,\dots,N}$ of $\mathbb{R}^N$ in which the $p$-vector takes the form $V=e'_1\wedge \cdots \wedge e'_N$, and $(ii)$ a unit vector, \ie $V\in \mathbb{S}^{\left(\substack{N\\p}\right)-1}\subset \mathbb{R}^{\left(\substack{N\\p}\right)}$ \cite{Kozlov_Gr}. By defining the $p$-vector $L$ as a wedge product of Bloch eigenvectors, \ie that are columns of an orthogonal matrix, both conditions are readily satisfied. It follows from the above considerations that $\mathsf{G}$ is a submanifold of the exterior power space \cite{Kozlov_Gr}. From now on we will use the short notations $\wedge_p  = \bigwedge^p \mathbb{R}^N$ and  $\mathsf{G}=\iota_P \left( \widetilde{\mathsf{Gr}}_{p,N}^{\mathbb{R}}\right)$ for the submanifold in $\wedge_p$.  

\subsubsection{From the local normal coordinates to the global parametrization of Grassmannians}\label{sec_norm_coo}
We now need to make a brief detour through the projector matrix representation, noted $\mathsf{G}_{\text{proj}}$, of the unoriented Grassmannian $\mathsf{Gr}_{p,N}^{\mathbb{R}}$ inherited from the Lie group $\mathsf{SO}(N)$. We follow \cite{Grassmannian_handbook} for this. We have the projection $\Phi_{P_0}:\mathsf{SO}(N)\rightarrow \mathsf{G}_{\text{proj}}: R\mapsto R_p R_p^{\top} =  RP_0R^{\tau} $, where $R_{p} = \left(u_1\cdots u_p \right)$ is the rectangular matrix of the $p$ first columns of $R$, and the projector 
$P_0 = \bigl[ \begin{smallmatrix} \mathbb{1}_p & 0 \\ 0 & 0 \end{smallmatrix} \bigr] $. Then, the differential of the projection gives a map from the tangent space of $\mathsf{SO}(N)$ at $R$, $T_R\mathsf{SO}(N)$, to the tangent space of the Grassmannian at $ R_pR^{\tau}_p$, $T_{R_pR^{\tau}_p}\mathsf{G}_{\text{proj}}$. Given a generic tangent vector $X = \bigl[ \begin{smallmatrix} A & -B^{\top} \\ B & C \end{smallmatrix} \bigr]\in \mathsf{so}(N) = T_{\mathbb{1}_N}\mathsf{SO}(N) $, \ie $X$ is a real skew-symmetric matrix with in particular an arbitrary rectangular matrix $B\in \mathbb{R}^{(N-p)\times p}$, the corresponding tangent direction at $R$ is $L_{R*}X=RX$. We then find $d\Phi_{P_0}: T_R\mathsf{SO}(N)\rightarrow T_{R_pR^{\tau}_p}\mathsf{G}_{\text{proj}} : RX\mapsto \left.d/dt\right\vert_{t=0} \gamma(t)P_0 \gamma(t)^{\top} = R \bigl[ \begin{smallmatrix} 0 & B^{\tau} \\ B & 0 \end{smallmatrix} \bigr] R^{\top}$, with $\gamma(t)$ a curve in $\mathsf{SO}(N)$ starting at $\gamma(0)=R$ with a tangent vector $\dot{\gamma}(0) = RX$ \cite{Grassmannian_handbook}. The expression $R \bigl[ \begin{smallmatrix} 0 & B^{\tau} \\ B & 0 \end{smallmatrix} \bigr] R^{\top}$ implies that the angular momentum matrices $ \{L_{ij}\}_{i=1,\dots,p}^{j=p+1,\dots,N}$ acting as a basis for $B$, \ie
\begin{equation}
\label{eq_basis_TG}
    \bigl[ \begin{smallmatrix} 0 & -B^{\top} \\ B & 0 \end{smallmatrix} \bigr] = 
    \sum\limits_{i=1}^p \sum\limits_{j=p+1}^{N}  \theta^{ij} L_{ij}\,,
\end{equation}
with the variables $\left\{\theta^{ij}\right\}_{i=1,\dots,p}^{j=p+1,\dots,N}$, equivalently act as a basis for the tangent space of the Grassmannian at any point $R_pR_p^{\top}$. 

A geodesic in $\mathsf{SO}(N)$ starting at $R$ with a tangent vector $X = \bigl[ \begin{smallmatrix} A & -B^{\top} \\ B & C \end{smallmatrix} \bigr] \in T_{R} \mathsf{SO}(N)$ is defined via the exponential map through $\gamma(t)=R e^{t X}$. Projecting onto the Grassmannian, the corresponding geodesic in $\mathsf{G}_{\text{proj}}$ is then obtained through \cite{Grassmannian_handbook}
\begin{equation}
\label{eq_geodesic_gras}
    \begin{aligned}
        \Phi_{P_0}(\gamma(t)) &= \Phi_{P_0}\left( R e^{t \bigl[ \begin{smallmatrix} 0 & -B^{\top} \\ B & 0 \end{smallmatrix} \bigr]} \right) \\
        &= R e^{t \bigl[ \begin{smallmatrix} 0 & -B^{\top} \\ B & 0 \end{smallmatrix} \bigr]} P_0  e^{t \bigl[ \begin{smallmatrix} 0 & B^{\top} \\ -B & 0 \end{smallmatrix} \bigr]}R^{\top}\,.    
    \end{aligned}
\end{equation}

Combining now Eq.\,(\ref{eq_basis_TG}) and Eq.\,(\ref{eq_geodesic_gras}), we arrive at the local parametrization of the Grassmannian \cite{Grassmannian_handbook}
\begin{equation}
\begin{aligned}
     \rho^{\mathsf{G}_{\text{proj}}} : \mathbb{R}^{(N-p)\times p} &\rightarrow \mathsf{G}_{\text{proj}} : \\
     \left[\theta^{ij}\right]_{ij} &\mapsto \Phi_{P_0}(\gamma(1))  \,,
\end{aligned}
\end{equation}
where the angles $\{\theta^{ij}\}_{ij}$ are called the {\it normal coordinates} of the Grassmannian at $R_pR_p^{\top}$. The lifting of the parametrization from $\mathsf{G}$ to $\mathsf{SO}(N)$ is then simply given by
\begin{equation}
\begin{aligned}
    \rho^{\mathsf{SO}(N)} : \mathbb{R}^{(N-p)\times p} &\rightarrow \mathsf{SO}(N) : \\
    \left[\theta^{ij}\right]_{ij} &\mapsto \gamma(1) = R e^{ 
        \sum\limits_{i=1}^p \sum\limits_{j=p+1}^{N}  \theta^{ij} L_{ij}
    }\,.
\end{aligned}
\end{equation}

In the following, we label the normal coordinates through $\{\theta^{\mu_{ij}}\}_{\mu_{ij}=1,\dots,p(N-p)}$ after setting a one-to-one correspondence between the sets $\{\mu_{ij}\}_{i=1,\dots,p}^{j=p+1,\dots,N}$ and $\{1,\dots,p(N-p)\}$. Choosing $R=\mathbb{1}_N$ as the origin of the parametrization, we deduce the following global ansatz for the $N$-band frame of eigenvectors that diagonalizes a $p+ (N-p)$-gapped Bloch Hamiltonian
\begin{equation}
\begin{aligned}
\label{eq_global_param}
    R^{NB}(\bs{\theta}) &= 
    \prod\limits_{i=1}^p \prod\limits_{j=p+1}^{N} 
    e^{\theta^{\mu_{ij}} L_{ij}}\,,\\
    \bs{\theta}&= \left(\theta^{\mu_{ij}}\right)_{\mu_{ij}}
 =\left(\theta^1,\dots,\theta^{p(N-p)}\right)\,,
\end{aligned}
\end{equation}
where the angle variables now play the role of curvilinear coordinates that parameterize the Grassmannian globally. The rationale for choosing this form is the agreement with the normal coordinates in the first order of the Taylor expansion around each angle variable taken separately. This is the ansatz used in Section \ref{sec_4B_model} for the 4-band case and in \ref{sec_5B_model} for the 5-band case. It is clear that this framework can be generalized to an arbitrary number of bands, which is a direction we will pursue in subsequent works. While the ordering of the matrix exponentials is not important {\it per se}, once chosen it determines the explicit expressions of the homotopy analysis, as in the previous section. In the following, we drop the $ij$-indices and label the angle variables simply as $\{\theta^\mu\}_{\mu=1,\dots,p(N-p)}$. We will also use the Einstein summation convention over repeated indices.

\subsubsection{Pl{\"u}cker tangent and cotangent bundles}

Now that we have a global and intrinsic coordinate system for the Grassmannians, it is straightforward to express all the derived geometric quantities, \ie from the tangent bundle to the metric, then to Riemannian tensor and the sectional curvature. We start here with the definition of the tangent and cotangent bundles of the Grassmannian in the Pl{\"u}cker framework. 

The parametrization Eq.\,(\ref{eq_global_param}) readily equips the Pl{\"u}cker $p$-vector with a global parametrization, \ie
\begin{equation}
\begin{aligned}
    \iota_P([R^{NB}(\bs{\theta})]) &= V(\bs{\theta})\\
    & = V^m(\bs{\theta}) \,\partial_m \in \mathsf{G} \subset \wedge_p \,,
\end{aligned}
\end{equation}
where the frame $(\partial_m)_{m}$ gives a Cartesian basis of $\wedge_p\cong \mathbb{R}^{\left(\substack{N\\p}\right)}$. 

Let us define the coordinate map $\varphi:\mathbb{R}^{p(N-p)}\rightarrow \widetilde{G}_{p,N}^{\mathbb{R}}:\bs{\theta}=(\theta^1,\dots,\theta^{p(N-p)}) \mapsto [R(\bs{\theta})]$, we then define the intrinsic tangent vectors of the Grassmannian induced by $\varphi$, \ie 
\begin{equation}
    \partial_\mu^{\text{int}} = \partial_{\theta^{\mu}}^{\text{int}}
    = \varphi_*\left(\dfrac{\partial~~\;}{\partial\theta^{\mu}}\right)
    = \partial_{\theta^{\mu}} [R(\bs{\theta})]
    \,,~\text{for\;every}~\mu\,. 
\end{equation}
Then, we define the tangent vectors induced by the push-forward by the Pl{\"u}cker embedding, \ie $\iota_{P*}:T_{\bs{\theta}}\widetilde{\mathsf{G}}_{p,N}^{\mathbb{R}}\rightarrow T_{V(\bs{\theta})}\mathsf{G} \subset T_{V(\bs{\theta})}\wedge_p $ (in the following we write $T_{\bs{\theta}}\mathsf{G}$, since there is a one-to-one correspondence $\bs{\theta}\leftrightarrow V(\bs{\theta})$ and there is no ambiguity), which gives
\begin{equation}
\label{eq_frame_TG}
\begin{aligned}
    \partial_\mu = \partial_{\theta^\mu} &\equiv \iota_{P*}(\partial_\mu^{\text{int}}) = \partial_{\mu} V^m(\bs{\theta}) \,e_m \,, \\
    v_\mu^m &= \partial_{\mu} V^m\,,
\end{aligned}
\end{equation}
for $\mu=1,\dots,p(N-p)$, where $\{e_m\}_{m}$ is a Cartesian basis for $\wedge_p$. The frame $(\partial_{\mu})_{\mu}$ forms a basis of $T\mathsf{G}$ that is parameterized by the intrinsic angle coordinates $\{\theta^{\mu}\}_{\mu}$. Furthermore, the form of Eq.\,(\ref{eq_frame_TG}), \ie a change of coordinates, implies a zero Lie bracket, \ie $[\partial_{\mu},\partial_{\nu}]=0$ for every pair $(\mu,\nu)$. 

We then write the dual basis, \ie the basis of the cotangent bundle $T^*\mathsf{G}$, still parametrized by $\{\partial_{\mu}\}_{\mu}$ as
\begin{equation}
    \left\{d\theta^\mu \right\}_{\mu} \,,\;\text{such\;that}\;\left\langle d\theta^\mu,\partial_{\nu} \right\rangle = d\theta^\mu(\partial_{\nu}) = \delta^\mu_{\nu}\,, 
\end{equation}
(Note that the angle variables parametrize the Grassmannian globally, but the basis formed by the vectors $v_\mu$ is called a {\it local basis} because it is not constant.)

\subsubsection{Pl{\"u}cker induced metric and volume form}

The Pl{\"u}cker embedding induces a metric on the Grassmannian submanifold $\mathsf{G}$ from the Euclidean metric of the space $\wedge_p$, \ie $g_{\wedge_p}=\delta_{mn} dx^m\otimes dx^n$. That is given by the pullback of $g_{\wedge_p}$ by the inclusion $\iota:\mathsf{G}\hookrightarrow \wedge_p $, \ie [see \eg \cite{riemann_petersen}]
\begin{equation}
\begin{aligned}
    g_{\mathsf{G}} &= \iota^* \,g_{\wedge_p} 
    = \delta_{mn} \, \iota^* \left(dx^m \otimes dx^n\right) \\
    &= g_{\mathsf{G},\mu\nu}  \; d\theta^\mu \otimes d\theta^{\nu}\,,
\end{aligned}
\end{equation}
with, since $dx^m(\partial_{\mu}) = (\partial_{\mu}V^n)\delta_{nm}$,
\begin{equation}
\begin{aligned}
    g_{\mathsf{G},\mu\nu} 
    &=\quad g_{\mathsf{G}} \left(\partial_\mu,\partial_{\nu}\right)= \sum_m (\partial_{\mu} V^m) (\partial_{\nu} V^m)\,,\\
    &= \sum_m v^m_{\mu} v^m_{\nu} = \left\langle v^{\top}_{\mu} , v_{\nu} \right\rangle  \,.
\end{aligned}
\end{equation}

Given that the angle variables $\{\theta^{\mu}\}_{\mu=1,\dots,p (N-p)}$ define a coordinate system, the volume form reads
\begin{equation}
    d\text{Vol}_{g_{\mathsf{G}}} = 
    \sqrt{\vert \det M_{g_{\mathsf{G}}} \vert}\, d\theta^1 \wedge \cdots \wedge d\theta^{p(N-p)}\,,
\end{equation}
with the metric matrix
\begin{equation}
    M_{g_{\mathsf{G}}} = \left[\begin{array}{ccc}
      g_{\mathsf{G},1,1} & \cdots  & g_{\mathsf{G},1,p(N-p)} \\
      \vdots   &  & \vdots \\
      g_{\mathsf{G},p(N-p),1} & \cdots  & g_{\mathsf{G},p(N-p),p(N-p)} 
    \end{array} 
    \right]\,.
\end{equation}

\subsubsection{Pl{\"u}cker sectional curvature}
The Christoffel symbols define the behavior of the tangent vectors under a covariant derivative, which we need to the definition of the sectional curvature. These are obtained from the metric through
\begin{equation}
\label{eq_christoffel}
    \Gamma^{\kappa}_{\mu\nu} =  \dfrac{1}{2} g_{\mathsf{G}}^{\kappa \lambda} \left( \partial_{\mu} g_{\mathsf{G},\nu\lambda}+
     \partial_{\nu} g_{\mathsf{G},\mu\lambda}-
      \partial_{\lambda} g_{\mathsf{G},\mu\nu}
    \right) \,,
\end{equation}
where $g_{\mathsf{G}}^{\kappa \lambda}$ is the inverse $[g_{\mathsf{G}}^{-1}]_{\kappa \lambda}$. 

We use below the Cartesian inner product in $\wedge_p$, namely $\langle V^{\top}, V'\rangle = V^m V'_m$ with $V^{\top}$ the dual (transpose) of $V$ and where $m=1,\dots,\left(\substack{N\\p}\right)$.

The sectional curvature is defined as
\begin{equation}
\begin{aligned}
     &sec :  T_{\bs{\theta}}\mathsf{G} \times
    T_{\bs{\theta}}\mathsf{G} \rightarrow
    \mathbb{R}: \\
     &(v,w) \mapsto 
    sec(v,w) =
    \dfrac{g_{\mathsf{G}}( R(w,v)v,w)}{g_{\mathsf{G}}(v,v)  g_{\mathsf{G}}(w,w) - g_{\mathsf{G}}(v,w)^2}\,,
\end{aligned}
\end{equation}
with the directional curvature operator
\begin{equation}
\begin{aligned}
     &R(\cdot,v)v :  T_{\bs{\theta}}\mathsf{G}
    \rightarrow  T_{\bs{\theta}}\mathsf{G}:\\ 
     &w \mapsto  
    R(w,v)v = \nabla_{w,v}^2 v - \nabla_{v,w}^2 v\,,
\end{aligned}
\end{equation}
where $\nabla$ is the Riemannian affine connection defining the covariant derivative $\nabla_{\partial_\mu} \partial_{\nu} = \Gamma_{\mu\nu}^{\kappa} \partial_{\kappa}$ where $\partial_{\{\mu,\nu,\kappa\}} = \partial_{\theta^{\{\mu,\nu,\kappa\}}} $ and $\mu,\nu,\kappa=1,\dots,p(N-p)$. Using the torsion-freeness of the connection \cite{riemann_petersen}, the directional curvature operator reads
\begin{equation}
    R(w,v)v = [\nabla_{w},\nabla_{v}]v-\nabla_{[w,v]}v\,.
\end{equation}

In the following, we will only consider the sectional curvature of (linearly independent) pairs of Pl{\"u}cker tangent vectors of the basis $\{\partial_{\mu}\equiv\partial_{\theta^{\mu}}\}_{\mu}$, that is, for $\mu\neq \nu$,
\begin{widetext}
\begin{equation}
\begin{aligned}
    R(\partial_{\mu},\partial_{\nu})\partial_{\nu} =& \left[\nabla_{\partial_{\mu}},\nabla_{\partial_{\nu}}\right] \partial_{\nu}\\
    =& \left[(\partial_{\mu} \Gamma^{\beta}_{\nu\nu} )  + \Gamma^\alpha_{\nu\nu} 
    \Gamma^\beta_{\mu\alpha} \partial_\beta
    - (\partial_{\nu} \Gamma^{\alpha}_{\mu\nu} ) \partial_{\alpha} - 
    \Gamma^\alpha_{\mu\nu}  \Gamma^\beta_{\nu\alpha} \right] \partial_\beta
    \,,
\end{aligned}
\end{equation}
where we only sum over $\alpha$ and $\beta$, and we used $\nabla_{[\partial_\mu,\partial_\nu]} \partial_\nu = 0$, since $[\partial_\mu,\partial_\nu]=0$ for the system of angles coordinates $\{\theta^{\mu}\}_{\mu}$. The corresponding sectional curvature then reads,
\begin{equation}
\label{eq_sec_curv}
    sec(\partial_{\mu},\partial_\nu) = \dfrac{
    \left(\partial_{\mu} \Gamma^{\alpha}_{\nu\nu} -
    \partial_{\nu} \Gamma^{\alpha}_{\mu\nu}
    \right) g_{\mathsf{G}}(\partial_{\alpha},\partial_\mu) + \left(\Gamma^\alpha_{\nu\nu} 
    \Gamma^\beta_{\mu\alpha} -
    \Gamma^\alpha_{\mu\nu}  \Gamma^\beta_{\nu\alpha}
    \right) g_{\mathsf{G}}(\partial_{\beta},\partial_\mu)}{g_{\mathsf{G}}(\partial_{\mu},\partial_\mu)
    g_{\mathsf{G}}(\partial_{\nu},\partial_\nu)-g_{\mathsf{G}}(\partial_{\mu},\partial_\nu)^2} \,,
\end{equation}
where again the summation is taken over $\alpha$ and $\beta$. 
\end{widetext}

\subsubsection{Gauss-Bonnet-Chern theorem}
In the case when the oriented Grassmannian is two-dimensional, the Gauss-Bonnet theorem states that the Euler characteristic is simply given by
\begin{equation}
\label{eq_GB}
    \chi[\mathsf{G}] = \dfrac{1}{2\pi} \int_{\mathsf{G}} \left.sec(\partial_{\theta^1},\partial_{\theta^2})\right\vert_{V} \left.d\text{Vol}_{\mathsf{G}}\right\vert_{V}\,,
\end{equation}
where the rank-2 vector bundle $\bigcup\limits_{V\in\mathsf{G}} \langle\partial_{\theta^1},\partial_{\theta^2}\rangle $ is nothing but the tangent bundle $T\mathsf{G}$, where $\{\theta^1,\theta^2\}$ are global curvilinear coordinates of the Grassmannian. Since the only 2D Grassmanians are $\mathsf{G} = \iota_P(\widetilde{\mathsf{G}}^{\mathbb{R}}_{2,3})\cong \mathbb{S}^2$, and $\mathsf{G} =\iota_P(\widetilde{\mathsf{G}}^{\mathbb{C}}_{1,2})\cong \mathbb{S}^2$ and taking the spherical coordinates, we get (see derivation in Section \ref{sec_3B_Euler_riemann})
\begin{equation}
    \chi[\mathsf{G}]_{\dim\mathsf{G}=2} = \chi[\mathbb{S}^2] = 2\,.
\end{equation}

The Gauss-Bonnet-Chern theorem generalizes this to arbitrary even-dimensional oriented manifolds as
\begin{equation}
\label{eq_GBC}    \chi[\mathsf{G}]_{\dim\mathsf{G}=2m} = 
    \int_{\mathsf{G}} e\left(
        \Omega
    \right)\,,
\end{equation}
where the Euler form (a differential $2m$-form)
\begin{equation}
    e\left(
        \Omega
    \right) = \dfrac{1}{(2\pi)^m} \text{Pf}\,\Omega\,,
\end{equation}
is defined as the Pfaffian of the $2m\times2m$-skew symmetric matrix of two-forms
\begin{equation}
    [\Omega]_{\mu\nu} = -\sum\limits_{\alpha<\beta} R_{\mu\nu\alpha\beta} \;d\theta^\alpha\wedge d\theta^\beta
    \,,~\text{for}\;\mu,\nu=1,\dots,2m\,,
\end{equation}
where the Riemannian curvature tensor is defined through
\begin{equation}
    R_{\mu\nu\alpha\beta} = g_{\mathsf{G}}\left(\partial_{\mu},R(\partial_{\alpha},\partial_{\beta}\right)\partial_{\nu})\,.
\end{equation}
It is straightforward to show that Eq.\,(\ref{eq_GBC}) reduces to Eq.\,(\ref{eq_GB}) when $m=1$ \cite{nicolaescu2022lectures}. Applying this general framework to the analytical few-band models representing the real Grassmannians [Section \ref{sec_real_fewband}], we compute the Euler characteristic $\chi[\widetilde{\mathsf{G}}^{\mathbb{R}}_{2,4}]=2$ and $\chi[\widetilde{\mathsf{G}}^{\mathbb{R}}_{3,5}]=2$ in Section \ref{sec_riemann_fewband}. We emphasize that these results characterize the topology of the whole Grassmannians. In the following, we will be interested in the topology of the winding Bloch Hamiltonian map $H^W$ that covers a two-dimensional sub-cell within the Grassmannian (possibly with a multiple wrapping number, \ie in the way of a branched covering) which is the image of the two-dimensional torus Brillouin zone. Fundamentally, the question is thus whether the sub-cell can be shrunk to a point, \ie if it is null-homotopic, or if it wraps unavoidable holes. 

\subsection{Pullback to the sphere}
So far, we have derived the Riemannian structures of the whole Grassmannians. Our aim, instead, is to characterize the Bloch Hamiltonian mapping from the torus Brillouin zone to a sub-region of the Grassmannian. As advocated in Sections \ref{sec_complex_fewband} and \ref{sec_real_fewband}, it is convenient to split the winding Bloch Hamiltonian map $H^W:\mathbb{T}^2\rightarrow \mathsf{G}$ into two, \ie $H^W = f^W_{SG}\circ f_{TS}$, such that it is the second map, $f^W_{SG}:\mathbb{S}^2_0 \rightarrow \mathsf{G}$, that determines the winding, while the first map, $f_{TS}:\mathbb{T}^2 \rightarrow \mathbb{S}^2_0$, has a fixed degree of 1. We do this because the complete topological features (\ie the {\it global} features) of the phases can be characterized completely analytically through the pullback of the second map, $f^W_{SG}$, to the sphere $\mathbb{S}^2_0$ (see below), \ie the topology is independent of the details of the first map $f_{TS}$. On the contrary, the details of $f_{TS}$ lead to {\it local} geometric features as we show in Section \ref{sec_riemann_fewband}. 

Here we focus on the winding map $f^W_{SG}$ from the two-sphere parameterized by the spherical coordinates $(\theta,\phi)$ to the Grassmannian parameterized by the global angle variables $(\theta^{\mu})_{\mu=1,\dots,p(N-p)}$, \ie
\begin{equation}
    f^W_{SG}:\mathbb{S}^2_0\rightarrow \mathsf{G}:\bs{n}(\theta,\phi) \mapsto 
    V(\bs{\theta}(\theta,\phi))\,.
\end{equation}
As argued in Section \ref{sec_Nband}, in the most general multi-gap context the map $f^W_{SG}$ determines one winding number per band-subspace of rank $r\geq 2$, that we write in a vector $\bs{W}=(W_1,\dots,W_{n_W})$. Nevertheless, as discussed in Section \ref{sec_Nband}, as long as we consider a single energy gap, the Hamiltonian classifying space (of orientable phases) is a (oriented) Grassmannian that always includes two sphere-like submanifolds, \ie two two-dimensional CW subcomplexes that are each a copy of $\widetilde{\mathsf{G}}_{2,3}^{\mathbb{R}}\cong \mathbb{S}^2$, such that two winding numbers are sufficient to model every two-dimensional phase. In the following, we restrict to single-gap phases and we label the winding map with $\bs{W}=(W_+,W_-)$ (similarly to the four-band model above). We then define the Pl{\"u}cker vector
\begin{equation}
    V(\bs{\theta}(\theta,\phi)) = V(\bs{\theta}_{\bs{W}}(\theta,\phi))\,,
\end{equation}
through the map between the coordinates
\begin{equation}
    \tilde{f}^{\bs{W}}_{SG}:(\theta,\phi)\mapsto \theta^{\mu}_{\bs{W}}(\theta,\phi)\equiv 
    [\tilde{f}^{\bs{W}}_{SG}(\theta,\phi)]^{\mu}\,,
\end{equation}
for $\mu=1,\dots,p(N-p)$. For instance, in the case of the real Grassmannian $\widetilde{\mathsf{G}}_{2,3}^{\mathbb{R}}\cong \mathbb{S}^2$, we set $(\theta^1,\theta^2)=(\theta,W\phi)$, such that the Pl{\"u}cker vector $V(\theta^1,\theta^2) = V(\theta,W\phi)$ wraps the Grassmannian $W$ times when $\bs{n}(\theta,\phi)$ wraps $\mathbb{S}^2_0$ one time (see next Section).

Crucially, the restriction of $\mathsf{G}$ to the image of $\mathbb{S}^2_0$ by $f^{\bs{W}}_{SG}$ defines a (maximally) two-dimensional submanifold of the Grassmannian (assuming that the Bloch Hamiltonian map is smooth). We are thus seeking the restriction of the Riemannian structures to the submanifold $\mathcal{M}\equiv f^{\bs{W}}_{SG}(\mathbb{S}^2_0)\subset \mathsf{G}$. In general, $f^{(W_+,W_-)}_{SG}(\mathbb{S}^2_0) = S_1 \cup S_2 \cong \mathbb{S}^2_+ \cup \mathbb{S}^2_-$ (we mean that each part is diffeomorphic to a two-sphere), with $W_+$, and $W_-$, the numbers of times the CW subcomplexes $S_1(\cong \mathbb{S}^2_+)$, and $S_2(\cong \mathbb{S}^2_-)$, are wrapped, respectively (similarly to the four-band model above). 

A simple ansatz for the modeling of all the topological phases is \eg given by the parametrization  
\begin{equation}
\begin{aligned} 
    %f^{\bs{W}}_{SG}:\bs{n}(\theta,\phi)  &\mapsto    \bs{\theta}([1-\delta_{W,0}]\theta,W \phi)\,,\\
    &\tilde{f}^{\bs{W}}_{SG}:(\theta,\phi)  \mapsto    
    \\
    & \left\{
\begin{aligned}    
    &~~\begin{alignedat}{3}
        \theta^1_{\bs{W}}(\theta,\phi) &= [1-\delta_{W_+,0}]\theta\,,~ && \theta^2_{\bs{W}}(\theta,\phi) &&= W_+ \phi\,,\\
        \theta^3_{\bs{W}}(\theta,\phi) &= [1-\delta_{W_-,0}]\theta\,,~ && \theta^4_{\bs{W}}(\theta,\phi) &&= W_- \phi\,,
    \end{alignedat}\\
        &\left(\theta^5_{\bs{W}}(\theta,\phi),\dots,\theta^{p(N-p)}_{\bs{W}}(\theta,\phi)\right) = \left(\theta^5,\dots,\theta^{p(N-p)}\right)\,.
\end{aligned}\right.
\end{aligned}
\end{equation}
where the four coordinates $\{\theta^1_{\bs{W}},\theta^2_{\bs{W}},\theta^3_{\bs{W}},\theta^4_{\bs{W}}\}$ are in general given by linear combinations of the global angle coordinates of Section \ref{sec_norm_coo}, and the remaining coordinates are fixed freely. We note that owing to the K{\"a}hler structure that can be defined on the two-dimensional submanifold defined by $f^{\bs{W}}_{SG}$, such an ansatz can be interpreted as a branched covering with ramification orders of $W_+-1$ and $W_--1$, such that we can use the Riemann-Hurwitz generalization of the Gauss-Bonnet theorem \cite{jost2006compact}. Namely, the Euler characteristic computed from the sectional curvature is corrected by a term with the total ramification order of $f^{\bs{W}}_{SG}$ (see Section \ref{sec_riemann_fewband} for the concrete examples provided by the few-band models).

In the following we describe the necessary steps to obtain the Riemannian structure. This general approach is then elucidated 
upon utilizing the perspective in the concrete setting of specific models in the next Section.

\subsubsection{Riemannian structures induced by the map $f^{\bs{W}}_{SG}$}

We first address the tangent vectors on the restricted region of $\mathsf{G}$ defined by the map $f^{\bs{W}}_{SG}$. Since $\mathcal{M}=f^{\bs{W}}_{SG}(\mathbb{S}^2_0)$ is (maximally) two-dimensional, the restriction to $\mathcal{M}$ gives a rank-2 tangent bundle. We first define the intrinsic tangent vectors to the sphere in the spherical coordinates $(\partial_{\theta}^{\text{int}},\partial_{\phi}^{\text{int}})$. Then, we obtained the tangent vectors of the Grassmannian obtained through the differential (or push-forward) of tangent vector fields $f^{\bs{W}}_{SG*}:T\mathbb{S}^2_0\rightarrow T\mathcal{M}\subset T\mathsf{G}$, which gives
\begin{equation}
\label{eq_push-forward}
\left\{
\begin{aligned}
    \partial_{\theta}& \equiv f^{\bs{W}}_{SG*}(\partial_{\theta}^{\text{int}
}) =  \left(\partial_{\theta} \theta^{\mu}_{\bs{W}}\right)\, \partial_{\mu} \,,\\
    \partial_{\phi}& \equiv f^{\bs{W}}_{SG*}(\partial_{\phi}^{\text{int}
}) =  
    \left(\partial_{\phi} \theta^{\mu}_{\bs{W}}\right)\, \partial_{\mu} \,.
\end{aligned}\right.
\end{equation}

As a next step, the metric induced by $f^{\bs{W}}_{SG}$ on $\mathcal{M}$, that is parameterized by $(\theta,\phi)$, is readily obtained through the pullback
\begin{equation}
\label{eq_metric_pullback_sphere}
\begin{aligned}
    g_{\mathcal{M}} &= f^{\bs{W}*}_{SG} g_{\mathsf{G}} \\
    &= g_{\mathsf{G},\mu\nu}(\bs{\theta}_{\bs{W}}(\theta,\phi)) \\
    &\quad\quad\quad~~\big[\left(\partial_{\theta} \theta^{\mu}_{\bs{W}}(\theta,\phi) \right)
    \left(\partial_{\theta} \theta^{\nu}_{\bs{W}}(\theta,\phi)\right) \,d\theta\otimes d\theta \\
    &\quad\quad~~\;+ \left(\partial_{\phi} \theta^{\mu}_{\bs{W}}(\theta,\phi) \right)
    \left(\partial_{\phi} \theta^{\nu}_{\bs{W}}(\theta,\phi)\right) \,d\phi\otimes d\phi \\
    &\quad\quad~~\;+ \left(\partial_{\theta} \theta^{\mu}_{\bs{W}}(\theta,\phi) \right)
    \left(\partial_{\phi} \theta^{\nu}_{\bs{W}}(\theta,\phi)\right) \,d\theta\otimes d\phi\\
    &\quad\quad~~\;+ \left(\partial_{\phi} \theta^{\mu}_{\bs{W}}(\theta,\phi) \right)
    \left(\partial_{\theta} \theta^{\nu}_{\bs{W}}(\theta,\phi)\right) \,d\phi\otimes d\theta \big]\,.
\end{aligned}
\end{equation}
Then, the volume form is simply given as
\begin{equation}
    d\text{Vol}_{g_{\mathcal{M}}} = \sqrt{\vert\det M_{g_{\mathcal{M}}}\vert} \;d\theta\wedge d\phi \,,
\end{equation}
with $M_{g_{\mathcal{M}}}$ the two-by-two matrix formed by the coefficients of Eq.\,(\ref{eq_metric_pullback_sphere}).

Since $\{\theta,\phi\}$ are curvilinear coordinates for $\mathcal{M}$ their Lie bracket vanishes, \ie $[\partial_{\theta},\partial_{\phi}]=0$, such that the sectional curvature takes the same form as in Eq.\,(\ref{eq_sec_curv}), simply by substituting $\{\partial_{\mu},\partial_{\nu}\}$ by $\{\partial_{\theta},\partial_{\phi}\}$, where we use the Christoffel symbols given by the form Eq.\,(\ref{eq_christoffel}) expressed for the new metric $g_{\mathcal{M}}$.

\subsection{Pullback to the torus Brillouin zone}\label{sec_pullback_BZ}
We finally address the last step that involves the pullback by $f_{TS}$ from the sphere $\mathbb{S}_0^2$ back to the torus Brillouin zone $\mathbb{T}^2$. Because of the non-triviality of the fundamental group of the torus ($\pi_1[\mathbb{T}^2]=\mathbb{Z}^2$) while the sphere is simply connected ($\pi_1[\mathbb{S}^2_0]=e$), the map $f_{TS}$ fails to be a Riemannian immersion \cite{meraozawa}, \ie the push-forward (differential) $f_{TS*}=df_{TS}$ is not injective \cite{Leeriemmanifolds2018}. As a consequence, the metric induced by $f_{TS}$, $g_{\mathbb{T}^2}$, must be degenerate, \ie there must be at least one point of the Brillouin zone where $\det g_{\mathbb{T}^2}$ \cite{meraozawa}.

In this section, we show that while the geometric structures pulled-back on the torus Brillouin zone fail to be Riemannian strictly speaking, they are nonetheless well defined and can be evaluated.

We first start with the tangent vectors of $\mathcal{M}$ parameterized by the points of the torus Brillouin zone $\bs{k}=(k^1,k^2)\in\mathbb{T}^2$, where the coordinates $\{k^1,k^2\}$ are associated with the basis $(\partial_{k^1}^{\text{int}},\partial_{k^2}^{\text{int}})$ of the intrinsic tangent bundle. The tangent vectors of the sphere $\mathbb{S}^2_0$ parameterized by the points of the Brillouin zone are then obtained through the push-forward by $f_{TS}:\bs{k}\mapsto (\theta(\bs{k}),\phi(\bs{k}))$, \ie
\begin{equation}
    \begin{aligned}
        \left.f_{TS*}(\partial_{k^i}^{\text{int}})\right\vert_{\bs{k}} = (\partial_{k^i}\theta(\bs{k}))\partial_{\theta} + (\partial_{k^i}\phi(\bs{k})) \partial_{\phi} \,,~\text{for}~i=1,2\,.
    \end{aligned}
\end{equation}
Then, together with Eq.\,(\ref{eq_push-forward}), we get the tangent vectors on $\mathcal{M}$ parameterized by the coordinates $\{k^1,k^2\}$ of the torus Brillouin zone, \ie
\begin{equation}
\begin{aligned}
    &\left\{
        \begin{aligned}
            \partial_{k^1} &\equiv 
            \left[(\partial_{k^1}\theta(\bs{k})) (\partial_{\theta}\theta_{\bs{W}}^\mu)  + 
            (\partial_{k^1}\phi(\bs{k})) (\partial_{\phi}\theta_{\bs{W}}^\mu) \right]\partial_{\mu} \,,\\
            \partial_{k^2} &\equiv \left[(\partial_{k^2}\theta(\bs{k})) (\partial_{\theta}\theta_{\bs{W}}^\mu)  + 
            (\partial_{k^2}\phi(\bs{k})) (\partial_{\phi}\theta_{\bs{W}}^\mu) \right]\partial_{\mu}\,,
        \end{aligned}
    \right.\\
    &~ \{ \partial_{k^1}, \partial_{k^2}\} \in T\mathsf{G}\,,
\end{aligned}
\end{equation}
where the summation over $\mu$ is assumed. 

Then, the metric on the Brillouin zone induced by $f_{TS}$ is 
\begin{multline}
    g_{\mathbb{T}^2} = f_{TS}^* g_{\mathcal{M}}
    = g_{\mathsf{G},\mu\nu}(\bs{\theta}_{\bs{W}}(\theta(\bs{k}),\phi(\bs{k}))) \\
    (\partial_{\alpha} \theta^{\mu}_{\bs{W}}) (\partial_{\beta} \theta^{\mu}_{\bs{W}}) (\partial_{k^i} \alpha(\bs{k}))
    (\partial_{k^j} \beta(\bs{k})) \;dk^i \otimes dk^j\,,
\end{multline}
where we assume the summation over $\alpha, \beta \in \{\theta,\phi\}$ and $i,j\in\{1,2\}$. Since the matrix of the metric in the $\{k^1,k^2\}$-basis is again 2-by-2, namely
\begin{equation}
    M_{g_{\mathbb{T}^2}} = \left[
        \begin{array}{cc}
          g_{\mathbb{T}^2,k^1k^1}   & g_{\mathbb{T}^2,k^1k^2} \\
          g_{\mathbb{T}^2,k^2k^1}   & g_{\mathbb{T}^2,k^2k^2}
        \end{array}
    \right]\,,
\end{equation}
the volume form takes the same form as above
\begin{equation}
    d\text{Vol}_{g_{\mathbb{T}^2}} = \sqrt{\vert \det M_{g_{\mathbb{T}^2}}\vert} \;dk^1\wedge dk^2\,.
\end{equation}

If we adopt the parametrization Eq.\,(\ref{eq_torus2sphere}), and using $[\partial_{\theta},\partial_{\phi}]=0$, we again find that the Lie bracket vanishes, \ie $[\partial_{k_1},\partial_{k_2}]=0$. Therefore, similarly to the basis $\{\partial_{\theta},\partial_{\phi}\}$, the expression of the sectional curvature is simply given by Eq.\,(\ref{eq_sec_curv}) after substituting $\{\partial_{\mu},\partial_{\nu}\}\rightarrow\{\partial_{k^1},\partial_{k^2}\}$, and where the Christoffel symbols Eq.\,(\ref{eq_christoffel}) is computed for the above metric $g_{\mathbb{T}^2}$.

Contrary to the parametrization Eq.\,(\ref{eq_torus2sphere}) that is not smooth everywhere, our procedure of tight-binding approximation [Section \ref{sec_TB_model}], where each matrix element of the Bloch Hamiltonian is obtained as a truncated Fourier series, readily provides a smooth mapping $H^W$ to the (maximally) two-dimensional sub-manifold $\mathcal{M}\subset \mathsf{G}$. Since first principles band structures, \eg obtained from density functional theory, can virtually always be approached through tight-binding modeling (\eg through the downfolding to optimized localized Wannier functions), the Bloch Hamiltonian map can always be assumed to be smooth for all practical purposes. We will further illustrate this with concrete tight-binding examples elsewhere, see also~\cite{bouhonGeometric2020,Bouhon2022braiding2}.

\subsection{Numerics}
When dealing directly with Bloch Hamiltonians of real systems, \eg in the form of multi-band tight-binding models, we can reduce the above expressions by skipping the intermediary maps. While these play an important role in our work to pinpoint the qualitative origin of nontrivial geometry in topological phases, they are cumbersome for direct evaluation. These are however numerical tractable, see also the detailed exposition in Refs.~\cite{bouhonGeometric2020, Bouhon2022braiding2}, and form the reduced expressions that may be applied in  case studies of various tight-binding models representing the settings as described below.

The tangent vectors on $\mathsf{G}$ parametrized by the momenta are given by
\begin{equation}
        \partial_{k^i} \equiv \partial_{k^i} V^m(\bs{k}) \,e_m  = v^m_i e_m \,,~i=1,2\,.
\end{equation}
We remark, generally speaking, that this expression is meant to be evaluated numerically with the Pl{\"u}cker vector simply given from the Bloch eigenvectors by $V(\bs{k}) = u_1(\bs{k})\wedge \cdots \wedge u_p(\bs{k})$. One advantage of using the Pl{\"u}cker vector $V(\bs{k})$ is that there is no ambiguity of choosing consistent (continuous) gauge phases for the each Bloch eigenvectors. 

The metric is then
\begin{equation}
\begin{aligned}
    g_{\mathbb{T}^2} &= \sum\limits_{m} (\partial_{k^i}V^m) (\partial_{k^j}V^m) \,dk^i\otimes dk^j \\
    &= v^m_i v^m_j \,dk^i\otimes dk^j\,,
\end{aligned}
\end{equation}
where we sum over $i,j\in\{1,2\}$. The volume form takes the same above as above. Also, since $[\partial_{k^1},\partial_{k^2}]=0$, the form of the sectional curvature Eq.\,(\ref{eq_sec_curv}) is preserved, and we only need to make substitution $(\partial_{\mu},\partial_{\nu})\rightarrow (\partial_{k^1},\partial_{k^2})$. 

Euler form = sec dVol.
\begin{equation}
    (\partial_{k^1}v_1)^{\top} (\partial_{k^2}v_2) - (\partial_{k^2}v_1)^{\top} (\partial_{k^1}v_2)
\end{equation}

\section{Riemannian structures of few-band models}\label{sec_riemann_fewband}
A virtue of the outlined Pl\"ucker framework is that we can address all Riemannan structures in a tractable manner, furnishing a tool that can be widely applied. We showcase this in the remaining parts of the paper. The analytical few-band models derived in the previous section prove a particularly precious framework for which we derive explicitly and analytically the successive geometric structures that are relevant in realistic many-band systems. We will in particular focus on the real Grassmannians due the rich multi-gap topological features, but we emphasize that the results directly translate to the complex cases. 

As in the previous section, it will be convenient to decompose the Bloch Hamiltonian mapping from the Brillouin zone to the Grassmannian into two steps, \ie $H^{W} = f^{W}_{SG}\circ f_{TS}$, due to the motivation that the topology of the system (\ie the homotopy class of the Bloch Hamiltonian) may be chosen to only depend on the winding of the second map $f^{W}_{SG}$. Using the perspective and line of thought of the previous Section we directly compute the Riemannian structures for concrete cases.

\subsection{Three-band Euler phases}\label{sec_3B_Euler_riemann}
We first turn to the three-band Euler phases. We recall that the Grassmannian here is the embedding sphere in $\mathbb{R}^3$, such that the global angle coordinates can be readily taken as
\begin{equation}
    \tilde{f}^W_{SG}:(\theta,\phi)\rightarrow(\theta^1,\theta^2)=(\theta,W\phi)\,,
\end{equation}
corresponding to the Pl{\"u}cker vector $V(\theta,W\phi) = \check{\bs{e}}^{\top}\cdot \bs{n}(\theta,W\phi)$, where $W\in\mathbb{Z}$ is the degree of the map. We then define the basis of the tangent bundle by
\begin{equation}
\begin{aligned}
    v_{\theta} &= \partial_{\theta} V(\theta,W\phi) =  (\cos W\phi\cos\theta,\sin W\phi\cos\theta,-\sin\theta)  \,,\\
    v_{\phi} &= \partial_{\phi} V(\theta,W\phi) =  W\sin\theta(-\sin W\phi,\cos W\phi,0) \,.
\end{aligned}
\end{equation}
We need to be cautious however that this basis is not normalized. Defining $e_{\theta}=v_{\theta}$ and $e_{\phi}=v_{\phi}/(W\sin\theta)$, the unit vectors $(e_{\theta},e_{\phi})$ span the tangent space at any point of the sphere. By including $e_r=V$, they simply correspond to the spherical coordinates frame. 

Since $v_{\theta}^{\top}\cdot v_{\theta} = 1$, $v_{\phi}^{\top}\cdot v_{\phi} = \sin\theta^2 $, and $v_{\theta}^{\top}\cdot v_{\phi}=0$, the induced metric
\begin{equation}
    g_{\mathbb{S}^2} = d\theta \otimes d\theta + W^2\sin^2\theta\, d\phi \otimes d\phi \,,
\end{equation}
which for $W=1$ is nothing but the usual metric of the unit sphere. Then the only nonzero Christoffel symbols are $\Gamma_{\theta,\phi}^{\phi} = \Gamma_{\phi,\theta}^{\phi} = \cot \theta$, and $\Gamma_{\phi,\phi}^{\theta} = -W^2 \cos\theta\sin\theta$, leading to the directional Riemannian operator
\begin{equation}
    R(\partial_{\theta},\partial_{\phi})\partial_{\phi} = W^2 \sin^2\theta \;\partial_{\theta}\,,
\end{equation}
and the sectional curvature is
\begin{equation}
    sec(\partial_{\theta},\partial_{\phi})  = 1\,.
\end{equation}
For the volume form, we find
\begin{equation}
    d\text{Vol}_{g_{\mathbb{S}^2}} = \vert W\vert \vert \sin\theta\vert\, d\theta\wedge d\phi\,.
\end{equation}

Evaluating now the integral of the Gauss-Bonnet theorem associated with the metric $g_{\mathbb{S}^2}$, we find 
\begin{equation}
\label{eq_RH}
\begin{aligned}
    \chi_{g_{\mathbb{S}^2}} &= \dfrac{1}{2\pi} \int_{\mathbb{S}^2} sec(\partial_{\theta},\partial_{\phi}) \, d\text{Vol}_{g_{\mathbb{S}^2}} \\
    &= \dfrac{1}{2\pi} \int_{\mathbb{S}^2} 1\cdot  \vert W\vert \vert \sin\theta\vert d\theta\wedge \; d \phi\,,\\
    &= 2 \vert W\vert \in 2\mathbb{Z}\,.
\end{aligned}
\end{equation}
Since the map $f^{\bs{W}}_{SG}$ is not an immersion (when $W\neq 1$, the differential of the map fails to be globally injective), the above result can more rigorously be interpreted as the Rieman-Hurwitz relation. Indeed, interpreting $f^{\bs{W}}_{SG}$ as a holomorphic map between two Riemann surfaces, it defines a branched covering with two ramification points, located at $\theta=0$ and $\pi$, each with the ramification order of $W-1$ \cite{jost2006compact}. The Rieman-Hurwitz theorem says, given that the genus of the sphere is zero, that the total ramification order of a map of degree $W$ is $v_f = 2(W-1)$ \cite{jost2006compact}.

Since $\widetilde{\mathsf{G}}^{\mathbb{C}}_{1,2} \cong \mathbb{S}^2$, the topology of the Chern phases of the complex $1+1$-phase can be obtained similarly.

% \subsubsection{Three-band tight-binding models}\label{sec_TB_3B}

%  \textcolor{orange}{[HERHERE]}

\subsection{Four-band Euler phases}

\subsubsection{Whole Grassmannian}

We now address the Riemannian structures for the whole of $\widetilde{\mathsf{Gr}}_{2,4}^{\mathbb{R}}$.

From Eq.\,(\ref{eq_pluckerfour}), we readily derive the tangent vectors
\begin{subequations}
\label{eq_tg_vect}
\begin{equation}
    \begin{aligned}
        \partial_{\theta_{+}} &\equiv \iota_{P*} (\partial_{\theta_{+}}^{\text{int}}) =    \partial_{\theta_{+}}V_{I}' = \dfrac{1}{\sqrt{2}} \left( e_{\theta_{+}} \oplus \bs{0} \right)\,,\\
        \partial_{\phi_{+}} &\equiv \iota_{P*} (\partial_{\phi_{+}}^{\text{int}}) =\partial_{\phi_{+}}V_{I}' = \dfrac{\sin\theta_+}{\sqrt{2}} \left( e_{\phi_{+}} \oplus \bs{0} \right)\,,\\
        \partial_{\theta_{-}} &\equiv \iota_{P*} (\partial_{\theta_{-}}^{\text{int}}) = \partial_{\theta_{-}}V_{I}' = \dfrac{1}{\sqrt{2}} \left( \bs{0} \oplus
        e_{\theta_{-}}   \right)\,,\\
        \partial_{\phi_{-}} &\equiv \iota_{P*} (\partial_{\phi_{-}}^{\text{int}}) = \partial_{\theta_{-}}V_{I}' = \dfrac{\sin\theta_-}{\sqrt{2}} \left( \bs{0} \oplus
 e_{\phi_{-}}   \right)\,,
    \end{aligned}
\end{equation}
with 
\begin{equation}
\left\{
\begin{aligned}
    e_{\theta_{\pm}} &= (\cos\phi_{\pm}\cos\theta_{\pm} ,\sin\phi_{\pm}\cos\theta_{\pm},-\sin\theta_{\pm})    ,\\
    e_{\phi_{\pm}} &= (-\sin\phi_{\pm} ,\cos\phi_{\pm},0)    ,
\end{aligned}\right.
\end{equation}
\end{subequations}
are the unit vectors of the spherical frame of reference $(e_{\theta}(\theta,\phi),e_{\phi}(\theta,\phi),e_r(\theta,\phi)=\bs{n}(\theta,\phi))$ on a two-sphere. We then readily find that the frame $(\partial_{\theta_+}\;\partial_{\phi_+}\;\partial_{\theta_-}\;\partial_{\phi_-})$ has vanishing Lie brackets, \ie $[\partial_{\mu},\partial_{\nu}]=0$ for all $\mu,\nu\in\{\theta_+,\phi_+,\theta_-,\phi_-\}$.

From the ansatz Eq.\,(\ref{eq_frame_parametrization_main}) and the change of variables Eq.\,(\ref{eq_Gr24_param_main}), we find the metric
\begin{multline}
\label{eq_g_G24_full}
    g_{\mathsf{G}_{2,4}} = \dfrac{1}{2}\left(d\theta_+ \otimes  d\theta_+ + \sin^2\theta_+ \,d\phi_+ \otimes  d\phi_+ \right.\\
    \left. + d\theta_- \otimes  d\theta_- + \sin^2\theta_- \, d\phi_- \otimes  d\phi_-
    \right)\,,
\end{multline}
as was expected from Eq.\,(\ref{eq_4B_diff}). The nonvanishing Christoffel symbols are then
\begin{equation}
\begin{aligned}
    \Gamma_{\theta_+\phi_+ }^{\phi_+} = 
    \Gamma_{\phi_+\theta_+ }^{\phi_+} &= \cot\theta_+\,,\\
    \Gamma_{\phi_+\phi_+ }^{\theta_+} &=-\cos\theta_+\sin\theta_+\,,\\
    \Gamma_{\theta_-\phi_- }^{\phi_-} = 
    \Gamma_{\phi_-\theta_- }^{\phi_-} &= \cot\theta_-\,,\\
    \Gamma_{\phi_-\phi_- }^{\theta_-} &=-\cos\theta_-\sin\theta_-\,,
\end{aligned}
\end{equation}
and the non-vanishing elements of the directional curvature tensor are
\begin{equation}
\begin{aligned}
R(\partial_{\theta_+},\partial_{\phi_+})\partial_{\phi_+} &= 
    \sin^2\theta_+ \,\partial_{\theta_+}\,,\\
    R(\partial_{\theta_-},\partial_{\phi_-})\partial_{\phi_-} &= 
    \sin^2\theta_- \,\partial_{\theta_-}\,.
\end{aligned}
\end{equation}
We finally find a non-vanishing sectional curvature only for the pairs $\{\theta_+,\phi_+\}$ and $\{\theta_-,\phi_-\}$,
\begin{equation}
    sec_{g_{\mathsf{G}_{2,4}}}(\partial_{\theta_+},\partial_{\phi_+})=sec_{g_{\mathsf{G}_{2,4}}}(\partial_{\theta_-},\partial_{\phi_-}) = 2 \,,
\end{equation}
(this agrees with \cite{Kozlov_Gr}) and the volume form
\begin{equation}
    d\text{Vol}_{g_{\mathsf{G}_{2,4}}} = \dfrac{1}{2} \vert \sin\theta_+\vert
    \vert \sin\theta_-\vert \, 
    d\theta_+\wedge d\phi_+ \wedge d\theta_-
    \wedge d\phi_-
    \,.
\end{equation}

\subsubsection{Restriction to $\mathcal{M}=f^{(W_+,W_-)}_{SG}(\mathbb{S}^2_0)$}\label{sec_G24_restriction}

We have defined the map $\tilde{f}^{(W_+,W_-)}_{SG}$ in Eq.\,(\ref{eq_winding_ansatz}) of Section \ref{sec_4B_model}, 
\begin{subequations}
\label{eq_winding_ansatz}
\begin{equation}
    \left\{\begin{alignedat}{3}
        \theta_+ & = C_+ \theta\,,\quad && \phi_+ && = W_+ \phi\,,\\
        \theta_- & = C_- \theta \,,&& \phi_- && = W_- \phi \,,
    \end{alignedat}
    \right.
\end{equation}
that can rewritten in terms of the Euler numbers $(\chi_I,\chi_{II})$ according to Eq.\,(\ref{eq_euler_winding}), \ie using
\begin{equation}
    \left\{\begin{alignedat}{3}
        W_+ & =\frac{-\chi_{I}-\chi_{II}}{2} \,,\quad && C_+ && =  [1-\delta_{\chi_{II},-\chi_{I}}]  \,,\\
        W_- &=\frac{\chi_{I}-\chi_{II}}{2} \,,&& C_- && =  [1-\delta_{\chi_{II},\chi_{I}}] \,.
    \end{alignedat}
    \right.
\end{equation}
\end{subequations}
From Eq.\,(\ref{eq_g_G24_full}), and through Eq.\,(\ref{eq_metric_pullback_sphere}), the induced metric is found
\begin{multline}
    g_{\mathcal{M}_{2,4}} = 
        \dfrac{1}{2} \left(C_++C_-\right) \;d\theta\otimes d\theta
          \\
           +
        \dfrac{1}{8}\big[
            (\chi_I+\chi_{II})^2 \sin^2  C_+\theta \\
            +
            (\chi_I-\chi_{II})^2 \sin^2  C_- \theta
        \big]\;d\phi\otimes d\phi \,.
\end{multline}

Let first assume $\vert \chi_{II}\vert=\vert\chi_I \vert= \chi$ and $\chi_{I(II)}\neq0$. We call this the {\it balanced} condition since the absolute Euler number is the same below and above the energy gap. We find the metric
\begin{equation}
    g^{(\text{bal})}_{\mathcal{M}_{2,4}} =
    \dfrac{1}{2} \; d\theta\otimes d\theta + \dfrac{1}{2} \chi^2 \sin^2\theta\; d\phi\otimes d\phi\,,
\end{equation} 
the sectional curvature
\begin{equation}
    sec(\partial_{\theta},\partial_{\phi}) =  2  \,,
\end{equation}
and the volume form
\begin{equation}
    d\text{Vol}_{\mathcal{M}_{2,4}} = \dfrac{1}{2}  \chi \vert \sin\theta\vert \; d\theta\wedge d\phi\,.
\end{equation}
Integrating the sectional curvature, we get an effective Euler characteristic
\begin{equation}
    \chi_{g_{\mathcal{M}_{2,4}}} = 2 \chi = 2\vert \chi_{I}\vert =2\vert \chi_{II}\vert \in 2\mathbb{Z}\,.
\end{equation}
The interpretation for this result is that the balanced phases are realized whenever one of the winding numbers $(W_+,W_-)$ is zero, which means that the map $f^{\bs{W}}_{SG}$ only covers one sphere of $\mathbb{S}^2_+(\frac{1}{\sqrt{2}})\times \mathbb{S}^2_-(\frac{1}{\sqrt{2}})$. Therefore, we recover the Euler characteristic of the sphere, as derived above for the three-band model (when $\vert \chi_I\vert=\vert \chi_{II}\vert=1$), and the Riemann-Huwitz relation between two Riemann surfaces of genus zero (when $\vert \chi_I\vert=\vert \chi_{II}\vert>1$), see the discussion below Eq.\,(\ref{eq_RH}). 

When $\vert\chi_{II}\vert\neq\vert\chi_{I}\vert$, which we call the {\it imbalance} case, we have $C_+=C_-=1$ and the induced metric is
\begin{equation}
    g^{(\text{imb})}_{\mathcal{M}_{2,4}} =
    d\theta\otimes d\theta + \dfrac{1}{4} \left(\chi_I^2+\chi_{II}^2\right) \sin^2\theta \; d\phi\otimes d\phi\,.
\end{equation}
We then find the sectional curvature
\begin{equation}
    sec(\partial_{\theta},\partial_{\phi}) = 1\,,
\end{equation}
and the volume form
\begin{equation}
    d\text{Vol}_{g_{\mathcal{M}_{2,4}}} = \dfrac{1}{2} \sqrt{\chi_I^2+\chi_{II}^2} \;\vert \sin\theta\vert\;d\theta\wedge d\phi\,.
\end{equation}
Finally, the integration of the sectional curvature gives an effective Euler characteristic
\begin{equation}
    \chi_{g_{\mathcal{M}_{2,4}}} = \sqrt{\chi_I^2+\chi_{II}^2}\,.
\end{equation}
When the two winding numbers have equal norm one Euler number vanishes, \ie $\chi_{II}=0$ when $W_-=W_+$ such that 
\begin{equation}
    \chi^{(W_-=W_+)}_{g_{\mathcal{M}_{2,4}}} = \vert \chi_I\vert = \vert W_+\vert=\vert W_-\vert \in \mathbb{Z}\,,    
\end{equation}
and $\chi_I=0$ when $W_-=-W_+$ such that 
\begin{equation}
    \chi^{(W_-=-W_+)}_{g_{\mathcal{M}_{2,4}}} = \vert \chi_{II}\vert = \vert W_+\vert=\vert W_-\vert \in \mathbb{Z}\,.    
\end{equation}
Interestingly, when $\vert W_+\vert=\vert W_-\vert = 1$, the Euler characteristic of $\mathcal{M}$ in such imbalanced phases is $1$, \ie the same as for a two-dimensional disc. When $W_+,W_-\neq 0$ and $\vert W_-\vert\neq \vert W_+\vert$, which implies that both Euler numbers are nonzero, the Euler characteristic of $\mathcal{M}$ is not even an integer. While this can be used as a geometric indication of this special imbalanced topology, the topological interpretation of the non-integer value is not clear, 
but poses an intriguing direction of investigation.

\subsubsection{Euler form and Euler number of $T\mathcal{M}$}\label{sec_Euler_submannifold}

Alternatively, we can define the Euler form from the tangent subbundle $T\mathcal{M}$, since it is of rank 2, and compute an Euler number similarly to Section \ref{sec_real_fewband}. It turns out that this quantity is identical to the effective Euler characteristic computed above from the sectional curvature, and this other derivation can be seen as short cut to the effective Euler characteristic. In order not to confuse it with the topological Euler numbers derived in Section \ref{sec_real_fewband}, we will continue to call it an effective Euler characteristic.

From Eq.\,(\ref{eq_push-forward}) and Eq.\,(\ref{eq_tg_vect}), we obtain the globally defined basis of the tangent subbundle 
\begin{equation}
\left\{
\begin{aligned}
    \partial_{\theta} &= \dfrac{1}{\sqrt{2}}\big[C_+ \,e_{\theta}(C_+\theta,W_+\phi)\oplus C_- \,e_{\theta}(C_-\theta,W_-\phi)\big]\,,\\
    \partial_{\phi} &= \dfrac{1}{\sqrt{2}}\big[W_+ \sin C_+\theta \,e_{\phi}(C_+\theta,W_+\phi) \\
    &\quad\quad\quad \quad \oplus W_- \sin C_-\theta \,e_{\phi}(C_-\theta,W_-\phi)\big]\,.
\end{aligned}\right.
\end{equation}
Defining the unit tangent vectors $\widetilde{e}_{\theta} = \partial_{\theta}/\vert\partial_{\theta}\vert$ and $\widetilde{e}_{\phi} = \partial_{\phi}/\vert\partial_{\phi}\vert$, we can now compute the Euler form [Eq.\,(\ref{eq_euler_form})]
\begin{equation}
\begin{aligned}
    \mathtt{Eu} &= \left[\left(\partial_{\theta}\widetilde{e}_{\theta} \right)^{\top}  \left(\partial_{\phi}\widetilde{e}_{\phi} \right) - \left(\partial_{\phi}\widetilde{e}_{\theta} \right)^{\top}  \left(\partial_{\theta}\widetilde{e}_{\phi} \right) \right]d\theta\wedge d\phi\,,
\end{aligned}
\end{equation}
from which we can compute the Euler number, $1/(2\pi)\int \mathtt{Eu} $, using Eq.\,(\ref{eq_euler_number}). In the case of balanced phases, setting $\chi=\vert \chi_I\vert =\vert \chi_{II}\vert$, and $(C_+,C_-)=(1,0)$ or $(C_+,C_-)=(0,1)$, we find $\mathtt{Eu} = \chi \sin\theta \;d\theta\wedge d\phi$, such that the effective Euler characteristic is $ 2\chi $. When the phase is imbalanced, setting $(C_+,C_-)=(1,1)$, we get $1/2 \sqrt{(\chi_I+\chi_{II})^2 }\sin\theta$, such that the effective Euler characteristic is $\sqrt{\chi_{I}^2+\chi_{II}^2}$. These results are identical to the above derivation via the sectional curvature, which shows the consistency of the Riemanniann structures. Again, the topological interpretation of the non-integer value of the effective Euler characteristic of $T\mathcal{M}$ poses an interesting future pursuit.

% \subsubsection{Four-band tight-binding models}\label{sec_TB_4B}

% \textcolor{orange}{[HEREHERE]}

\subsection{Five-band Euler and second Stiefel-Whitney phases}\label{sec:5-band and sw}
We finally address the generalization to five bands. We start with the Riemannian structures of the whole of $\widetilde{\mathsf{G}}^{\mathbb{R}}_{3,5}$. Moreover, we then derive the geometric structures induced by the restriction from the whole Grassmannian to $\mathcal{M}$, the image of the two-sphere, obtained via our pullback construction. There, we essentially retrieve the above results for the restricted band subspaces. As such these tractable examples set the stage for full generalizations to arbitrary $N$-band systems and isolated band sub-spaces thereof, 
showing the universal power of the Pl\"ucker framework.

\subsubsection{Whole Grassmannian}
We here evaluate the Riemannian structures for the whole of $\widetilde{\mathsf{G}}^{\mathbb{R}}_{3,5}$. From the ansatz Eq.\,(\ref{eq_R5_ansatz}), together with the change of variables Eq.\,(\ref{eq_Gr24_param_main}) and the parametrization Eq.\,(\ref{eq_winding_ansatz}), we find the metric
\begin{subequations}
\begin{equation}
    g_{\mathsf{G}_{3,5}} = 
    d\bs{\theta}^{\top} \cdot M_{g_{\mathsf{G}_{3,5}}}^{(\otimes)} \cdot d\bs{\theta}\,,
\end{equation}
with
\begin{equation}
\begin{aligned}
    d\bs{\theta} &= \left(d\theta^1,d\theta^2,d\theta^3,d\theta^4,d\theta^5,d\theta^6\right)
    \,,\\
    &= \left(d\theta_+,d\phi_+,d\theta_-,d\phi_-,d\theta^5,d\theta^6\right) \,,
\end{aligned}
\end{equation}
\end{subequations}
where we combine the matrix and tensor products as  
\begin{equation}
    d\theta^{\mu}\left[M_{g_{\mathsf{G}_{3,5}}}^{(\otimes)}\right]_{\mu\nu} d\theta^{\nu}= g_{\mathsf{G}_{3,5},\mu\nu} \; d\theta^{\mu}\otimes d\theta^{\nu}\,,
\end{equation}
and evaluate only the non-vanishing elements given by
\begin{equation}
\begin{aligned}
&\begin{alignedat}{3}
    g_{\mathsf{G}_{3,5},11} &= \dfrac{1}{2} \,,\quad && 
    g_{\mathsf{G}_{3,5},22} &&= \dfrac{1}{2} \sin^2\theta_+ \,,\\
    g_{\mathsf{G}_{3,5},33} &= \dfrac{1}{2} \,,\quad && 
    g_{\mathsf{G}_{3,5},44} &&= \dfrac{1}{2} \sin^2\theta_- \,,
\end{alignedat}\\
& g_{\mathsf{G}_{3,5},55} = \dfrac{1+ n_+^1n_-^1-n^3_+n^3_-  + n_+^2n_-^2 }{2} ,\\
&\begin{alignedat}{3}   
    g_{\mathsf{G}_{3,5},16} &= g_{\mathsf{G}_{3,5},61} && 
    g_{\mathsf{G}_{3,5},26} &&=  g_{\mathsf{G}_{3,5},62} \,,\\
    &=  -\dfrac{\cos\phi_+ \sin\theta^5}{2}\,, &&
    &&=  \dfrac{n_+^2 n_+^3 \sin\theta^5}{2}\,,\\
    g_{\mathsf{G}_{3,5},36} &= g_{\mathsf{G}_{3,5},63} &&
    g_{\mathsf{G}_{3,5},46} &&= g_{\mathsf{G}_{3,5},64}\,,\\
    &= -\dfrac{\cos\phi_- \sin\theta^5}{2}\,, &&
    &&= \dfrac{n_-^2 n_-^3 \sin\theta^5}{2}\,,
\end{alignedat}\\
&g_{\mathsf{G}_{3,5},56} = g_{\mathsf{G}_{3,5},65} = -\cos\theta_5 \dfrac{n_+^3 n_-^1 + n_-^3 n_+^1}{2}\,,\\
\end{aligned}
\end{equation}
where we substituted the elements of the unit vectors defined in Eq.\,(\ref{eq_unit_vec}). Putting this all together we verify that
\medskip
\begin{widetext}
\begin{equation}
\begin{aligned}
    &g_{\mathsf{G}_{3,5},66} = \frac{1}{64} \left[-32 \cos^2\theta^5 \sin \theta_- \sin \theta_+ \cos (\phi_-+\phi_+) +4 \cos (2 \theta^5-\theta_--\theta_+) +4 \cos (2 \theta^5+\theta_- -\theta_+)  \right.\\
    & +4 \cos (2 \theta^5 -\theta_- +\theta_+)+4 \cos (2 \theta_a+\theta_-+\theta_+)+
    \cos 2 (\theta^5-\theta_--\phi_-) +\cos 2 (\theta^5+\theta_- -\phi_-) +\cos 2 (\theta^5 -\theta_-+\phi_-)\\
    & +\cos 2 (\theta^5+\theta_-+\phi_-) -2 \cos 2 (\theta^5-\theta_-) -2 \cos 2
   (\theta^5+\theta_-) + 16 \sin ^2\theta^5
   \sin ^2\theta_+ \cos 2 \phi_+ - 2 \cos 2
   (\theta^5-\theta_+) \\
   & -2 \cos  2 (\theta^5 + \theta_+) - 2 \cos 2 (\theta^5-\phi_-) -2 \cos 2 (\theta^5+\phi_-) - 8 \cos 2
   \theta^5 + 8 \cos (\theta_--\theta_+)+8
   \cos (\theta_-+\theta_+) \\
   &\left. -2 \cos 2 (\theta_--\phi_-) -2 \cos 2 (\theta_-+\phi_- ) +4 \cos 2 \theta_- +4 \cos 2 \theta_+ +4
   \cos  2 \phi_- +40\right]\,.
\end{aligned}
\end{equation}
\end{widetext}

While the Christoffel symbols can now be readily derived, their expressions are cumbersome and we do not write them here. We just note that we find the volume form
\begin{multline}
    d\text{Vol}_{g_{\mathsf{G}_{3,5}}} = 
    \dfrac{1}{8} \vert \sin\theta_+\sin\theta_- \cos\theta^5
        \left(n^2_++n^2_-\right)
    \vert \\
    d\theta_+\wedge d\phi_+\wedge
    d\theta_-\wedge d\phi_-\wedge
    d\theta^5\wedge d\theta^6
    \,.
\end{multline}

We obtain the constant nonzero sectional curvature
\begin{equation}
    sec_{g_{\mathsf{G}_{3,5}}}(\partial_{\theta_+},\partial_{\phi_+}) = sec_{g_{\mathsf{G}_{3,5}}}(\partial_{\theta_-},\partial_{\phi_-}) = 2\,,
\end{equation}
and
\begin{equation}
    sec_{g_{\mathsf{G}_{3,5}}}(\partial_{\theta^{\mu}},\partial_{\theta^{\nu}}) = \dfrac{1}{2} \,,
\end{equation}
for the pairs of Pl{\"ucker} tangent vectors, $(\theta^\mu,\theta^\nu)\in\{(\theta_+,\theta_5),(\phi_+,\theta_5), (\theta_-,\theta_5),(\phi_-,\theta_5)\}$. As in the four-band case, the sectional curvature is zero for $(\theta^\mu,\theta^\nu)\in\{(\theta_+,\theta_-),(\theta_+,\phi_-), (\phi_+,\theta_-),(\phi_+,\phi_-)\}$.

Finally, we obtain a variable and bounded sectional curvature
\begin{equation}
    0\leq sec_{g_{\mathsf{G}_{3,5}}}(\partial_{\theta^{\mu}},\partial_{\theta^{\nu}}) \leq 2 \,,
\end{equation}
for the remaining pairs $(\theta^\mu,\theta^\nu)\in\{(\theta_+,\theta^6),(\phi_+,\theta^6), (\theta_-,\theta^6),(\phi_-,\theta^6),(\theta^5,\theta^6)\}$. 

We note that these results are in agreement with the classical result \cite{wong1968sec} derived from the projector matrix representation of the Grassmannians \cite{Grassmannian_handbook}. We are not aware of any other source where the Pl{\"u}cker representation of Grassmannians is directly used to obtain global analytical expressions of the Riemannian structures beyond the case $\mathsf{G}^{\mathbb{R}}_{2,4}$. We thus here demonstrate with the example $\mathsf{G}^{\mathbb{R}}_{3,5}$ that our approach is tractable and can be generalized. Moreover, our approach provides a very efficient framework for the design of new material phases with non-trivial geometric signatures. 

\subsubsection{Restriction to $\mathcal{M}=f^{(W_+,W_-)}_{SG}(\mathbb{S}^2_0)$}
We now take the pullback by $f^{(W_+,W_-)}_{SG}$, still using Eq.\,(\ref{eq_winding_ansatz}), \ie we assume that the angles $\{\theta^5, \theta^6\}$ are independent of $\{\theta,\phi\}$. The latter assumption is justified since we have shown that these directions do not affect the two-dimensional topologies. As a consequence, the Riemannian metric and other structures on $\mathcal{M}$ are identically the same as in Section \ref{sec_G24_restriction} for the four-band systems.

Very interestingly, as for the four-band case, the tangent subbundle $T\mathcal{M}$ is of rank 2, which allows us to compute the associated Euler form and integrate it as an Euler number. Again making the same assumptions on $\{\theta^5,\theta^6\}$, we find the same expressions of the effective Euler characteristics as in Section \ref{sec_Euler_submannifold}. 

We emphasize that this result has a non-trivial consequence. Namely, while the (topological) Euler number introduced in Section \ref{sec_real_fewband} is only defined for rank-2 Bloch vector bundles, \ie strictly for a two-band subspace (see the discussion on the Euler-to-Stiefel-Whitney reduction in Section \ref{sec_ESSW_reduction}), we now can characterize the topology of the three-band subspace (\ie a rank-3 vector bundle) in term of an effective Euler characteristic of the two-dimensional submanifold $\mathcal{M}$, that is computed either via the sectional curvature, or via the Euler form of the rank-2 tangent bundle [Section \ref{sec_Euler_submannifold}].

% \subsubsection{Five-band tight-binding models}\label{sec_TB_5B}

% We here consider another type of extension of the four-band model within the five-band context. We simply introduce hybridization terms in the Bloch Hamiltonian between the new atomic orbital and the four others, namely
% \begin{equation}
%     H^{5B} = jkjk
% \end{equation}

% \textcolor{orange}{[HEREHERE]}

\subsection{Towards a fully general description}
We stress that our  discussion can be readily extended to systems with arbitrary many bands. Indeed, the restriction to the sub-manifold $\mathcal{M}$, image of the sphere $\mathbb{S}^2_0$ within an arbitrary large Grassmannian $\mathsf{G}^{\mathbb{R}}_{p,N}$ (for $p\geq 2$ and $N-p\geq 2$), always induces a (maximally) rank-2 tangent subbundle, $T\mathcal{M}$, for which there is a well defined Euler form and Euler number. The obtained Euler number corresponds to the effective Euler characteristic obtained from the integration of sectional curvature over $\mathcal{M}$. Contrary to the topological Euler number introduced in Section \ref{sec_real_fewband}, the effective Euler characteristic exists for band-subspaces with an arbitrary number of bands.

\section{Physical applications}\label{sec_applications}
We already noted the quantum geometric tensor has recently been appearing in multiple physical contexts that range from bounding superfluid densities~\cite{peotta2015superfluidity,torma2021superfluidity,rhim2020quantum,herzog2022superfluid, parker2021field,julku2016,peri2021} to probing topological band invariants as responses to perturbations such as light, quench dynamics and in other metrological setups~\cite{ozawa_2018,Tran_2017,deJuan_2017,meraozawa,YuUnal_2022,Li_2022_metrology}. While the application of the introduced Pl\"ucker technology within these contexts opens up many routes for new research initiatives that deserve separate treatments, we here outline some preliminary points of view that already firmly underpin the promised potential.

\subsection{Operators and Response functions}
We first consider how operators are defined within the Pl\"ucker setting. As a result, we can then directly relate to perturbations and response theories, setting the stage for a universal 
framework that will be applicable for a wide range of physical settings as alluded to above.

\subsubsection{Operator in the Pl\"ucker setting}
As a first step we outline how to map operators $O$ from the original Hilbert space to operators $\check{O}$ in the Pl\"ucker embedding. Evidently, such a map needs to preserve expectation values over the occupied manifold
\begin{equation}
\Bra{V}\check{O}\Ket{V} = \tr{U^\dagger O U}. 
\end{equation}
Analyzing $\inner{V}{\partial_i V} = \tr{U^\dagger \partial_i U}$ subsequently gives an insight on how to accomplish this.  Replacing $\partial_i$ with $O$, it can be deduced that we obtain the desired relation if $\check{O}$ obeys a Leibniz rule on the wedge product of $u_i$.
\begin{equation*}
\check{O}\Ket{V} = Ou_1\wedge u_2 \wedge \ldots \wedge u_k + \ldots + u_1 \wedge \ldots \wedge Ou_k.
\end{equation*}
We can now write $\check{O}$ explicitly in the $V_i$ basis
\begin{equation}
 \check{O}_{nm} = \sum_{I_n / i_j = I_m / i_\ell} (-1)^{j-\ell} O_{i_j i_\ell},   
\end{equation}
where $n,m$ index the energy eigenbasis of the Pl\"ucker embedding, $I_n, I_m$ are their corresponding ordered index sets, and the sum runs over all $i_j, i_\ell$ labelling the $j$th ($\ell$th) element of $I_n$ ($I_m$) for which the relation $I_n / i_j = I_m / i_\ell$ is true.  The number of terms in the sum will either be zero or one for $n\neq m$, and $k$ for $n=m$.  Specifically, for the diagonal elements of $\check{O}$, we simply obtain
\begin{equation}
\check{O}_{nn} = \sum_{j = 1}^k O_{jj}.
\end{equation} 
Therefore, we can easily see that the groundstate expectation value amount to the desired value
\begin{equation}
    \Bra{V}\check{O}\Ket{V} = \sum_{j=1}^k \Bra{u_j}O\Ket{u_j}
\end{equation}
We note here that one can interpret the off-diagonal elements $n\neq m$ as single band excitations out of the occupied manifold. Indeed, suppose $V_m$ can be obtained from $V_n$ by removing $u_j$ and appending $u_\ell$.  Then $\check{O}_{nm}$ is precisely the amplitude (up to a sign) for $O$ to excite an electron from $u_j$ to $u_\ell$. That is,
\begin{equation}
\check{O}_{nm} = (-1)^{k-j} O_{j\ell}  
\end{equation}
If $V_n$ differs from $V_m$ by more than one $u_j$, then $\check{O}_{nm}$ is simply zero as expected.

\subsubsection{Perturbation theory and response functions}
Since the expression to map operators into the Pl\"ucker embedding is linear, we can directly analyze perturbations $\lambda H_1$ to the Hamiltonian $H$ that thus carry over as 
\begin{equation}
H + \lambda H' \mapsto \check{H} + \lambda \check{H}'.   
\end{equation} 
To make this more concrete let us consider the perturbations to the original states
\begin{equation}
\Ket{u_j} \rightarrow \Ket{u_j} + \lambda \sum_{\ell \neq j} \frac{\Ket{u_\ell}\Bra{u_\ell}H' \Ket{u_j}}{\epsilon_j - \epsilon_\ell}. 
\end{equation}
When we now take the wedge product of the perturbed $u_1,\ldots,u_k$, the term at first order in $\lambda$ amounts to
\begin{equation}
\lambda \sum_{j = 1}^k \sum_{\ell = k+1}^N u_1\wedge \ldots \wedge \hat{u}_{j} \wedge \ldots \wedge u_k \wedge u_\ell  (-1)^{k-\ell}\frac{\Bra{u_\ell}H'\Ket{u_j}}{\epsilon_j - \epsilon_\ell},
\end{equation}
where $\hat{u}_j$ denotes omission of $u_j$ and the minus sign arises from reordering $u_\ell$ from where it is inserted at the $j$th position to the end.  We see that $(-1)^{k-j}\Bra{u_\ell}H'\Ket{u_j}$ is the matrix element of $\check{H}'_{nm}$ for energy eigenstates $V_n$ and $V_m$ in the Pl\"ucker embedding that differ by exactly one band index.  In such a case, the energies take the form
\begin{equation}
\check{\epsilon}_n - \check{\epsilon}_m = \sum_{i_n \in I_n} \epsilon_{i_n} - \sum_{i_m \in I_m} \epsilon_{i_m} = \epsilon_j - \epsilon_\ell.
\end{equation}
When $I_n$ and $I_m$ differ by more than one index, we have $\check{H}'_{nm} = 0$.  Therefore we can rewrite the perturbation as a sum over all $n\neq m$
\begin{equation}
\Ket{V_n} \rightarrow \Ket{V_n} + \lambda \sum_{n \neq m} \frac{\Ket{V_m}\Bra{V_m}\check{H}'\Ket{V_n}}{\check{\epsilon}_n - \check{\epsilon}_m}
\end{equation}
This exactly corresponds to the expected shifts of the wavefunctions if we were to calculate the perturbation entirely in the Pl\"ucker embedding.  It follows that the linear order expression for a response function $R$ measured by $O$ to a perturbation $\lambda H'$ can be calculated in the Pl\"ucker embedding as
\begin{equation}
R=\lambda \sum_{n \neq m} \frac{\Bra{V_n}\check{O}\Ket{V_m}\Bra{V_m}\check{H}'\Ket{V_n}}{\check{\epsilon}_n - \check{\epsilon}_m} + c.c.
\end{equation}
Usually, however, we are interested in operators $O$ of the form $\frac{1}{i\hbar}[A,H]$, imposing a slight subtlety, since products $AH$ of operators do not in general map to products $\check{A}\check{H}$ under the Pl\"ucker embedding.  That is, we have
\begin{equation}
\check{O}_{nm} = \sum_{I_n/i_j = I_m/i_\ell} (-1)^{j-\ell} (A_{i_j i_p} H_{i_p i_\ell} - H_{i_j i_p} A_{i_p i_\ell}). 
\end{equation}
The only relevant matrix elements in the response function must have $I_n$ and $I_m$ differing by one band index.  Furthermore, we note that $H$ is diagonal. As a result, we thus arrive at
\begin{equation}
\check{O}_{nm} = \delta_{I_n/i_j,I_m/i_\ell} (-1)^{j-\ell} (A_{i_j i_\ell} H_{i_\ell i_\ell} - H_{i_j i_j} A_{i_j i_\ell}).
\end{equation}
We therefore directly infer the impact of the mentioned subtlety, being that  $[\check{A},\check{H}]$ would include a sum of $H_{ii}$ over the occupied sector, while there is only a single $H_{ii}$.  However, since $I_n$ and $I_m$ differ by only one band index, we can add and subtract these additional terms.  Since $(-1)^{j-\ell}A_{i_j i_\ell} = \check{A}_{nm}$ for $n\neq m$, we obtain
\begin{equation}
\check{O}_{nm} = \check{A}_{nm} \check{H}_{mm} - \check{H}_{nn} \check{A}_{nm} = [\check{A},\check{H}]_{nm}, \;\;\; n\neq m.
\end{equation}
As a last step,  we may follow the usual process for turning the response function into a curvature.  Concretely, if $A = i\partial_i$ and $H' = i\partial_j$, we get
\begin{eqnarray}
R&=&\sum_{n \neq m} \frac{\Bra{V_n}[\partial_j,\check{H}]\Ket{V_m}\Bra{V_m}i\partial_j\Ket{V_n}}{\check{\epsilon}_n - \check{\epsilon}_m} + c.c. \nonumber \\
&=&i(\inner{\partial_i V_n}{\partial_j V_n} - c.c.)
\end{eqnarray}
Since the Berry curvature within the Pl\"ucker embedding gives the sum of single-band Berry curvatures over the occupied manifold, we know that the curvature can be interpreted as the sum of band-diagonal responses.  With the above result we therefore confirm the anticipation that this response can furthermore be expressed in the form of a Kubo formula.

\subsubsection{Example of probing Chern numbers with dichroism}
%\textcolor{orange}{As a specific example of the above response theory we may connect to recent suggestion to probe the Chen number with polarized light~\cite{ozawa_2018,Tran_2017,deJuan_2017,meraozawa}; in this case we couple to gauge field get first order response to Chern number in Pl\"ucker it is as follows. add one paragraph. Under circular light we get $H\rightarrow H_0+\frac{2E}{\hbar\omega} [\sin(\omega t)\frac{\partial H_0}{\partial k_x}\mp\cos(\omega t)\frac{\partial H_0}{\partial k_y}]$ where I sue the the rotating frame transformation $R=exp(i\frac{2E}{\hbar\omega}(\sin(\omega t)\mp\cos(\omega t) )$ for $H=H_0+E(\sin(\omega t)\pm\cos(\omega t))$}

As a specific example of the above response theory we may connect to recent predictions that suggest how to probe Chern numbers with circularly polarized light~~\cite{ozawa_2018,Tran_2017,deJuan_2017,meraozawa}.  To couple with circularly polarized light, we consider a standard minimal coupling and take $H(\vec{k}) \rightarrow H(\vec{k}+\vec{A})$, where $\vec{A}= \frac{E}{\omega}(\cos(\omega t), \pm \sin(\omega t))$ in the plane of the material.  Viewing the original Hamiltonian $H(k)$ within a rotating frame, $R^\dagger_\pm H(k) R_\pm$ in terms of~\cite{Tran_2017}
\begin{equation}
\quad R_\pm = \exp\left(-i\tfrac{E}{\hbar\omega}(\cos(\omega t)\hat{x} \pm \sin(\omega t) \hat{y})\right), 
\end{equation}
we obtain a time-dependent perturbation that, to first order $E$, reads 
\begin{eqnarray}
    &R^\dagger_\pm H(k) R_\pm \approx \nonumber\\
    &H_0 + \tfrac{E}{\hbar\omega}\left(\cos(\omega t)\tfrac{\partial H}{\partial k_x} \pm \sin(\omega t)\tfrac{\partial H}{\partial k_y}\right).
\end{eqnarray}
The probability of transition from the ground state manifold to higher bands over long times is given by the sum of transition rates between every pair of occupied and unoccupied bands, which evaluates to
\begin{equation}
 \Gamma_\pm(\omega) = \sum_{n\in \text{gs}}\sum_{m \in \text{ex}} \frac{E}{2\hbar\omega} |\bra{m}(\partial_{k_x} \mp i\partial_{k_y})\ket{n}|^2 \delta(\epsilon_n - \epsilon_m -\omega).   
\end{equation}
This expression is now directly in terms of off-diagonal elements of $\partial_{k}$ in the Pl\"ucker embedding. A Kubo formula, i.e. the curvature form in the Pl\"ucker embedding, can be obtained by finding the integrated rate $\tilde{\Gamma} = \int \Gamma \dd \omega$ over a suitable range of $\omega$ and the taking difference of the two chiralities $\Delta \Gamma = \Gamma_+ - \Gamma_-$ to produce
\begin{equation}
  \Delta \tilde{\Gamma} = \sum_{n\in \text{gs}} \frac{iE}{2\hbar\omega}\left( \inner{\partial_{k_x} n}{\partial_{k_y} n} - \inner{\partial_{k_y} n}{\partial_{k_x} n}\right).  
\end{equation}
This physical interpretation provides a useful intuition for the important role that transitions between occupied and unoccupied bands take in the Pl\"ucker embedding. We moreover emphasize that the embedding precisely renders physical degrees of freedom as all gauge degrees of freedom have by construction been eliminated, meaning that for this complex $n$-band system we directly evaluate the describing Chern number and carry over no extra redudant information.

\subsection{Quantum volumes and bounds}
The metric directly affects various other physical quantities. These relations have notably resulted in revived perspectives on quantum metrology~\cite{YuUnal_2022,Li_2022_metrology}, for example in the context of quantum simulators, and Cramér-Rao information bounds. In addition, given the surge of interest in flatband physics, especially in the context of twisted multi-layered Van der Waals materials and superfluid (ultracold) systems, also relations that bound the superfluid density by general correspondences to the Fubini-Study metric at zero temperature have been receiving increasing interest~\cite{peotta2015superfluidity,torma2021superfluidity,bergholtz2013,rhim2020quantum,herzog2022superfluid, parker2021field,julku2016,peri2021}. In particular, using the quantum geometric tensor various trace and determinant identities that relate the imaginary (entailing generalized Berry curvatures) and real part (being the quantum distance part) can be derived. Indeed, using the relation between the geometric and arithmetic mean, one readily obtains $\text{Tr}g(\mathbf{k})\geq |\Omega (\mathbf{k})|$, paving a saturation condition that facilitates the formation of fractional Chern insulating states~\cite{parker2021field,bergholtz2013,repellin2020}. In addition, a similar application of such identities was shown to be applicable in superfluid systems. That is, it was found the integral over the Brillouin-zone of the quantum metric renders the superfluid weight in a flat band and hence is bounded by the presence of invariants (the integral of $\Omega$) such as Chern numbers or Euler invariants~\cite{peotta2015superfluidity,torma2021superfluidity,peri2021}. As such our technology of expressing quantum metric formulations and finding quantum volumes of {\it arbitrary} multi-band systems also promises a versatile approach in these contexts. We therefore close this section of applications of our perspective by evaluating such quantum volumes and bounds for two representative model settings. Namely, we derive analytically that the integrated quantum metric is bounded from below by the topological Euler number in the 3-band and the 4-band cases when pulled-back on the two-sphere $S^2_0$. We then close the discussion by addressing the effect of pulling-back the phase on the Brillouin zone in terms of the tight-binding models, whose numerical form is obtained as discussed above and in Refs.~\cite{bouhonGeometric2020,Bouhon2022braiding2}. We reemphasize that these two cases serve as an illustration that can be generlized to a varity of model settings.

As before, the Pl\"ucker approach is powerful in the intricately-related dual aspects that concern the modelling aspect as well as retrieving the full Riemannian structure. Given the generality of the framework we again focus on the real topological phases due to their rich multi-gap nature, while the complex counterparts can readily be derived analogously. In the subsequent it will be of use to unify the discussion. To this end we write $R_p = (u_1\cdots u_p)\in\mathbb{R}^{N}\times \mathbb{R}^{p}$ to denote the rectangular matrix with the occupied column-Bloch eigenvectors. The quantum geometric tensor may then be defined as 
\begin{equation}
    \bs{\sigma}_{ij} = \partial_i R_p^{\top} \left(
        \mathbb{1}_N - R_pR_p^{\top}
    \right) \partial_j R_p \,,
\end{equation}
that is an element of $\mathbb{R}^p\times \mathbb{R}^p$. The symmetric and the anti-symmetric parts accordingly read
\begin{equation}
    \bs{g}_{ij} = \dfrac{1}{2} (\bs{\sigma}_{ij}+\bs{\sigma}_{ij}^{\top})\,,~ \bs{\omega}_{ij} = \dfrac{1}{2} (\bs{\sigma}_{ij}-\bs{\sigma}_{ij}^{\top}) \,.
\end{equation}

An important feature of $\sigma^{mn}_{ij}$ is its positive definiteness, see e.g. \cite{peotta2015superfluidity,Xie2020}, \ie it satisfies
\begin{equation}
\label{eq_ineq}
    \sum_{ij} v_i^{\top} \bs{\sigma}_{ij} v_j \geq 0 \,,
\end{equation}
for any pair of real vectors $v_i,v_j\in \mathbb{R}^p$.

\subsubsection{Three-band $2+1$-Euler phases}

\begin{figure}[t!]
\centering
\begin{tabular}{l}
\begin{tabular}{ll}
    (a) & (b) \\
    \includegraphics[width=0.44\linewidth]{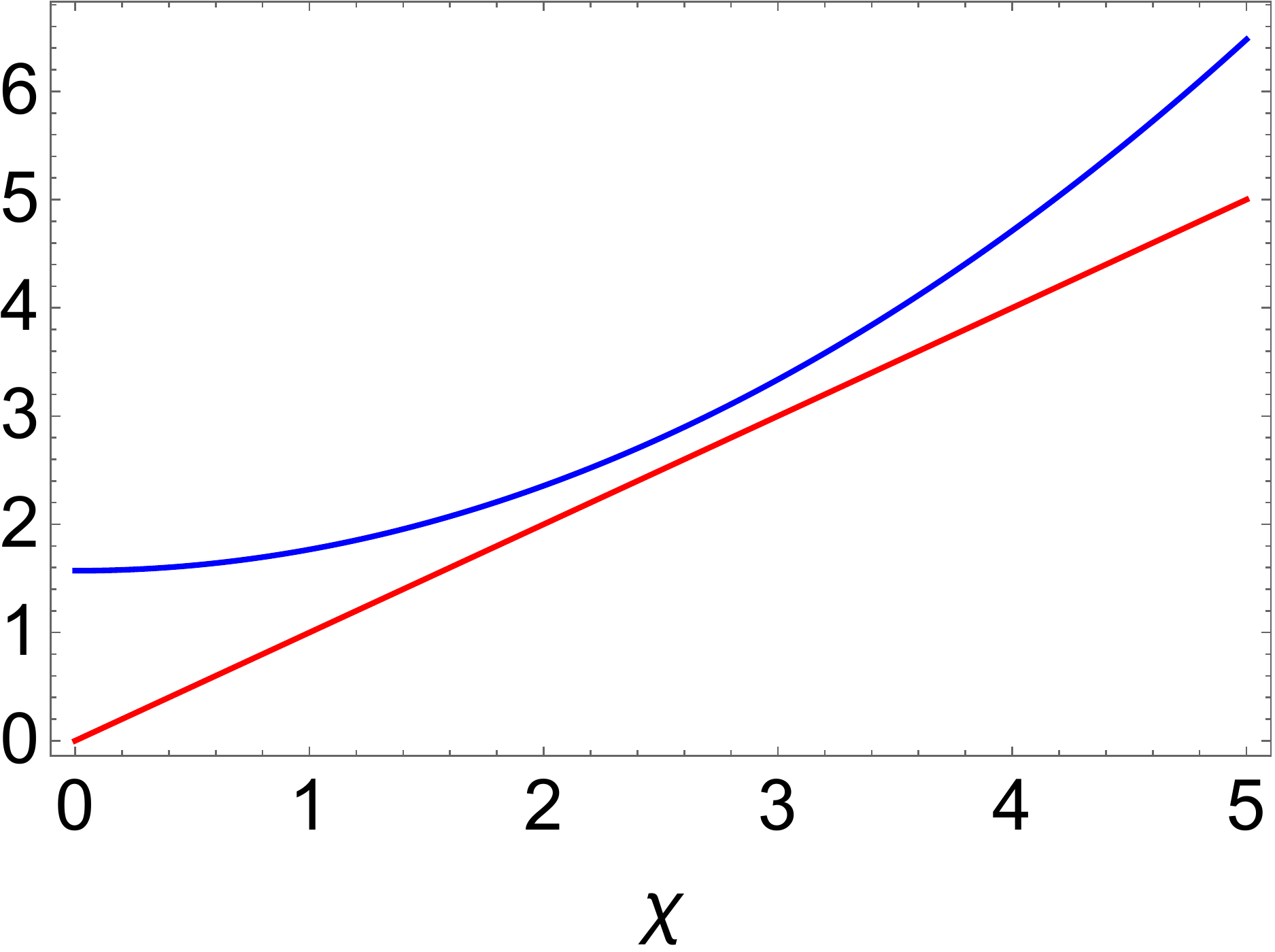}    & 
    \includegraphics[width=0.44\linewidth]{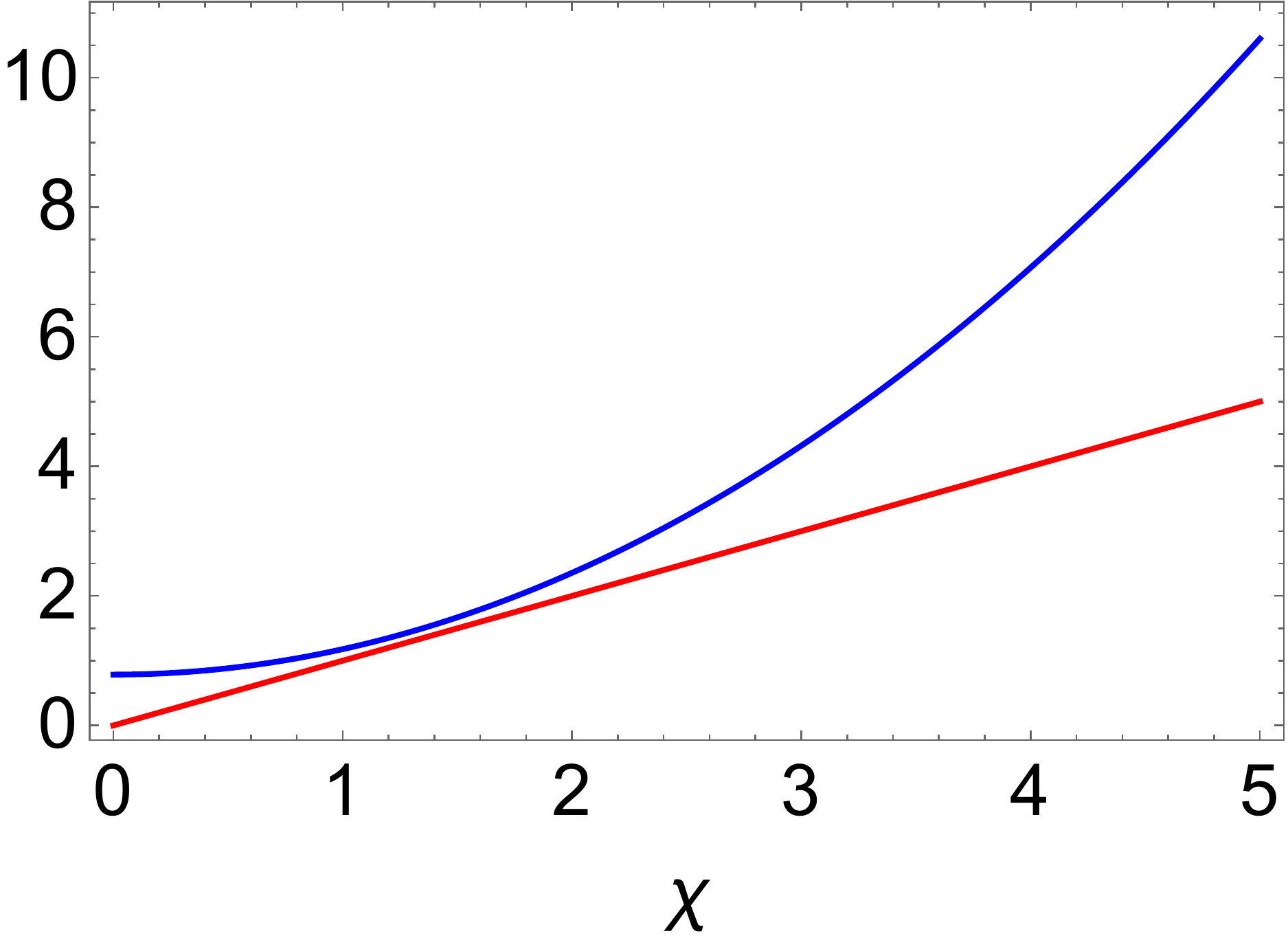} \\
\end{tabular}
    (c) \\
    \includegraphics[width=0.8\linewidth]{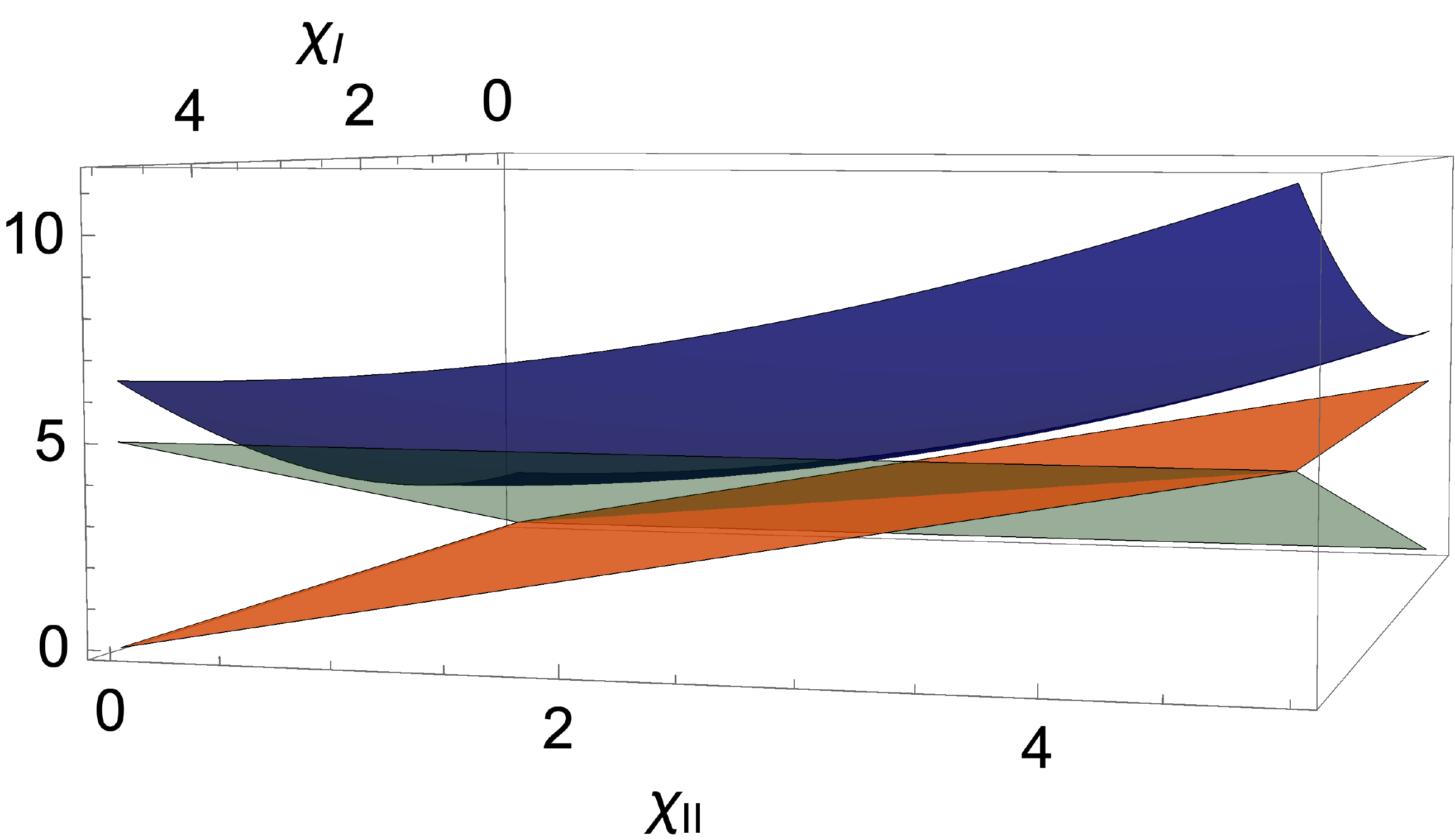}
\end{tabular}
\caption{\label{fig_1} Integrated quantum metric (blue) bounded by the topological Euler number $\chi \geq0 $ (red). (a) Three-band $2+1$-Euler phases. (b) Four-band $2+2$-Euler phases with balanced Euler numbers $\vert\chi_{I}\vert =\vert \chi_{II}\vert = \chi$. (c) Four-band $2+2$-Euler phases with the imbalanced Euler numbers $\vert\chi_{I}\vert$ (green),  $ \vert \chi_{II}\vert$ (orange).}
\end{figure}

We depart from the ansatz of the $2+1$-Euler phases of Section \ref{sec_3B_real}. Considering the two-band occupied subspace, $\bs{\sigma}_{ij}$ is a 2-by-2 matrix in the band space and the labels $\{i,j\}$ run through the coordinates $\{\theta,\phi\}$ of $\mathcal{M}=\mathsf{G}=\mathbb{S}^2$. Choosing the vectors $v_{\theta} = (1,1)$ and $v_{\phi} = (1,-1)$, the inequality Eq.\,(\ref{eq_ineq}) gives
\begin{equation}
    \text{tr}\, \bs{g} - 4 \bs{\omega}^{12}_{\theta\phi} \geq 0\,,
\end{equation}
where the trace is taken over all the degrees of freedom, \ie $\text{tr}\, \bs{g} = \bs{g}^{11}_{\theta\theta} +  \bs{g}^{22}_{\theta\theta} +\bs{g}^{11}_{\phi\phi} +  \bs{g}^{22}_{\phi\phi} $. If we take $v_{\theta} = (1,-1)$ and $v_{\phi} = (1,1)$ instead, we get $\text{tr}\, \bs{g} + 4 \bs{\omega}^{12}_{\theta\phi} \geq 0$, such that in general there a non-negative lower bound on the quantum metric
\begin{equation}
    \text{tr}\, \bs{g} \geq 4 \vert \bs{\omega}^{12}_{\theta\phi} \vert \,.
\end{equation}
From the analytic ansatz of Section \ref{sec_3B_real}, where $W$ fixes the winding of $f^W_{SG}$ and the Euler number $\chi = 2\vert W\vert$ of the phase, we find
\begin{equation}
\left\{
\begin{aligned}
    \text{tr}\, \bs{g} &=  1 + W^2 \sin^2\theta = 1 + \dfrac{\chi^2}{4} \sin^2\theta \,,\\
    \vert \bs{\omega}^{12}_{\theta\phi} \vert & = \vert 2W\sin\theta \vert =  \chi \vert\sin\theta \vert\,,
\end{aligned}\right.
\end{equation}
which after integration gives
\begin{equation}
\left\{
    \begin{aligned}
        I_1 & = \dfrac{1}{4\pi} \int \text{tr}\, \bs{g} \, d\theta\wedge d\phi
        = \dfrac{\pi}{16}(8+\chi^2)\,,\\
        I_2 & = \dfrac{1}{\pi} \int \vert \bs{\omega}^{12}_{\theta\phi}\vert \, d\theta\wedge d\phi = \chi \,,
    \end{aligned}\right.
\end{equation}
with the allowed values $\chi\in 2\mathbb{N}$. We plot in Fig.\,\ref{fig_1}(a) the two integrals as a function of the Euler number of the phase $\chi$, where we see that the ansatz actually gives a strict inequality $I_1>I_2$.

%%%%%%%%%%%%%% FIGURE %%%%%%%%%%%%%%%%%%%%%%%
%%%%%%%%%%%%%% FIGURE %%%%%%%%%%%%%%%%%%%%%%%

\subsubsection{Four-band $2+2$-Euler phases}

Considering again two-band subspaces, we find the same general bound on the metric
\begin{equation}
    \text{tr}\, \bs{g} \geq 4 \vert \bs{\omega}^{12}_{\theta\phi} \vert \,.
\end{equation}
As a next step, using the four-band ansatz of Section \ref{sec_4B_model} for the balanced phases ($\vert\chi_I\vert = \vert\chi_{II}\vert =\chi$), we find 
\begin{equation}
\left\{
\begin{aligned}
    \text{tr}\, \bs{g}^I = \text{tr}\, \bs{g}^{II} &=  \dfrac{1+ \chi^2 \sin^2\theta }{2} \,,\\
    \vert \bs{\omega}^{I,12}_{\theta\phi} \vert = \vert \bs{\omega}^{II,12}_{\theta\phi} \vert & = \chi \vert\sin\theta\vert \,,
\end{aligned}\right.
\end{equation}
where $\{\bs{g}^I\,\bs{\omega}^{I,12}_{\theta\phi}\}$ ($\{\bs{g}^{II}\,\bs{\omega}^{II,12}_{\theta\phi}\}$) are the quantities obtained for the occupied (respectively, unoccupied) two-band subspace, which after integration gives
\begin{equation}
\left\{
    \begin{aligned}
        I^I_1 &= I^{II}_1
        = \dfrac{\pi}{8}(2+\chi^2)\,,\\
        I^I_2 &= I^{II}_2 = \chi \,,
    \end{aligned}\right.
\end{equation}
where here $\chi\in \mathbb{N}$. We again have a strict bound $I_1>I_2$. 

If we however instead consider the imbalanced Euler phases ($\vert \chi_I \vert \neq \vert \chi_{II} \vert$), we find 
\begin{equation}
\left\{
\begin{aligned}
    \text{tr}\, \bs{g}^I &= \text{tr}\, \bs{g}^{II} =  1+ \dfrac{1}{4} (\chi_I^2+\chi_{II}^2) \sin^2\theta  \,,\\
    \vert \bs{\omega}^{I,12}_{\theta\phi} \vert & = \vert\chi_I\vert \vert\sin\theta\vert\,,~
    \vert \bs{\omega}^{II,12}_{\theta\phi}  \vert  = \vert\chi_{II}\vert \vert\sin\theta\vert \,,
\end{aligned}\right.
\end{equation}
which, after integration, gives
\begin{equation}
\left\{
    \begin{aligned}
        I^I_1 &= I^{II}_1
        = \dfrac{\pi}{16}(8+\chi_I^2+\chi_{II}^2)\,,\\
        I^I_2 &= \vert\chi_I\vert \,,~
        I^{II}_2 = \vert\chi_{II}\vert \,.
    \end{aligned}\right.
\end{equation}

\subsubsection{General numerical settings}

The above analytical study can now be brought to the context of tight-binding settings, where the geometric structures must be computed numerically. In that respect, the fact that the sectional curvature of the submanifold $\mathcal{M}$, which is defined numerically from the evaluations of the Pl{\"u}cker vector, can be computed most efficiently from the Euler form of the numerically obtained tangent bundle $T\mathcal{M}$, represents a great advantage. In upcoming future work, we will show that the numerically integrated metric is its self bounded from below by the analytical expressions obtained above. This can be interpreted as the consequence that the metric of tight-binding models is degenerate since the mapping of the Brillouin zone to $\mathcal{M}$ fails to be an immersion, see the discussion in Section \ref{sec_pullback_BZ}. This will be reported in detail elsewhere, but its general applicability shows the promise of the introduced perspective.

\section{Conclusions and Discussion}
We have shown that the Pl\"ucker embedding entails a versatile route to analyze geometric tensors and Riemannian structures for arbitrary $n$-band systems. Departing from simple two-band Chern models, we demonstrate how this embedding universally generalizes the quantum metric tensor in terms of a simple geometric viewpoint that formulates all relevant information. This concrete, non-redundant, approach however becomes even more far-reaching in evaluating general topological many-state systems hosting recently discovered multi-gap phases that arise by rather generic reality conditions. 
As such, our universal approach that manifests all relevant geometrical identities by appealing to a simple vector description, presents a powerful universal benchmark to evaluate physical topological systems.                                                     

The generic framework also provides a direct route towards a systematic modelling, highlighting the interplay of topology, geometry and direct descriptions in various situations. We outline how this approach  accordingly can by utilized in numerous physical contexts. Indeed, as preliminary applications, we highlighted its manifestation in general perturbation theory that in turn can be used to analyze optical responses of many-band systems and outlined its use in defining quantum volumes that are of active interest to derive bounds on superfluidity or formulate ideal conditions to host fractional Chern states.

The above directions promise eminent potential to use the presented framework and explore several novel directions. These not only include analyzing novel interplays between topology and optical responses in many-band systems or utilizing the generalized metric and its bounds in different settings, but even reach to quantum computation and information. We already mentioned the relation to Cramér-Rao information bounds and reemphasize that the active scene of quantum metrology may directly profit from our framework. Moreover, we anticipate that the direct handle of many-band systems, also in terms of modelling, could induce new holonomic approaches to quantum compuation~\cite{Pachos}. In such approaches one essentially encodes information in a degenerate eigenspace of a parametric family of Hamiltonians, which basically thrives on defining generalized non-Abelian Berry connections that can be readily computed within our approach. We therefore 
believe that our results can set a benchmark for a wide range of novel fundamental insights as well as concrete physically relevant pursuits.

\begin{acknowledgements}
A.~B. was funded by a Marie-Curie fellowship, grant no. 101025315. 
R.~J.~S acknowledges funding from a New Investigator Award, EPSRC grant EP/W00187X/1, as well as Trinity college, Cambridge. We thank F. Nur \"Unal for valuable discussions. 
\end{acknowledgements}

\newpage
\clearpage
\appendix
\section{Derivation of Equation \eqref{eq:Chern_genplucker}}\label{app:pluckercherneq}
We here detail the derivation of Eq. \eqref{eq:Chern_genplucker}. To define a notion of distance $D$ between two states $V_1$ and $V_2$ in the Pl\"ucker embedding is we consider the canonical expression 
\begin{equation}
D^2_\text{Pl} = 1 - \inner{ V_2}{ V_1 }\inner{ V_1}{ V_2}
\end{equation}
Assuming that $V_2$ differs from $V_1$ by an infinitesimal change $d\vec{x}$ in the parameters of the Hamiltonian, we then obtain
\begin{equation}
D^2_\text{Pl} = 1 - \inner{ V(\vec{x}+\dd \vec{x})}{ V(\vec{x}) } \inner{ V(\vec{x})}{ V(\vec{x}+\dd\vec{x}) } 
\end{equation}
Expressing this result in term of the eigenstates, the above implies
\begin{widetext}
  \begin{gather*}
    D^2_\text{Pl} = 1 - \inner{ u_1(\vec{x} + \dd \vec{x}) \wedge \ldots \wedge u_k(\vec{x} + \dd \vec{x})}{ u_1(\vec{x}) \wedge \ldots \wedge u_k(\vec{x}) } \\
    \inner{ u_1(\vec{x}) \wedge \ldots \wedge u_k(\vec{x})}{ u_1(\vec{x}+ \dd\vec{x}) \wedge \ldots \wedge u_k(\vec{x} + \dd\vec{x})}
\end{gather*}
Expanding to second order in $\dd \vec{x}$ then results in 
\begin{gather*} D^2_\text{Pl} = 1 - \left\langle \left(\textstyle \sum_{i,j} u_1 + \partial_i u_1 \dd x_i + \tfrac{1}{2} \partial_i \partial_j u_1 \dd x_i\dd x_j\right) \wedge \ldots \wedge \left(\textstyle \sum_{i,j} u_k + \partial_i u_k \dd x_i + \tfrac{1}{2} \partial_i \partial_j u_k \dd x_i \dd x_j\right)\right| \\
\left| u_1(\vec{x}) \wedge \ldots \wedge u_k(\vec{x}) \right\rangle
        \left\langle u_1(\vec{x}) \wedge \ldots \wedge u_k(\vec{x})\right| \\
        \left|\left(\textstyle \sum_{i,j} u_1 + \partial_i u_1 \dd x_i + \tfrac{1}{2} \partial_i \partial_j u_1 \dd x_i \dd x_j\right) \wedge \ldots \wedge \left(\textstyle \sum_{i,j} u_k + \partial_i u_k \dd x_i + \tfrac{1}{2} \partial_i \partial_j u_k \dd x_i \dd x_j\right) \right\rangle,
\end{gather*}
where $i$ and $j$ indices denote components of the parameters $\vec{x}$.  Multiplying out each inner product and keeping only terms to second order, we will have a few different categories of terms:
\begin{itemize}
        \item{The zeroth order term $\inner {u_1 \wedge \ldots \wedge u_k}{ u_1 \wedge \ldots \wedge u_k } = 1$}
        \item{First order terms containing a single $\partial_i u_n \dd x_i$.  A simple determinant calculation shows that these evaluate to $(\partial_i u_n)^\dagger u_n \dd x_i$}
        \item{Second order terms containing two instances of $\partial_i u_n \dd x_i$.  For these, we have not only $(\partial_i u_n)^\dagger u_n (\partial_j u_m)^\dagger u_m \dd x_i \dd x_j$ from the diagonal contribution to the determinant, but also an off-diagonal contribution $-(\partial_i u_n)^\dagger u_m (\partial_j u_m)^\dagger u_n \dd x_i \dd x_j$.  To get the correct count, we require $n > m$ for these.}
        \item{Second order terms containing one instance of $\frac{1}{2} \partial_i \partial_j u_n \dd x_i \dd x_j$.  This again has only a diagonal contribution to the determinant and thus evaluates to $\frac{1}{2} (\partial_i \partial_j u_n)^\dagger u_n \dd x_i \dd x_j$}
\end{itemize}
Applying the above, we arrive at
\begin{gather*}
        D^2_\text{Pl} = 1 - \left| 1 + \textstyle \sum_{i,j,n>m} (\partial_i u_n)^\dagger u_n \dd x_i +  (\partial_i u_n)^\dagger u_n (\partial_j u_m)^\dagger u_m \dd x_i \dd x_j - (\partial_i u_n)^\dagger u_m (\partial_j u_m)^\dagger u_n \dd x_i \dd x_j\right.\\
        \left.+ \tfrac{1}{2} (\partial_i \partial_j u_n)^\dagger u_n \dd x_i \dd x_j \right|^2.
\end{gather*}
Multiplying out the square and keeping only terms to second order, we then get
\begin{gather*}
        D^2_\text{Pl} = 1 - \left(1 + \textstyle \sum_{i,j,n,m} 2\re \left((\partial_i u_n)^\dagger u_n\right) \dd x_i + (\partial_i u_n)^\dagger u_n u_m^\dagger (\partial_j u_m) \dd x_i \dd x_j\right. \\
        + 2\re\left((\partial_i u_n)^\dagger u_n (\partial_j u_m)^\dagger u_m\right)_{n>m} \dd x_i \dd x_j - 2\re\left((\partial_i u_n)^\dagger u_m (\partial_j u_m)^\dagger u_n\right)_{n>m} \dd x_i \dd x_j \\
        + \left.\tfrac{1}{2} \cdot 2\re\left((\partial_i \partial_j u_n)^\dagger u_n\right) \dd x_i \dd x_j\right)
\end{gather*}
The first order terms vanish because $(\partial_i u_n)^\dagger u_n$ is pure imaginary.  We can drop the $\re$ on most of the second order terms because they are products of two such imaginary terms.  For the $n>m$ terms, we extend to a sum over all $n,m$ by absorbing a factor of $2$ and noticing that the $n=m$ terms internally cancel.
\begin{gather*}
    D^2_\text{Pl} = -\textstyle \sum_{i,j,n,m} \left((\partial_i u_n)^\dagger u_n u_m^\dagger (\partial_j u_m) + (\partial_i u_n)^\dagger u_n (\partial_j u_m)^\dagger u_m - (\partial_i u_n)^\dagger u_m (\partial_j u_m)^\dagger u_n\right. \\
    + \left.\re\left((\partial_i \partial_j u_n)^\dagger u_n\right) \right)\dd x_i \dd x_j
\end{gather*}
We also have that $(\partial_i u_n)^\dagger u_n u_m^\dagger (\partial_j u_m) = -(\partial_i u_n)^\dagger u_n (\partial_j u_m)^\dagger u_m$, so the first two terms cancel.  We are left with
$$D^2_\text{Pl} = -\textstyle \sum_{i,j,n,m} \left(- (\partial_i u_n)^\dagger u_m (\partial_j u_m)^\dagger u_n + \re\left((\partial_i \partial_j u_n)^\dagger u_n\right) \right)\dd x_i \dd x_j$$
Integrating by parts, this can be written
$$D^2_\text{Pl} = -\textstyle \sum_{i,j,n,m} \left((\partial_i u_n)^\dagger u_m u_m^\dagger \partial_j u_n + \re\left(\partial_j ((\partial_i u_n)^\dagger u_n) - (\partial_i u_n)^\dagger \partial_j u_n\right) \right)\dd x_i \dd x_j$$
The first term is pure real and the first term inside the $\re$ is pure imaginary, so this reduces to
$$D^2_\text{Pl} = \textstyle \sum_{i,j,n,m} \re\left((\partial_i u_n)^\dagger \partial_j u_n - (\partial_i u_n)^\dagger u_m u_m^\dagger \partial_j u_n\right) \dd x_i \dd x_j$$
Switching to Dirac notation, we discern the usual quantum metric 
$$ g_{ij} = \sum_{n,m \in \text{occ}} \re \bra {\partial_i u_n } \left(\mathbb{I} - \ket{u_m}\bra{ u_m}\right) \ket{\partial_j u_n }, \quad D^2_\text{Pl} = \sum_{i,j} g_{ij} \dd x_i \dd x_j $$
\end{widetext}
We thus observe that the usual notion of infinitesimal distance between two manifolds of states $D^2$ corresponds exactly to the distance we defined in the Pl\"ucker embedding,
\begin{eqnarray*}
D^2 &=& D^2_\text{Pl} \nonumber\\
&=& 1 - \inner{ V(x+\dd x)}{ V(x)}\inner{ V(x)}{ V(x+\dd x)}. 
\end{eqnarray*}
To write the quantum metric in the Pl\"ucker embedding, we would then go through the exercise of expanding $V(x+dx)$ and rearranging $D^2_\text{Pl}$ into the form $g_{ij} \dd x_i \dd x_j$. This however amounts to the same derivation one would do for the single band case in the standard formalism. Therefore we irectly infer that the usual quantum metric expressed in the Pl\"ucker embedding is
\begin{equation}
g_{ij} = \re\left(\inner{ \partial_i V }{ \partial_j V } - \inner{ \partial_i V}{ V }\inner{ V}{ \partial_j V }\right).
\end{equation}

\section{Alternative parametrization of $\widetilde{\mathsf{Gr}}_{2,4}^{\mathbb{R}}$}\label{app_SO4_B}

We start with the parametrization of a generic element of $\mathsf{SO}(4)$ as \cite{guise2018factorization}
\begin{multline}
    R^{4B}(\theta_1\cdots\theta_6) =  e^{\theta_1 L_{12}} 
      e^{\theta_2 L_{13}}
     e^{\theta_3 L_{14}}
      e^{\theta_4 L_{23}}
      e^{\theta_5 L_{24}}
     e^{\theta_6 L_{34}}\,,
\end{multline}
with the angular momentum matrices $[L_{ij}]_{\alpha\beta} = -\delta_{\alpha i}\delta_{\beta j}+\delta_{\alpha j}\delta_{\beta i}$ for all pairs $(i,j)\in I_2 = \{(a,b)\vert 1\leq a< b\leq 4\} = \{(1,2),(1,3),(1,4),(2,3),(2,4),(3,4)\}$, that form a basis of $\mathsf{so}(4)$ (dim\,$\mathsf{so}(4)$=6). A direct inspection shows that $\theta_6$ is a pure gauge sign and we thus set $\theta_6=0$. In the end, we need a reduction to a maximum of $4$ principal angles that parametrize the four-dimensional Grassmannian $\widetilde{\mathsf{Gr}}_{2,4}^{\mathbb{R}}$. For this we take the wedge products 
\begin{equation}
    V_{I} =u_1\wedge u_2 \,,\; V_{II} = u_3\wedge u_4 \,,
\end{equation}
that both define six-dimensional vectors in $\bigwedge^2\left(\mathbb{R}^4\right)$ and written in the basis $\{\check{\bs{e}}_{ij} = e_i\wedge e_j\}_{(i,j)\in I_2} $ (again with $\{e_i\}_{i=1}^4$ the Cartersian basis of $\mathbb{R}^4$). We now take the linear combinations 
\begin{equation}
    V_{\pm} = V_I \pm V_{II} \,,
\end{equation}
and write these as
\begin{equation}
    V_{\pm} = \check{\bs{e}}_{ij} \, [V_{\pm}]_{ij} = \check{\bs{e}}_{ij}' \, [V_{\pm}']_{ij} \,,
\end{equation}
in the new basis,
\begin{equation}
    \left[\begin{array}{c}
    \check{\bs{e}}_{12}'\\
    \check{\bs{e}}_{13}' \\
    \check{\bs{e}}_{14}' \\
    \check{\bs{e}}_{23}' \\
    \check{\bs{e}}_{24}' \\
    \check{\bs{e}}_{34}' 
    \end{array}\right]^{\top} =
    \left[\begin{array}{c}
    \check{\bs{e}}_{12}\\
    \check{\bs{e}}_{13} \\
    \check{\bs{e}}_{14} \\
    \check{\bs{e}}_{23} \\
    \check{\bs{e}}_{24} \\
    \check{\bs{e}}_{34} 
    \end{array}\right]^{\top}\cdot \dfrac{1}{2} \left[
        \begin{array}{rrrrrr}
            0&1&0&0&-1&0 \\
            0&0&1&1&0&0 \\
            1&0&0&0&0&1 \\
            0&1&0&0&1&0 \\
            0&0&1&-1&0&0 \\
            1&0&0&0&0&-1 
        \end{array}
    \right]\,.
\end{equation}
Then, by setting 
\begin{equation}
\label{eq_Gr24_param}
    \begin{aligned}
        (\theta_1,\theta_2)&=(0,0)\,, \\(\theta_3,\theta_4)&=\dfrac{1}{2}(\phi_++\phi_-,\phi_+-\phi_-) \,,\\
        \theta_5 &= -\theta+\pi/2\,,
    \end{aligned}
\end{equation}
we finally obtain
\begin{equation}
\begin{aligned}
    V_{+}' &= (\sin\phi_+\sin\theta,\cos\theta,\cos\phi_+ \sin\theta,0,0,0)\,,\\
    V_{-}' &= (0,0,0,-\sin\phi_-\sin\theta,\cos\theta,\cos\phi_- \sin\theta)\,,
\end{aligned}
\end{equation}
such that each Pl{\"u}cker vector ($V_{\pm}'$) defines a two-sphere ($\mathbb{S}^2_{\pm}$) within one three-dimensional half of the six-dimensional vector space $\bigwedge^2(\mathbb{R}^4) = V_+ \oplus V_- \cong \mathbb{R}^3 \oplus \mathbb{R}^3$, \ie $V_{\pm}'\in \mathbb{S}^2_{\pm}\in V_{\pm}$. We remark that by only keeping three angles, $V_{\pm}'$ do not cover the whole of $\mathsf{Gr}_{2,4}^{\mathbb{R}}$. Nevertheless, we do capture the two sub-dimensional spheres contained in the four-dimensional Grassmannian and this is sufficient to fully characterize the 2D Euler topology, as we now show. 

Defining the unit vectors 
\begin{subequations}
\begin{equation}
    \begin{aligned}
        \bs{n}_+ & = (\cos\phi_+\sin\theta,\sin\phi_+\sin\theta,\cos\theta) \in \mathbb{S}^2_+\,,\\
        \bs{n}_- & = (\cos\phi_+\sin\theta,\sin\phi_+\sin\theta,\cos\theta) \in \mathbb{S}^2_-\,,
    \end{aligned}
\end{equation}
and substituting the parameters Eq.\,(\ref{eq_Gr24_param}) in the $2+2$-Hamiltonian form Eq.\,(\ref{eq_H_22}), we get, after setting $E_1=E_2=-E_3=-E_4=-1$,
\begin{equation}
\begin{aligned}
    H^{\mathbb{R},2+2}[\bs{n}_+,\bs{n}_-] =&\,  
        n^3_{-}\left(-n^3_{+} \Gamma_{33} +
        n^2_{+} \Gamma_{31} 
        -n^1_{+} \Gamma_{10}\right) \\
         +&\,
        n^1_{-} \left(+ n^3_{+} \Gamma_{13} 
        -n^2_{+} \Gamma_{11} 
        -n^1_{+} \Gamma_{30}\right) \\
         +&\,
        n^2_{-} \left(+ n^3_{+} \Gamma_{01} +
        n^2_{+} \Gamma_{03} +
        n^1_{+} \Gamma_{22} \right) \,,\\
        =&\, 
        \bs{n}_+^{\top}\cdot \underline{\Gamma} \cdot \bs{n}_-\,,
\end{aligned}
\end{equation}
with the tensor
\begin{equation}
    \underline{\Gamma} = \left(\begin{array}{rrr}
        -\Gamma_{30} & \Gamma_{22} & -\Gamma_{10} \\
        -\Gamma_{11} & \Gamma_{03} & \Gamma_{31} \\
        \Gamma_{13} & \Gamma_{01} & -\Gamma_{33}
    \end{array}\right)\,.
\end{equation}
\end{subequations}
We note that the above derivation differs from the previous ones exposed in \cite{bouhonGeometric2020,Bouhon2022braiding2} by the initial choice of the parametrization  of the $\mathsf{SO}(4)$ matrix representing the frame of eigenvectors. 

If we set
\begin{equation}
\begin{aligned}
    (\phi_+,\phi_-) &= (q_+,q_-) \,\phi_0\,,\;
    \theta = \theta_0\,,\\
    q_+,q_-&\in \mathbb{Z} \,,
\end{aligned}
\end{equation}
such that $\bs{n}_+$ ($\bs{n}_-$) wraps the sphere $\mathbb{S}^2_+$ ($\mathbb{S}^2_-$) a number of times $q_+$ ($q_-$) whenever $(\phi_0,\theta_0)$ covers one time the base sphere $\mathbb{S}^2_0$. The direct computation of the Euler two-form for the occupied and unoccupied two-band subspaces then gives
\begin{equation}
    \begin{aligned}
        \mathtt{F}_{\mathbb{S}^2_0}[\{u_1,u_2\}] &= -\dfrac{q_++q_-}{2}\sin\theta_0 \,,\\
        \mathtt{F}_{\mathbb{S}^2_0}[\{u_3,u_4\}] &= -\dfrac{q_+-q_-}{2}\sin\theta_0 \,,
    \end{aligned}
\end{equation}
and accordingly Euler classes
\begin{equation}
    \begin{aligned}
        \chi_{I,\mathbb{S}^2_0} &= -(q_++q_-)\in \mathbb{Z}\,,\\ 
        \chi_{II,\mathbb{S}^2_0} &= -(q_+-q_-)\in \mathbb{Z}\,.
    \end{aligned}
\end{equation}

\section{{\it Orientability} of Euler homotopy classes}\label{sec_orientability}

The fact that we have used the oriented Grassmannian for the Pl{\"u}cker embedding must now be corrected since the Bloch Hamiltonian are only orientable, and the phases are classified by the {\it free} homotopy set (\ie no base point), together allowing the reversal of the orientation through an adiabatic transformation (automorphism of $\pi_2$ by the action of $\pi_1$) \cite{bouhonGeometric2020}. The strict homotopy classification gives the following equivalence 
\begin{equation}
    (\chi_I,\chi_{II}) \simeq (-\chi_I,-\chi_{II}) \,.
\end{equation}
Whenever the partial gap between two bands of the same subspace (that is closed by nodal points) closes completely (\ie at every momentum), or in other words, when a band inversion takes place between these two bands such that they become fully degenerate, there is yet a further homotopy equivalence of signed Euler classes, namely 
\begin{equation}
    \chi_I \simeq -\chi_I\,,\;
    \chi_{II} \simeq -\chi_{II}\,,
\end{equation}
see \cite{Bouhon2022braiding2} for a detailed discussion. 

There only remains to pullback the base sphere $\mathbb{S}^2_0$ to the torus Brillouin zone via the map Eq.\,(\ref{eq_pullback}) such that the resulting $2+2$-Bloch Hamiltonian has an Euler topology exhaustively captures by
\begin{equation}
\begin{aligned}
    (\chi_{I,\mathbb{T}^2},\chi_{II,\mathbb{T}^2}) = (\chi_{I,\mathbb{S}^2_0},\chi_{II,\mathbb{S}^2_0})\in \mathbb{Z}^2/\sim\,, 
\end{aligned}
\end{equation}
with $\sim$ for the above homotopy equivalences.

\bibliography{References}

\end{document}